\documentclass[twocolumn]{aastex631}

\makeatletter
\let\frontmatter@title@above\relax
\makeatother

\usepackage{amsmath,amssymb,amstext}

\usepackage[all]{hypcap}
\usepackage[T1]{fontenc}
\usepackage[caption=false]{subfig}
\usepackage{booktabs}

\usepackage{xcolor}
\usepackage{graphicx}

\usepackage{amsthm}
\usepackage{hyperref}
\usepackage{etoolbox}
\usepackage{algpseudocode}

\makeatletter
\newcounter{algorithm}
\newenvironment{algorithm}[1][!htbp]{%
  \begin{figure}[#1]%
  \addtocounter{figure}{-1}%
  \refstepcounter{algorithm}%
}{%
  \end{figure}%
}
\makeatother

\shortauthors{Giron, Krief \& Steinberg}
\numberwithin{equation}{section}

\begin{document}

\title{MACE: Multigroup-Multiangle Integrated Thermal Compton Scattering Kernel}
\shorttitle{MACE: Multigroup Compton Kernel Integration}

\author[0000-0003-2234-1319]{Itamar Giron}
\affiliation{Racah Institute of Physics, The Hebrew University, 9190401 Jerusalem, Israel}
\email{itamar.giron@mail.huji.ac.il}

\author[0000-0002-7093-6467]{Menahem Krief}
\email{menahem.krief@mail.huji.ac.il}
\affiliation{Racah Institute of Physics, The Hebrew University, 9190401 Jerusalem, Israel}

\author[0000-0003-0053-0696]{Elad Steinberg}
\email{elad.steinberg@mail.huji.ac.il}
\affiliation{Racah Institute of Physics, The Hebrew University, 9190401 Jerusalem, Israel}

\begin{abstract}
  We present a comprehensive set of accurate Compton-scattering tables for use in multigroup, multiangle radiation-transport calculations. The tables contain the exact thermal Compton kernel integrated over incident-energy groups, scattered-energy groups, and angular bins. To generate them, we develop a deterministic integration method in which the pointwise kernel is evaluated with a robust numerical treatment that remains accurate across challenging regions of parameter space. The multidimensional integrals are computed using nested quadratures whose panels adapt to the Compton recoil ridge, its thermal broadening, and the nearly elastic forward-scattering boundary layer. We validate the tabulated results against published kernel profiles, angular and energy moments, mean free paths, multigroup redistribution data, and an independent Monte Carlo estimator. The resulting tables provide the radiation-transport community with accurate, reproducible, and readily usable Compton redistribution matrices and angular moments.
\end{abstract}

\section{Introduction} \label{sec:introduction}

Compton scattering is the inelastic scattering of photons by electrons, changing both the photon energy and propagation direction. The phenomenon was established experimentally by \citet{Compton1923}, and the relativistic cross section for scattering by a stationary electron was derived by \citet{Klein1929}. In a thermal plasma, electron motion allows photons to either lose or gain energy during scattering. Repeated interactions can consequently modify the radiation spectrum and transfer energy between the radiation field and the plasma.

Comptonization is important in high-energy astrophysical environments such as accretion flows and hot coronae, neutron-star atmospheres, active galactic nuclei, and gamma-ray bursts \citep{Pozdnyakov1983,Canfield1987,Peer2005,Madej2017}. Inverse Compton upscattering is also important in tidal disruption events, where disc photons can gain energy from hot coronal electrons, and in galaxy clusters, where it produces the thermal Sunyaev--Zeldovich effect \citep{Mummery2021,BIRKINSHAW199997}. It also contributes to spectral-line broadening \citep{Sazonov2000,Sarkar2019}. These applications often involve broad photon spectra and dynamically evolving matter temperatures, requiring transport models that accurately represent both upscattering and downscattering.

Frequency-dependent radiation-transport calculations commonly discretize the photon spectrum into energy groups, while angularly resolved methods further represent the scattering direction through angular bins or moments. Multigroup transport is used in modern astrophysical codes such as \textsc{rich} \citep{Giron_2026,yalinewich_rich_2015,Steinberg+15,Steinberg+16}, \textsc{quokka} \citep{Krumholz_2024}, \textsc{castro} \citep{Zhang_2013}, \textsc{athena}++ \citep{Stone2020,Jiang2022}, and \textsc{SuperNu} \citep{Wollaeger_2013}, through methods including diffusion, discrete ordinates, and implicit Monte Carlo. Compton-specific treatments have also been developed for deterministic transport and implicit Monte Carlo \citep{Keady2017,TILL2020108990,mcgraw_till_warsa_2021,MCGRAW2023111980}. These methods require accurate group-to-group redistribution coefficients and, for angularly resolved calculations, their angular dependence.

Constructing these coefficients from the exact thermal Compton kernel is numerically difficult. The kernel depends on incident energy, scattered energy, scattering angle, and electron temperature. It is sharply localized near the thermally broadened recoil ridge, can become strongly concentrated near the nearly elastic forward-scattering boundary, and may vary by many orders of magnitude within a single integration interval. Straightforward product quadrature is therefore inefficient and may fail to resolve the narrow regions that dominate the integral.

The numerical evaluation of thermal Compton redistribution has a long history. Relativistic formulations were developed by \citet{POMRANING19721047,pomraning1973equations}, while \citet{Prasad1986} and \citet{KERSHAW1986273} derived reduced representations suitable for numerical evaluation. Later studies examined integrated cross sections, moments, approximations, and alternative numerical methods \citep{SHESTAKOV1988577,BEASON1991497,Sazonov2000,Belmont2009,Madej2017,Sarkar2019,WIENKE1985555}. Fokker-Planck approximations can reduce the computational cost when the energy exchange per scattering is small \citep{WINSLOW1995262}, but full-kernel treatments are needed when individual scatterings produce substantial energy shifts \citep{MCGRAW2023111980}. In particular, \citet{Sarkar2019} derived numerically stable expressions for the exact Compton kernel in isotropic media and its energy-transfer moments, implemented together with thermal averaging in the \textsc{CSpack} library \citep{CSpack}. \citet{Acharya2021} and \citet{Acharya2022} used these exact thermal kernels to build scattering matrices for repeated-scattering calculations, tabulating the matrix elements over temperature and interpolating between them. These treatments describe the kernel integrated over the scattering angle.

In this work, we develop a robust deterministic procedure for integrating the exact thermal Compton kernel over arbitrary incident-energy groups, scattered-energy groups, and scattering-angle bins. The multidimensional integrals are evaluated as nested one-dimensional quadratures whose panels track the kernel's sharply localized structure. The same framework is applied to the partial temperature derivative of the kernel and to kernel-weighted moments containing general smooth multipliers. Unlike these angle-integrated treatments, the resulting coefficients resolve the scattering angle, average the kernel over each incident-energy group with a prescribed spectral weight, and include the temperature derivatives needed by implicit transport schemes. The multipliers may depend on the scattering angle as well as on the incident and scattered energies. We validate the individual integration stages against published kernel profiles, angular moments, energy-transfer moments, mean free paths, and multigroup redistribution data. The complete calculation is further tested against an independent Monte Carlo estimator that samples the underlying scattering process directly. 

The method presented here has already been incorporated into a multigroup diffusion treatment of Compton scattering in \textsc{rich}\footnote{\url{https://gitlab.com/eladtan/RICH}} \citep{Giron_2026_compton} and forms a fundamental component of the ongoing development of \textsc{rich} for radiation-hydrodynamic simulations of tidal disruption events.

The integrator developed in this work is publicly available as open-source software, and a companion website provides pre-computed scattering matrices. Links to both are given in Section~\ref{sec:software}. To our knowledge, this integrator is the only open-source code that provides multigroup-multiangle thermal Compton coefficients.

The remainder of this paper is organized as follows. Section~\ref{sec:compton_kernel} defines the thermal Compton kernel and the interval integrals considered in this work. Section~\ref{sec:stable_kernel_evaluation} describes the stable pointwise evaluation of the closed-form kernel. Section~\ref{sec:mg_ma_quadrature} presents the nested multigroup-multiangle quadrature and validates each integration stage against published results. Section~\ref{sec:results} presents the resulting redistribution matrices, kernel-weighted moments, and comparisons with the independent Monte Carlo estimator. In Section~\ref{sec:conclusions} we summarize.

\section{The Compton Scattering Kernel} \label{sec:compton_kernel}

\begin{figure}[!htbp]
  \centering
  \includegraphics[width=\columnwidth]{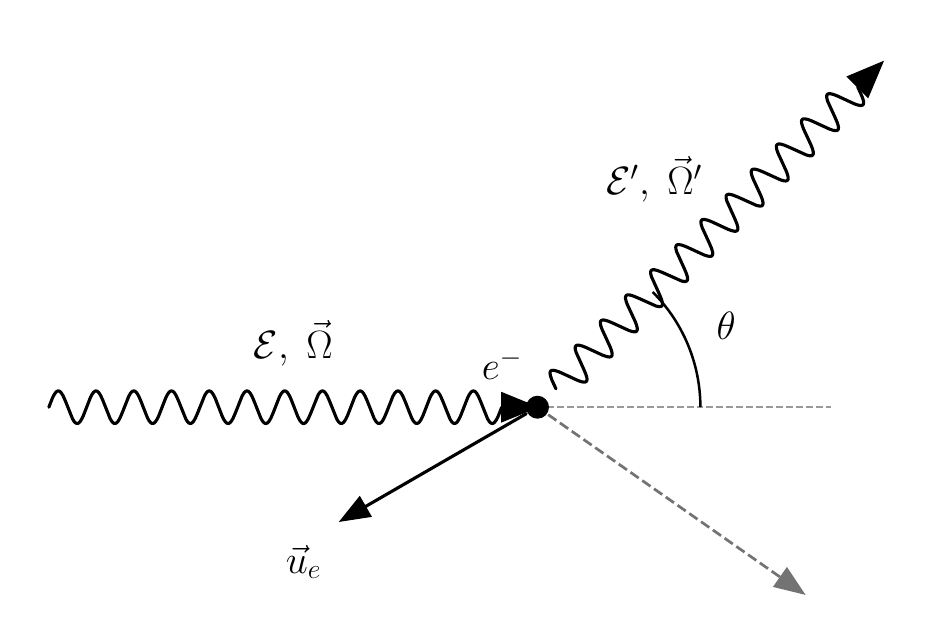}
  \caption{Schematic of Compton scattering. An incident photon with energy $\mathcal{E}$ and direction $\vec{\Omega}$ scatters off a free electron moving with velocity $\vec{u}_e$, producing a scattered photon with energy $\mathcal{E}'$ and direction $\vec{\Omega}'$. The scattering angle $\theta$ is the angle between the incident and scattered photon directions.}
  \label{fig:compton_scattering_schematic}
\end{figure}

Compton scattering is the inelastic scattering of a photon by a free electron, the basic scattering geometry is illustrated in Figure~\ref{fig:compton_scattering_schematic}.
We denote the incident photon state by its energy and direction
$(\mathcal{E},\vec{\Omega})$, and the scattered photon state by
$(\mathcal{E}',\vec{\Omega}')$. Primed quantities always refer to
the scattered photon, while unprimed quantities refer to the incident
photon. Quantities measured in the electron
rest frame carry a subscript $0$. For example, the lab-frame scattering-angle cosine is
\begin{equation}
    \xi \equiv \cos\theta = \vec{\Omega}\cdot\vec{\Omega}' ,
    \label{eq:xi_def}
\end{equation}
and the corresponding rest-frame cosine is therefore
\begin{equation}
    \xi_0 \equiv \vec{\Omega}_0\cdot\vec{\Omega}'_0 .
    \label{eq:xi0_def}
\end{equation}
In the electron rest frame, Compton scattering by a free electron is
described by the Klein--Nishina formula
\citep{Klein1929,pomraning1973equations,POMRANING19721047}. Written
differentially in both scattered direction and scattered photon energy
\begin{multline}
    \frac{d\sigma_{KN}}{d\Omega'_0 d\mathcal{E}'_0}
    \left(
    \mathcal{E}_0\rightarrow\mathcal{E}'_0,
    \vec{\Omega}_0\rightarrow\vec{\Omega}'_0
    \right)
    = \\
    \frac{r_e^2}{2}\left(\frac{\mathcal{E}'_0}{\mathcal{E}_0}\right)^2
    \left[
        \frac{\mathcal{E}'_0}{\mathcal{E}_0}
        +
        \frac{\mathcal{E}_0}{\mathcal{E}'_0}
        -
        (1-\xi_0^2)
    \right] \\
    \times\delta\left(\mathcal{E}'_0-\mathcal{E}'_{C,0}\right)
    \label{eq:KN_rest}
\end{multline}
where $r_e \approx 2.81\times10^{-13}\,\mathrm{cm}$ is the classical electron radius and
\begin{equation}
    \mathcal{E}'_{C,0}
    =
    \frac{\mathcal{E}_0}
    {1+\frac{\mathcal{E}_0}{m_e c^2}(1-\xi_0)}
    \label{eq:compton_recoil_rest}
\end{equation}
is the scattered photon energy fixed by energy and momentum conservation,
with $m_e$ the electron rest mass and $c$ the speed of light~\citep{Compton1923}. 
The rest-frame cross section in Eq.~\eqref{eq:KN_rest} can be
transformed to lab-frame variables for an electron moving with velocity
$\vec{u}_e$ \citep{pomraning1973equations}:
\begin{multline}
    \frac{d\sigma_{KN}}{d\Omega' d\mathcal{E}'}
    \left(
    \mathcal{E}\rightarrow\mathcal{E}',
    \vec{\Omega}\rightarrow\vec{\Omega}'
    \,;\,\vec{u}_e
    \right)
    = \\
    \frac{r_e^2}{2\lambda\gamma\mathcal{E}}
    \left[
        1+
        \left(
            1-\frac{1-\xi}{\lambda^2DD'}
        \right)^2
        +
        \frac{\gamma\gamma'}{\lambda^2DD'}(1-\xi)^2
    \right] \\
    \times
    \delta\!\left(
        \xi-1+\lambda\frac{D}{\gamma'}-\lambda\frac{D'}{\gamma}
    \right).
    \label{eq:KN_lab}
\end{multline}
Here $\gamma$, $\gamma'$, and $\lambda$ are the dimensionless incident
photon energy, dimensionless scattered photon energy, and electron
Lorentz factor, respectively:
\begin{equation}
    \gamma=\frac{\mathcal{E}}{m_e c^2},
    \qquad
    \gamma'=\frac{\mathcal{E}'}{m_e c^2},
    \qquad
    \lambda=\frac{1}{\sqrt{1-\beta^2}}.
    \label{eq:def_gamma}
\end{equation}
The quantities $D$ and $D'$ are the Doppler factors associated with the
incident and scattered photon directions:
\begin{equation}
    D=1-\vec{\Omega}\cdot\vec{\beta},
    \qquad
    D'=1-\vec{\Omega}'\cdot\vec{\beta},
    \qquad
    \vec{\beta}=\frac{\vec{u}_e}{c}.
    \label{eq:def_D}
\end{equation}
The Compton scattering kernel is the macroscopic differential cross section averaged over a thermal distribution of electrons (units of $\mathrm{cm}^{-1}\,\mathrm{keV}^{-1}\,\mathrm{sr}^{-1}$). It is obtained by averaging Eq.~\eqref{eq:KN_lab} over the velocity distribution for
a relativistic thermal gas of free electrons~\citep{KERSHAW1986273,BEASON1991497,SHESTAKOV1988577,Sarkar2019}:
\begin{multline}
    \mathcal{K}(\mathcal{E}\rightarrow\mathcal{E}',\xi;\tau,N_e)
    = \\
    N_e
    \int
    \mathrm{MJ}(\vec{u}_e,\tau)
    \frac{d\sigma_{KN}}{d\Omega' d\mathcal{E}'}
    \left(
    \mathcal{E}\rightarrow\mathcal{E}',
    \vec{\Omega}\rightarrow\vec{\Omega}'
    \,;\,\vec{u}_e
    \right)
    d^3u_e
    \label{eq:def_compton_kernel}
\end{multline}
Here $N_e$ is the electron number density. The electron velocities are distributed according to the Maxwell--Juttner
distribution \citep{Juttner1911},
\begin{equation}
    \mathrm{MJ}(\vec{u}_e,\tau)
    =
    \frac{\lambda^5 \exp(-\lambda/\tau)}
    {4\pi c^3 \tau K_2(1/\tau)},
    \qquad
    \tau \equiv \frac{k_{\rm B}T}{m_e c^2}.
    \label{eq:def_MJ}
\end{equation}
Here $\tau$ is the dimensionless electron temperature, $T$ is the
electron gas temperature, $k_{\rm B}$ is Boltzmann's constant, and $K_2$ is the
modified Bessel function of the second kind. Since the electron
distribution is isotropic, the thermally averaged kernel depends on the photon directions only through the scattering angle $\xi$. 

\begin{figure}[!htbp]
  \centering
  \includegraphics[width=\columnwidth]{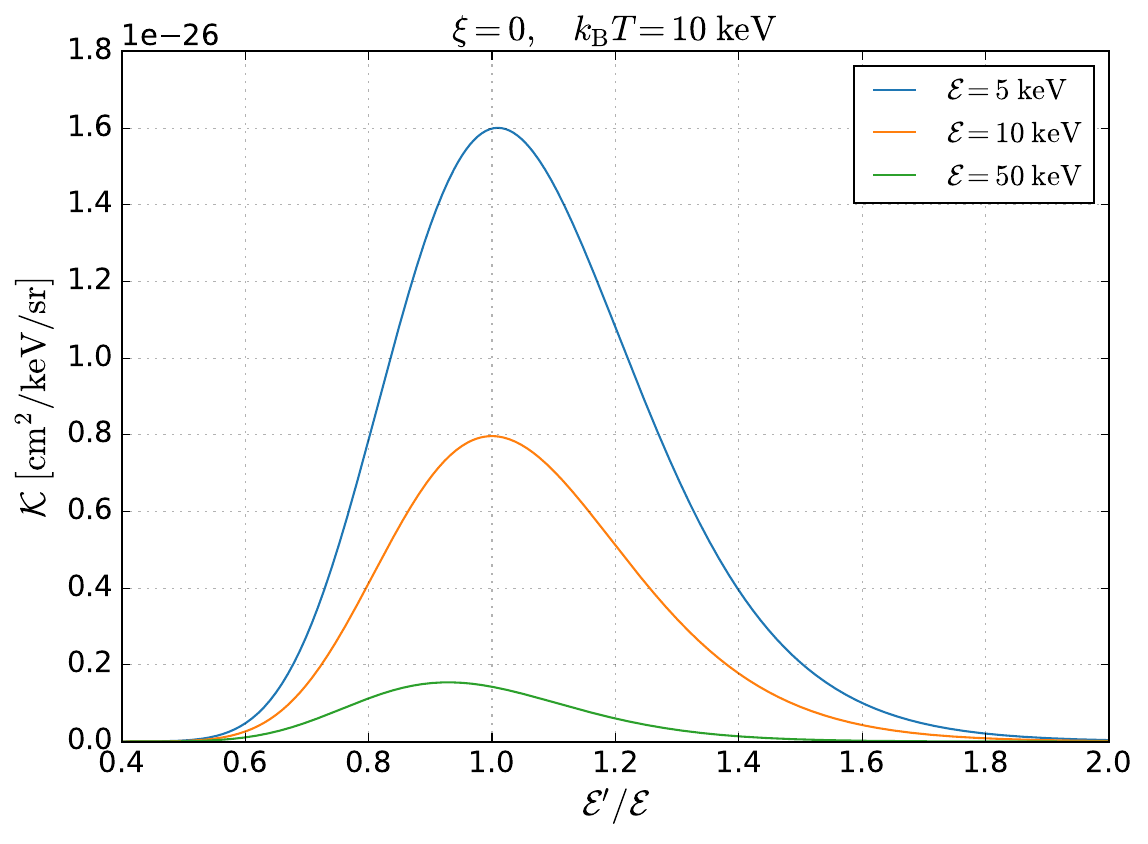}
  \caption{Energy profile of the Compton scattering kernel $\mathcal{K}$ at $k_{\rm B}T=10$\,keV and fixed scattering angle $\xi=0$ for three incident photon energies.}
  \label{fig:kernel_energy_profile}
\end{figure}

\begin{figure}[!htbp]
  \centering
  \includegraphics[width=\columnwidth]{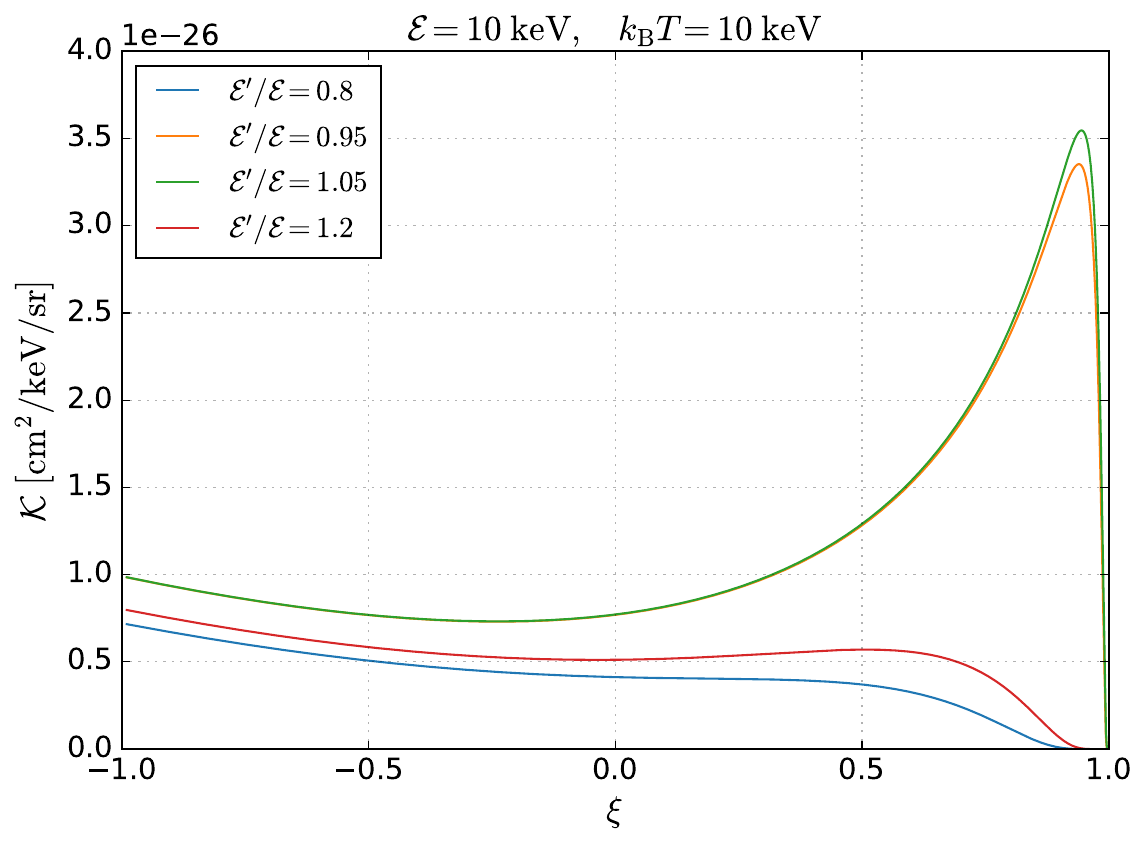}
  \caption{Angular profile of the Compton scattering kernel $\mathcal{K}$ at $k_{\rm B}T=10$\,keV and fixed $\mathcal{E}=10$\,keV for four outgoing-to-incoming energy ratios.}
  \label{fig:kernel_angle_profile}
\end{figure}

Throughout this paper we present the microscopic kernel in units of $\mathrm{cm}^{2}\,\mathrm{keV}^{-1}\,\mathrm{sr}^{-1}$, obtained by dropping the electron number density prefactor from Eq.~\eqref{eq:def_compton_kernel}. We therefore omit the explicit
dependence on $N_e$ from the notation unless stated otherwise. Figures~\ref{fig:kernel_energy_profile} and~\ref{fig:kernel_angle_profile}
show representative profiles of the kernel $\mathcal{K}$ at
$k_{\rm B}T=10\,{\rm keV}$, illustrating both the redistribution in scattered
photon energy and the dependence on scattering angle for several
representative photon energies.

The aim of this work is to construct a robust deterministic procedure
for integrating the Compton kernel $\mathcal{K}$, and its temperature
derivative $\partial\mathcal{K}/\partial T$, over prescribed
incident-energy, scattered-energy, and scattering-angle intervals.
To make this precise, for a quantity
$X(\mathcal{E},\mathcal{E}',\xi)$, we define the angular
kernel-weighted integral over an interval $\Delta\xi$ by
\begin{equation}
    \mathcal{I}_{\Delta\xi}[X](\mathcal{E}, \mathcal{E}';\tau)
    =
    \int_{\Delta\xi}
    X(\mathcal{E},\mathcal{E}',\xi)
    \mathcal{K}(\mathcal{E}\rightarrow\mathcal{E}',\xi;\tau)
    \,d\xi .
    \label{eq:I_xi}
\end{equation}
Here $X$ can represent any additional factor to be included in the
integral. The same integration scheme should apply without modification for any
sufficiently smooth $X$ that does not introduce features sharper than
those of the kernel itself.
We extend the same notation to scattered-energy and incident-energy
intervals. For a scattered-energy interval $\Delta\mathcal{E}'$,
\begin{multline}
    \mathcal{I}_{\Delta\mathcal{E}'\Delta\xi}[X](\mathcal{E};\tau)
    = \\
    \int_{\Delta\mathcal{E}'}
    \int_{\Delta\xi}
    X(\mathcal{E},\mathcal{E}',\xi)
    \mathcal{K}(\mathcal{E}\rightarrow\mathcal{E}',\xi;\tau)
    \,d\xi\,d\mathcal{E}' .
    \label{eq:I_Ep_xi}
\end{multline}
For an incident-energy interval $\Delta\mathcal{E}$,
\begin{multline}
    \mathcal{I}_{\Delta\mathcal{E}\Delta\mathcal{E}'\Delta\xi}[X](\tau)
    = \\
    \int_{\Delta\mathcal{E}}
    \int_{\Delta\mathcal{E}'}
    \int_{\Delta\xi}
    X(\mathcal{E},\mathcal{E}',\xi)
    \mathcal{K}(\mathcal{E}\rightarrow\mathcal{E}',\xi;\tau)
    \,d\xi\,d\mathcal{E}'\,d\mathcal{E}.
    \label{eq:I_E_Ep_xi}
\end{multline}
For integrals over the temperature derivative, we use the same notation
with a superscript $\partial T$. For example,
\begin{multline}
    \mathcal{I}^{\partial T}_{\Delta\mathcal{E}\Delta\mathcal{E}'\Delta\xi}[X](\tau)
    = \\
    \int_{\Delta\mathcal{E}}
    \int_{\Delta\mathcal{E}'}
    \int_{\Delta\xi}
    X(\mathcal{E},\mathcal{E}',\xi)
    \frac{\partial\mathcal{K}}{\partial T}
    (\mathcal{E}\rightarrow\mathcal{E}',\xi;\tau)
    \,d\xi\,d\mathcal{E}'\,d\mathcal{E}.
    \label{eq:I_der_E_Ep_xi}
\end{multline}
Analogous definitions apply when the integration is only over
$\Delta\xi$, or over $\Delta\mathcal{E}'$ and $\Delta\xi$.

\begin{figure}[!htbp]
  \centering
  \includegraphics[width=\columnwidth]{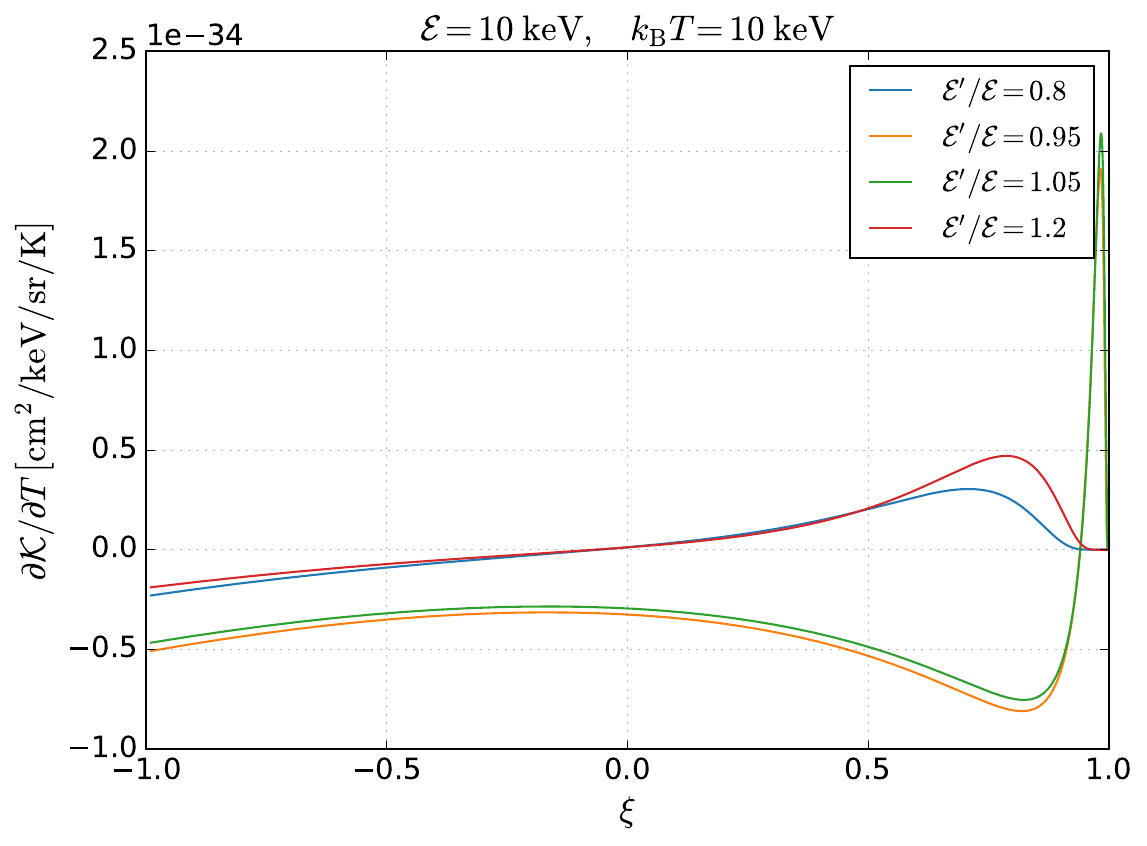}
  \caption{Angular profile of the temperature derivative $\partial\mathcal{K}/\partial T$ of the Compton kernel at $k_{\rm B}T=10$\,keV and fixed $\mathcal{E}=10$\,keV for four outgoing-to-incoming energy ratios.}
  \label{fig:kernel_angle_deriv}
\end{figure}

In the numerical implementation, the integral in
Eq.~\eqref{eq:def_compton_kernel} is not evaluated directly. Instead,
we evaluate the pointwise kernel using the reduced closed-form
expressions derived by \citet{KERSHAW1986273}. These expressions provide
two complementary representations of the same kernel: the power-series
form given in Appendix~\ref{sec:power_series_appendix},
Eq.~\eqref{eq:power_series}, and the asymptotic-series form given in
Appendix~\ref{sec:asymptotic_series_appendix},
Eq.~\eqref{eq:asymptotic_series}. The corresponding temperature
derivatives are given by Eqs.~\eqref{eq:dpower_series_dT}--\eqref{eq:dP_pm_dtau}
for the power series and by Eq.~\eqref{eq:dasymptotic_dT} for the
asymptotic series.
We also provide an explicit fifth-order approximation whose derivation, accuracy validation, and temperature derivative are given in Appendix~\ref{sec:g5_kernel_appendix}.
Figures~\ref{fig:kernel_energy_deriv} and~\ref{fig:kernel_angle_deriv}
show representative profiles of the microscopic temperature derivative
$\partial\mathcal{K}/\partial T$ at $k_{\rm B}T=10\,{\rm keV}$.
\begin{figure}[!htbp]
  \centering
  \includegraphics[width=\columnwidth]{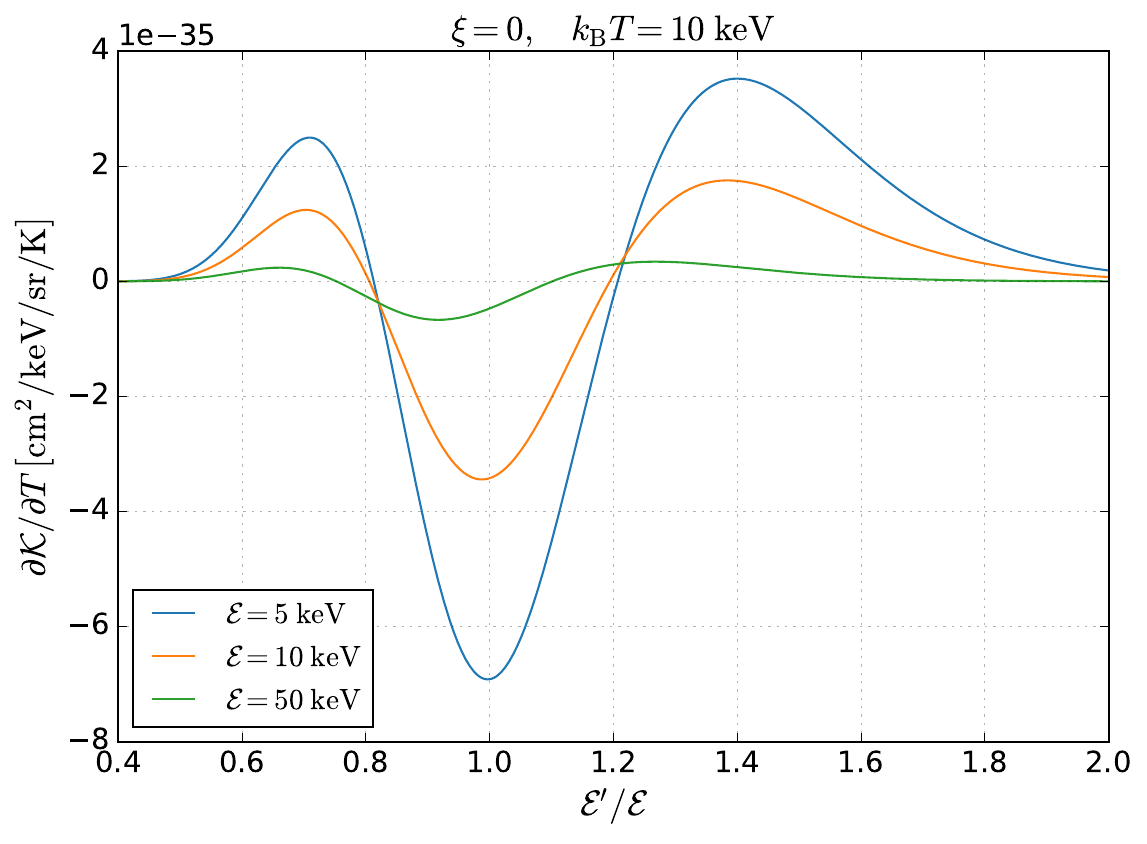}
  \caption{Energy profile of the temperature derivative $\partial\mathcal{K}/\partial T$ of the Compton kernel at $k_{\rm B}T=10$\,keV and fixed scattering angle $\xi=0$ for three incident photon energies.}
  \label{fig:kernel_energy_deriv}
\end{figure}

\section{Stable Evaluation of the Closed-Form Kernel}
\label{sec:stable_kernel_evaluation}

\subsection{Special-Function Evaluation}
\label{sec:special_function_evaluation}

The reduced kernel expressions used by the pointwise solver contain
special-function factors that must be evaluated in scaled form to avoid
overflow and underflow. The normalization factor
$\exp(-\lambda_+/\tau)/K_2(1/\tau)$ in Eq.~\eqref{eq:def_Sigma_E}
underflows at small $\tau$ if $K_2(1/\tau)$ is evaluated directly. We
therefore use the exponentially scaled modified Bessel function
\begin{equation}
  \widetilde{K}_n(x) = e^x K_n(x),
  \qquad
  \frac{e^{-\lambda_+/\tau}}{K_2(1/\tau)}
  =
  \frac{e^{-(\lambda_+ - 1)/\tau}}{\widetilde{K}_2(1/\tau)}
  \label{eq:scaled_K_2}
\end{equation}
The same scaling is used for the Bessel-function ratio that enters the
temperature derivative,
\begin{equation}
  \kappa(\tau)
  =
  \frac{K_1(\tau^{-1})}{K_2(\tau^{-1})}
  =
  \frac{\widetilde{K}_1(\tau^{-1})}
       {\widetilde{K}_2(\tau^{-1})} .
  \label{eq:scaled_kappa}
\end{equation}
For $x<50$ we evaluate $K_1(x)$ and $K_2(x)$ using
\citet{boost} and form $\widetilde{K}_n(x)$ by multiplication by $e^x$.
For $x\geq 50$, we evaluate the scaled functions directly using a
five-term large-$x$ asymptotic expansion
\citep{NIST:DLMF_Knu_Hankel}.\footnote{The transition point $x=50$ was
chosen so that the relative error of the five-term asymptotic evaluation
of the scaled Bessel functions remains below $10^{-8}$ in the regime
where it is used.}
The generalized exponential integrals $E_n(x)$ appearing in
Eqs.~\eqref{eq:power_series} and \eqref{eq:dpower_series_dT} are also evaluated in scaled form
\begin{equation}
  \widetilde{E}_n(x) = e^x E_n(x)
  \label{eq:scaled_E_n}
\end{equation}
The scaled exponential integrals are calculated primarily by recurrence \citep{NIST:DLMF_Ep_recurrence}. 
When the recurrence must be initialized or restarted, the required $\widetilde{E}_n(x)$
is evaluated directly using a continued fraction
\citep{NIST:DLMF_Ep_cf}. For the special case $n=1$ and $x<4$, $\widetilde{E}_1(x)$
is instead evaluated from its series expansion
\citep{NIST:DLMF_E1_series}. The recurrence is restarted whenever the
estimated amplification error becomes too large.

\subsection{Cancellation Control}
\label{sec:cancellation_control}

The power-series representation is numerically delicate because the
kernel is obtained from the combination $\Psi+P_+-P_-$ in
Eq.~\eqref{eq:power_series}. In parts of the kinematic domain, $P_+$
and $P_-$ are individually much larger than their difference, and the
remaining difference can then partially cancel against $\Psi$. The
temperature derivative in Eq.~\eqref{eq:dpower_series_dT} contains the
same cancellation channel, together with an additional cancellation in
$\partial P_+/\partial\tau-\partial P_-/\partial\tau$. To decide when
these cancellations make the double-precision result unreliable, we
monitor the roundoff-error estimates derived in
Appendix~\ref{sec:power_series_roundoff}, denoted
$\epsilon_{\rm ro}$ for the kernel and
$\epsilon_{\rm ro}^{\partial T}$ for its temperature derivative.

The asymptotic representation has a different cancellation channel. In
the raw form of Eq.~\eqref{eq:asymptotic_series}, the $+$ and $-$
contributions are accumulated separately and combined only at the end.
At ultra-low photon energies, the factor
$G$ in Eq.~\eqref{eq:def_G} is dominated by the term
$2/(\gamma\gamma'a^2)$ and can become very large, while
$\alpha_+ \simeq \alpha_-$ and $\chi_+ \simeq \chi_-$. The large
$G$-dependent parts of the two sign contributions therefore nearly
cancel. To avoid forming two large partial sums whose difference is
small, we evaluate a rearranged per-order form derived in
Appendix~\ref{sec:grouped_rearrangement}. In this form, the two signs
are grouped at each order before accumulation, so the large $G$ factor
enters only through the controlled differences $D_n$ and $F_{n+1}$ in
Eqs.~\eqref{eq:def_Dn}--\eqref{eq:def_Fn}. The roundoff-error
estimates for the rearranged series, also denoted
$\epsilon_{\rm ro}$ and $\epsilon_{\rm ro}^{\partial T}$, are derived in
Appendix~\ref{sec:asymptotic_roundoff}.

\subsection{Error Estimates}
\label{sec:error_estimates}

We separate the numerical error into two contributions. The first is the
roundoff error discussed above, which comes from subtracting nearly equal
quantities in the power-series and asymptotic representations. The
second is the truncation error, which comes from retaining only a finite
number of terms in either series. We estimate both contributions
for each attempted evaluation and use them to decide whether the result
can be accepted or whether the same method must be retried with double-double
precision.\footnote{We use the \texttt{doubledouble} C++ type of \citet{weckesser_doubledouble} for this fallback.}

For the power-series, the truncation estimate is based on the
tail of the convergent series in Eq.~\eqref{eq:power_series}.
For the temperature derivative, the tail estimate accounts for the
additional prefactor derivative term in Eq.~\eqref{eq:dpower_series_dT}. This gives the
power-series truncation indicators $\epsilon_{\rm tr}$ and
$\epsilon_{\rm tr}^{\partial T}$ for the kernel and its temperature
derivative, respectively.

For the asymptotic series, the truncation estimate is based on optimal
truncation rather than convergence. The expansion in
Eq.~\eqref{eq:asymptotic_rearranged} is asymptotic, so its terms
initially decrease but eventually grow. At each order $n$, we
record the magnitude of the rearranged combined contribution,
$|T_n^+ + T_n^-|$, together with the corresponding partial sum. Once
the term magnitudes begin to increase, we return the partial sum
associated with the smallest retained term. This gives the asymptotic
truncation indicators $\epsilon_{\rm tr}$ and $\epsilon_{\rm tr}^{\partial T}$ for the
kernel and temperature derivative, respectively. Both series are accepted as converged only after two consecutive contributions fall below a relative tolerance of $10^{-8}$, because the $+$ and $-$ terms of a single order can nearly cancel while the series is still far from converged.

The reported relative error estimate combines the truncation and
roundoff contributions from whichever backend (power series or
asymptotic) produced the result. For the kernel,
\begin{equation}
  \epsilon_{\rm rel}
  =
  \max\!\left(\epsilon_{\rm tr},\epsilon_{\rm ro}\right),
  \label{eq:kernel_error_estimate}
\end{equation}
and for the temperature derivative as
\begin{equation}
  \epsilon_{\rm rel}^{\partial T}
  =
  \max\!\left(\epsilon_{\rm tr}^{\partial T},\epsilon_{\rm ro}^{\partial T}\right).
  \label{eq:kernel_derivative_error_estimate}
\end{equation}

\subsection{Kernel Dispatch Logic}
\label{sec:kernel_dispatch_logic}

\begin{figure}[!htbp]
  \centering
  \includegraphics[width=\columnwidth]{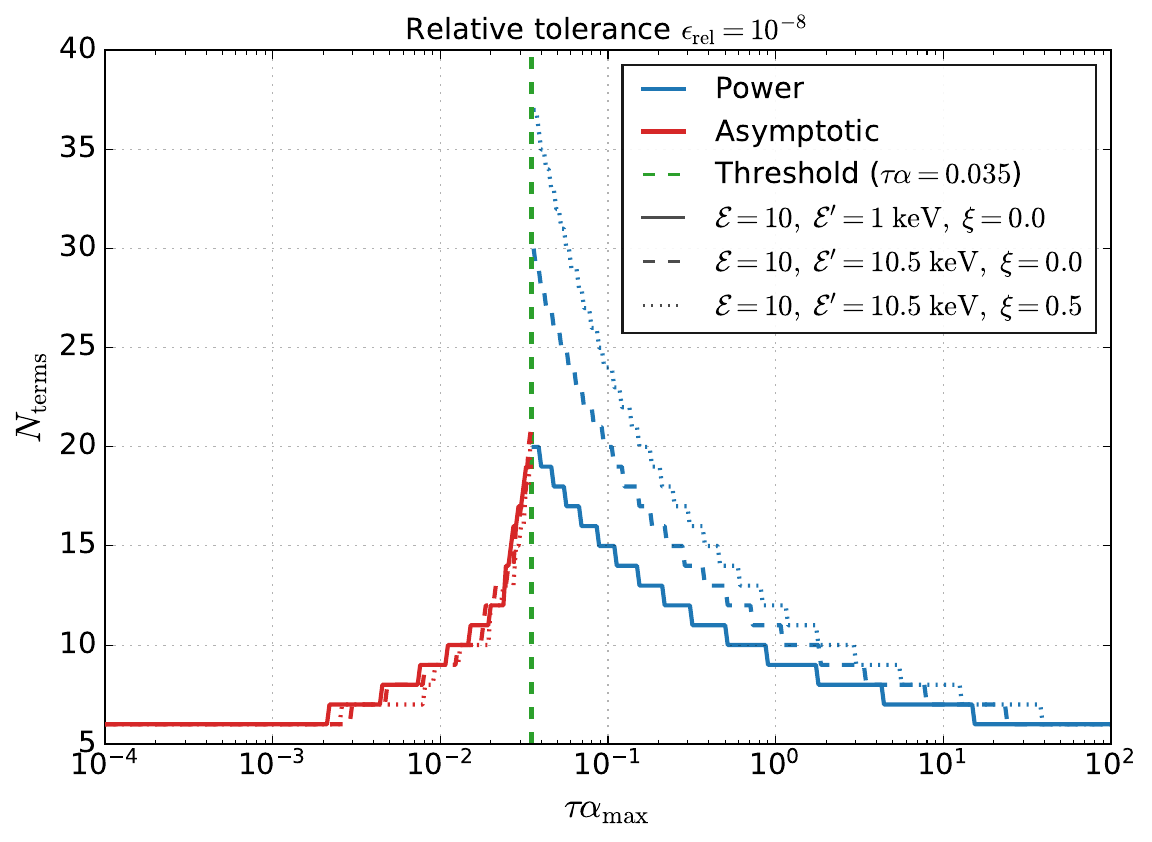}
  \caption{Number of series terms required as a function of
  $\tau\alpha_{\max}$ at a relative tolerance
  $\epsilon_{\rm rel}=10^{-8}$ for the power series
  \eqref{eq:power_series} (blue) and the asymptotic series
  \eqref{eq:asymptotic_rearranged} (red). Each line style corresponds
  to a different representative configuration. The vertical
  dashed line marks the dispatch threshold $\tau\alpha_{\max}=0.035$.}
  \label{fig:series_terms_vs_tau_alpha}
\end{figure}

The choice between the two closed-form representations is controlled by
the dimensionless parameter
$\tau\alpha_{\max}=\tau\max(\alpha_+,\alpha_-)$ where $\alpha_\pm$
are defined in Eq.~\eqref{eq:def_rho_pm_omega2}. When
$\tau\alpha_{\max}$ is small, the asymptotic representation reaches its
useful truncation in only a few terms, while the power series requires
increasingly many terms. When $\tau\alpha_{\max}$ is large, the
asymptotic expansion is no longer accurate and the convergent power series
is used instead. Figure~\ref{fig:series_terms_vs_tau_alpha} compares the
number of terms required by the two representations at a relative
tolerance of $10^{-8}$. As $\tau\alpha_{\max}$ increases, the
asymptotic series (in its useful asymptotic regime) requires more
terms while the power series requires fewer. The crossover occurs near
$\tau\alpha_{\max}\approx0.035$, and this value is used as the dispatch
threshold in Algorithm~\ref{alg:solver_dispatch}.
\begin{algorithm}
  \hrule\smallskip
  \begin{algorithmic}[1]
  \If{$\tau\alpha_{\max} < 0.035$} \Comment{asymptotic regime}
    \State $\mathcal{K},\,\epsilon_{\rm rel} \gets$ asymptotic series in double precision
    \State \textbf{if} $\epsilon_{\rm rel} < \epsilon_{\rm double}$ \textbf{return} $\mathcal{K}$
    \State $\mathcal{K},\,\epsilon_{\rm rel} \gets$ asymptotic series in double-double precision
    \State \textbf{if} $\epsilon_{\rm rel} < \epsilon_{\rm dd}$ \textbf{return} $\mathcal{K}$
  \Else \Comment{power-series regime}
    \State $\mathcal{K},\,\epsilon_{\rm rel} \gets$ power series in double precision
    \State \textbf{if} $\epsilon_{\rm rel} < \epsilon_{\rm double}$ \textbf{return} $\mathcal{K}$
    \State $\mathcal{K},\,\epsilon_{\rm rel} \gets$ power series in double-double precision
    \State \textbf{if} $\epsilon_{\rm rel} < \epsilon_{\rm dd}$ \textbf{return} $\mathcal{K}$
  \EndIf
  \State \textbf{error:} kernel evaluation failure
  \end{algorithmic}
  \smallskip\hrule
  \caption{Kernel solver dispatch. Each backend returns
  $\mathcal{K}$ together with its estimated relative error
  $\epsilon_{\rm rel}$. The same dispatch applies to the temperature
  derivative $\partial\mathcal{K}/\partial T$ with
  $\epsilon_{\rm rel}^{\partial T}$ in place of $\epsilon_{\rm rel}$.
  The dispatch threshold $\tau\alpha_{\max}=0.035$ corresponds to
  the crossover in Figure~\ref{fig:series_terms_vs_tau_alpha}.
  The values of $\epsilon_{\rm double}$ and
  $\epsilon_{\rm dd}$ are listed in
  Table~\ref{tab:dispatch_tolerances}.}
  \label{alg:solver_dispatch}
\end{algorithm}

\section{Multigroup-Multiangle Quadrature of the Kernel} \label{sec:mg_ma_quadrature}

This section describes the quadrature of the kernel
$\mathcal{K}(\mathcal{E}\to\mathcal{E}',\xi;\tau)$ and its
temperature derivative $\partial\mathcal{K}/\partial T$ over the
interval integrals defined in
Eqs.~\eqref{eq:I_xi}--\eqref{eq:I_der_E_Ep_xi}, on a
multigroup-multiangle grid.
We divide the photon-energy domain into $G$ groups, with boundaries at 
\begin{equation}
  \mathcal{E}_{1/2}
  <
  \mathcal{E}_{3/2}
  <
  \cdots
  <
  \mathcal{E}_{G+1/2},
  \label{eq:groups_boundaries}
\end{equation}
and denote group $g$ by
\begin{equation}
  \Delta\mathcal{E}_g
  =
  [\mathcal{E}_{g-1/2},\mathcal{E}_{g+1/2}].
  \label{eq:energy_group_definition}
\end{equation}
The scattering-angle cosine is divided into $M$ bins, with boundaries at 
\begin{equation}
  -1
  =
  \xi_{1/2}
  <
  \xi_{3/2}
  <
  \cdots
  <
  \xi_{M+1/2}
  =
  1,
  \label{eq:angle_boundaries}
\end{equation}
with
\begin{equation}
  \Delta\xi_i
  =
  [\xi_{i-1/2},\xi_{i+1/2}].
  \label{eq:angle_bin_definition}
\end{equation}
In this work the angular bins are always uniformly spaced,
$\xi_{i+1/2}-\xi_{i-1/2}=2/M$ though in general they don't have to be.
For a group-to-group transition $g\to g'$ and an angular bin
$\Delta\xi_i$, the kernel integrals defined in
Eqs.~\eqref{eq:I_xi}--\eqref{eq:I_E_Ep_xi} are, in increasing
dimensionality,
\begin{equation}
  \mathcal{I}_{\Delta\xi_i}[X],
  \qquad
  \mathcal{I}_{\Delta\mathcal{E}_{g'}\Delta\xi_i}[X],
  \qquad
  \mathcal{I}_{\Delta\mathcal{E}_g
  \Delta\mathcal{E}_{g'}
  \Delta\xi_i}[X].
  \label{eq:mgma_kernel_integral_objects}
\end{equation}
The corresponding temperature-derivative integrals
(cf.\ Eq.~\eqref{eq:I_der_E_Ep_xi}) are
\begin{equation}
  \mathcal{I}^{\partial T}_{\Delta\xi_i}[X],
  \qquad
  \mathcal{I}^{\partial T}_{\Delta\mathcal{E}_{g'}\Delta\xi_i}[X],
  \qquad
  \mathcal{I}^{\partial T}_{\Delta\mathcal{E}_g
  \Delta\mathcal{E}_{g'}
  \Delta\xi_i}[X].
  \label{eq:mgma_kernel_derivative_integral_objects}
\end{equation}
The localized and rapidly varying structure of the kernel makes these integrals challenging.
A simple product quadrature would therefore
be inefficient and, in narrow regimes, unreliable. We instead evaluate
the integrals as nested one-dimensional quadratures whose panels are
chosen from the local structure of the kernel. For fixed
$(\mathcal{E},\mathcal{E}')$, we first construct the angular quadrature
over $\xi$. Then for a fixed $\mathcal{E}$ and angle bin $\Delta\xi_i$, we
construct the scattered-energy quadrature over $\mathcal{E}'$. The
remaining incident-energy integral is performed over the source group
$\Delta\mathcal{E}_g$.

\subsection{\texorpdfstring{$\xi$}{xi} Integration}
\label{sec:xi_integration}

\begin{figure*}[!htbp]
  \centering
  \subfloat[$\mathcal{E}'/\mathcal{E}=0.95$, $k_{\rm B}T=0.01$\,keV]{%
    \includegraphics[width=0.49\textwidth]{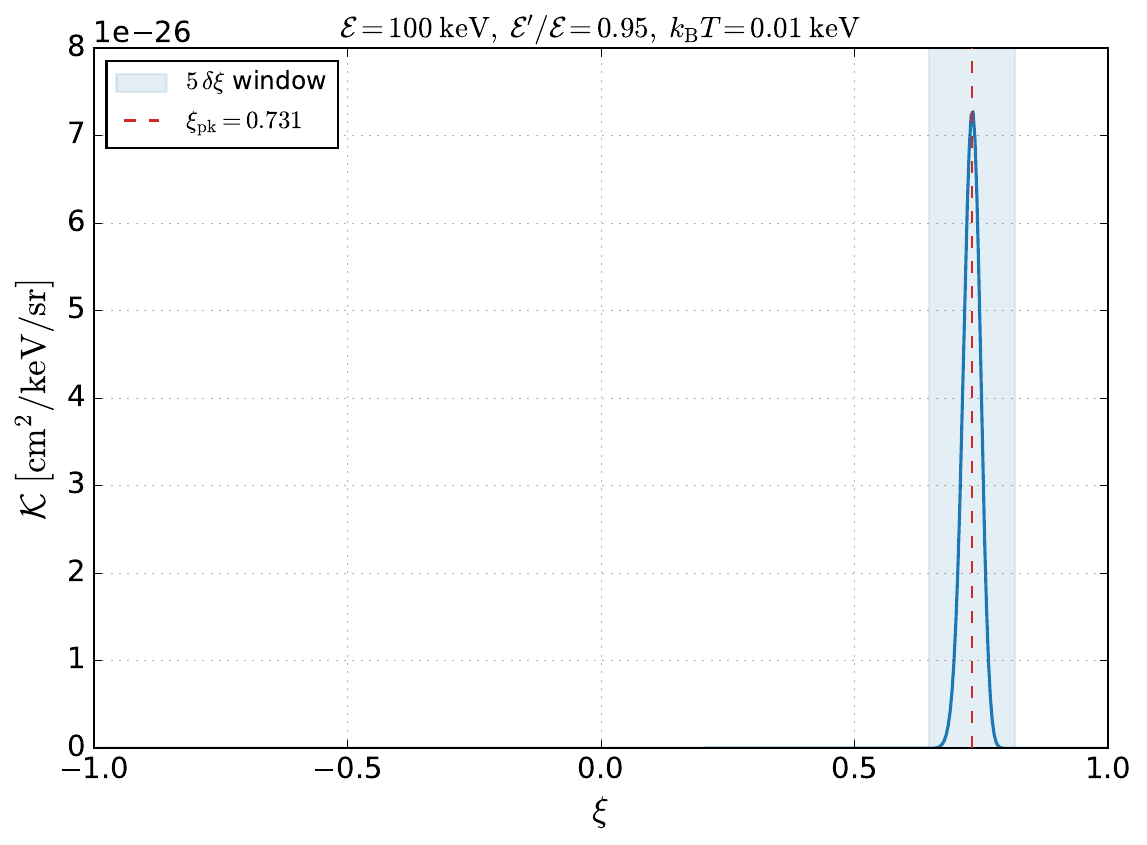}}
  \hfill
  \subfloat[$\mathcal{E}'/\mathcal{E}=1.05$, $k_{\rm B}T=1.0$\,keV]{%
    \includegraphics[width=0.49\textwidth]{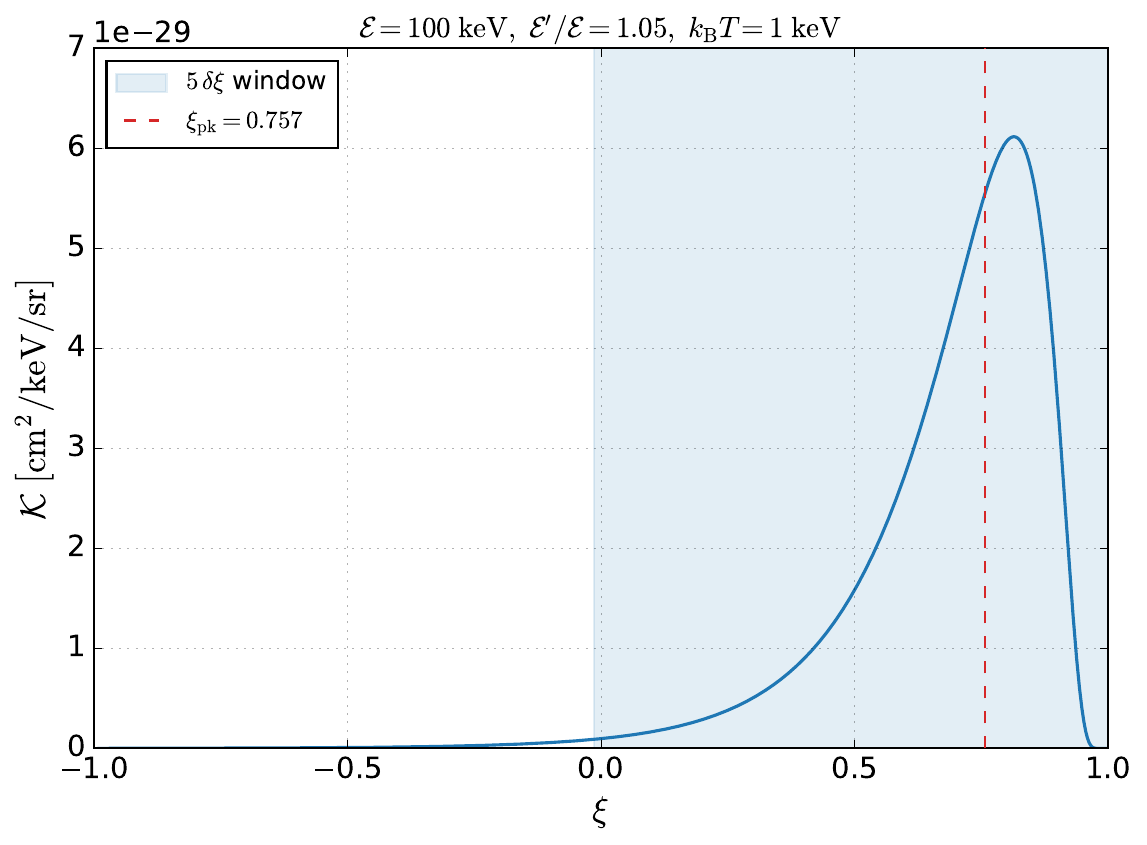}}
  \caption{The Compton kernel $\mathcal{K}(\xi)$ at
  $\mathcal{E}=100$\,keV for representative downscattering and
  upscattering energy ratios. The dashed line marks the analytic peak
  location $\xi_{\rm pk}$ from Eq.~\eqref{eq:xi_pk}, and the shaded
  band shows the $5\delta\xi$ main integration window used in
  Algorithm~\ref{alg:xi_integration}.}
  \label{fig:kernel_peak_structure}
\end{figure*}

\begin{figure*}[!htbp]
  \centering
  \subfloat[$\mathcal{E}'/\mathcal{E}=0.95$, $k_{\rm B}T=0.01$\,keV]{%
    \includegraphics[width=0.49\textwidth]{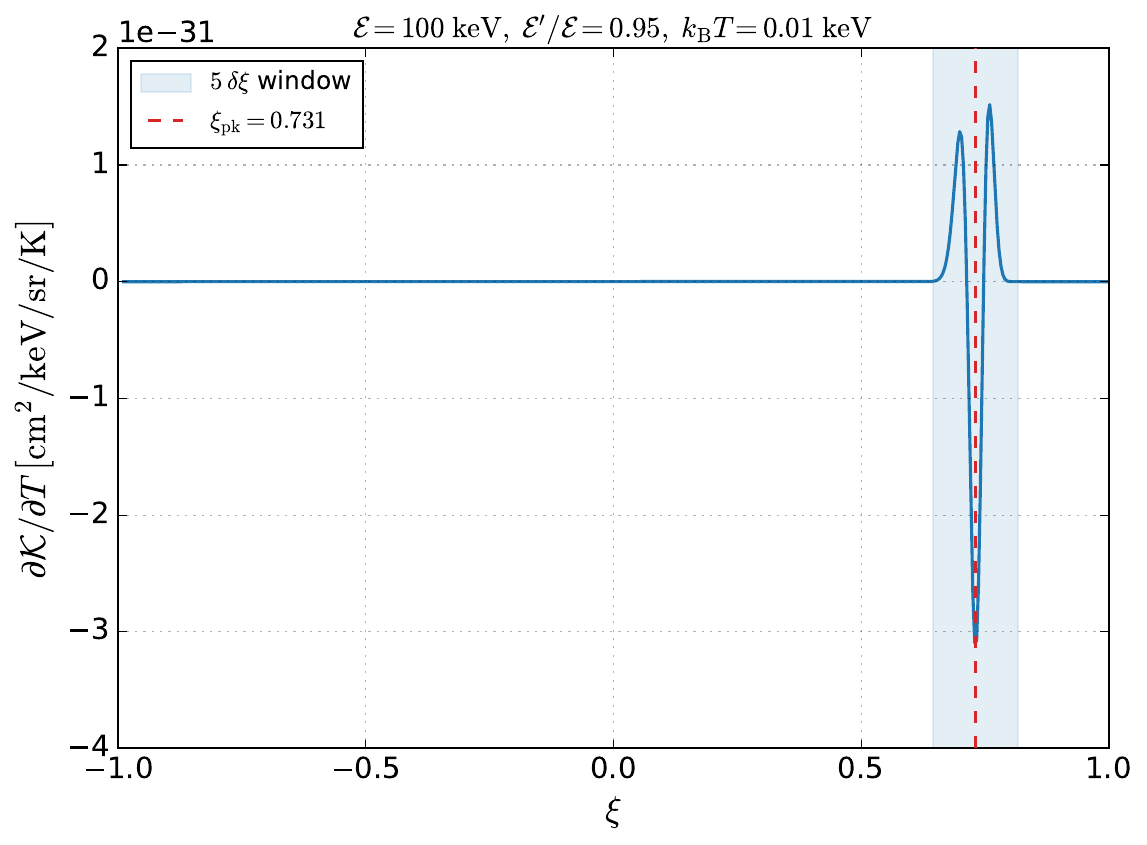}}
  \hfill
  \subfloat[$\mathcal{E}'/\mathcal{E}=1.05$, $k_{\rm B}T=1.0$\,keV]{%
    \includegraphics[width=0.49\textwidth]{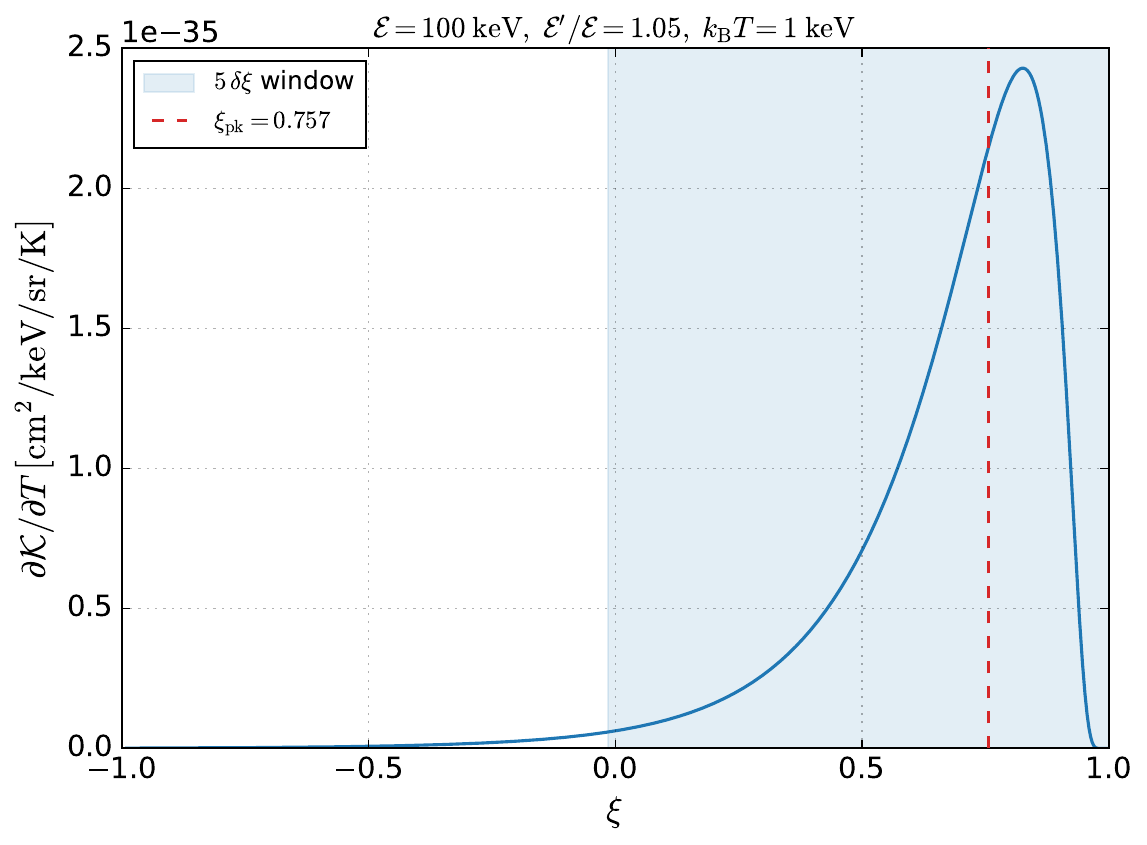}}
  \caption{Temperature derivative $\partial\mathcal{K}/\partial T$
  evaluated at the same parameters as
  Figure~\ref{fig:kernel_peak_structure}. The dashed line marks the
  same analytic peak location $\xi_{\rm pk}$ from Eq.~\eqref{eq:xi_pk},
  and the shaded band shows the same $5\delta\xi$ integration window.}
  \label{fig:kernel_peak_structure_deriv}
\end{figure*}

For fixed $\mathcal{E}$ and $\mathcal{E}'$, the angular structure of the
kernel is controlled primarily by the exponential factor involving
$\lambda_+$, defined in Eq.~\eqref{eq:def_lambda_+}.
Away from the nearly elastic case
$\mathcal{E}\simeq\mathcal{E}'$, where the kernel diverges as
$\xi\to 1$, the angular peak can be estimated by minimizing
$\lambda_+$ with respect to $\xi$
\citep{KERSHAW1986273,BEASON1991497}.
Differentiating $\lambda_+$ with respect to $\xi$ at fixed $\gamma$
and $\gamma'$ gives the peak scattering angle,
\begin{equation}
  \xi_{\rm pk}
  =
  1 - \frac{|\Delta\gamma|}{\gamma\gamma'} ,
  \label{eq:xi_pk}
\end{equation}
where $\Delta\gamma \equiv \gamma' - \gamma$. Expanding $\lambda_+$ to second order about $\xi_{\rm pk}$
\begin{equation}
  \lambda_+(\xi)
  \simeq
  \lambda_+(\xi_{\rm pk})
  +
  \frac{(\gamma\gamma')^2}
  {2|\Delta\gamma|(2+|\Delta\gamma|)}
  (\xi-\xi_{\rm pk})^2 .
  \label{eq:lambda_plus_xi_expansion}
\end{equation}
Since the dominant thermal factor is proportional to
$\exp[-\lambda_+/\tau]$, the peak can be approximated as a Gaussian with width
\begin{equation}
  \delta\xi
  =
  \frac{\sqrt{\tau\,|\Delta\gamma|(2+|\Delta\gamma|)}}{\gamma\gamma'} .
  \label{eq:delta_xi}
\end{equation}
Because the Gaussian approximation captures the dominant angular
variation of the kernel, the interval
$[\xi_{\rm pk}-5\delta\xi,\,\xi_{\rm pk}+5\delta\xi]$ encloses
essentially all of the peak contribution. When the estimated peak
intersects the current angle bin
$[\xi_{\rm low},\xi_{\rm high}]$, we clip this window to the bin
boundaries,
\begin{equation}
\begin{split}
  \xi_- &= \max(\xi_{\rm low},\,\xi_{\rm pk} - 5\delta\xi), \\
  \xi_+ &= \min(\xi_{\rm high},\,\xi_{\rm pk} + 5\delta\xi),
\end{split}
\label{eq:xi_pm}
\end{equation}
and designate $[\xi_-,\xi_+]$ as the peak panel. A higher-order
Gauss--Legendre rule is applied on this panel than on the remaining
tail regions of the bin (Algorithm~\ref{alg:xi_integration}).
Figures~\ref{fig:kernel_peak_structure}-\ref{fig:kernel_peak_structure_deriv} illustrate this structure
for representative energy ratios and temperatures, showing both the
analytic peak location and the $5\delta\xi$ main integration window for
$\mathcal{K}$ and $\partial\mathcal{K}/\partial T$, respectively.

When $\xi_{\rm pk}$ approaches the endpoint $\xi=1$, the Gaussian
peak-centered split is no longer the most useful angular rule. In this
regime the kernel becomes endpoint dominated. In the limit
$\mathcal{E}'=\mathcal{E}$, the kernel at the forward limit $\xi\to 1$ diverges, 
and for small but nonzero $|\Delta\gamma|$ the dominant variation is
a steep rise toward the upper edge of the bin. To handle this we therefore
use a reflected-log quadrature rule (Algorithm~\ref{alg:xi_integration}), which clusters nodes near the upper
integration bound. This branch is triggered when
\begin{multline}
\left(
\frac{|\Delta\gamma|}{\gamma\gamma'} \le \delta\xi
\ \text{and}\
\delta\xi \le 0.1
\right)
\ \text{or}\ \\
\left(
\frac{|\Delta\gamma|}{\gamma} \le 0.1
\ \text{and}\ \frac{|\Delta\gamma|}{\gamma\gamma'} \le \delta\xi
\right)
\label{eq:rlog_xi_condition}
\end{multline}
The first condition captures the cold-to-moderate regime where the peak width is
small and the peak merges with the forward boundary. The second
condition handles near-elastic scattering in which
$\delta\xi$ may be order unity, while the forward peak at $\xi\to 1$ still
controls the angular shape. Its last clause excludes high-energy photons, whose narrow angular peak lies well below $\xi=1$. Otherwise, the bin is split into the
peak panel $[\xi_-,\xi_+]$ from Eq.~\eqref{eq:xi_pm} and up to two
tail panels, as described in Algorithm~\ref{alg:xi_integration}.
\begin{algorithm}
  \hrule\smallskip
  \begin{algorithmic}[1]
  \Require $\mathcal{E}$, $\mathcal{E}'$, $[\xi_{\rm low},\xi_{\rm high}]$
  \Ensure $\mathcal{I}_{\Delta\xi}[X] \approx \int_{\xi_{\rm low}}^{\xi_{\rm high}} X\mathcal{K}\,d\xi$
  \State $\mathcal{I}_{\Delta\xi}\gets 0$
  
  \If{Eq.~\eqref{eq:rlog_xi_condition} is true}
    \State $\mathcal{I}_{\Delta\xi}\mathrel{+}=
    \mathrm{RLogGL}(X\mathcal{K},N_{\rm peak},\xi_{\rm low},\xi_{\rm high})$
    \State \Return $\mathcal{I}_{\Delta\xi}$
  \EndIf
  
  \If{$\xi_+ \leq \xi_{\rm low}$ \textbf{or} $\xi_- \geq \xi_{\rm high}$} \Comment{peak outside bin}
    \State $\mathcal{I}_{\Delta\xi}\mathrel{+}=
    \mathrm{GL}(X\mathcal{K},N_{\rm peak},\xi_{\rm low},\xi_{\rm high})$
    \State \Return $\mathcal{I}_{\Delta\xi}$
  \EndIf
  
  \State $\mathcal{I}_{\Delta\xi}\mathrel{+}=
  \mathrm{GL}(X\mathcal{K},N_{\rm tail},\xi_{\rm low},\xi_-)$
  \Comment{left tail}
  
  \State $\mathcal{I}_{\Delta\xi}\mathrel{+}=
  \mathrm{GL}(X\mathcal{K},N_{\rm peak},\xi_-,\xi_+)$
  \Comment{peak panel}
  
  \State $\mathcal{I}_{\Delta\xi}\mathrel{+}=
  \mathrm{GL}(X\mathcal{K},N_{\rm tail},\xi_+,\xi_{\rm high})$
  \Comment{right tail}
  
  \State \Return $\mathcal{I}_{\Delta\xi}$
  \end{algorithmic}
  \smallskip\hrule
  \caption{Structure-aware $\xi$-integration for fixed
  $(\mathcal{E},\mathcal{E}')$. The quadrature rules $\mathrm{GL}$ and
  $\mathrm{RLogGL}$ are defined in
  Appendix~\ref{sec:quadrature_rules_appendix}. The reflected-log rule is
  used when the endpoint condition in
  Eq.~\eqref{eq:rlog_xi_condition} is satisfied. If the peak lies outside
  the angular bin, the full bin is integrated with a single
  $\mathrm{GL}$ rule. Otherwise, the bin is split into a peak panel
  $[\xi_-,\xi_+]$ and two tail panels. Default quadrature orders for
  all panels are listed in Table~\ref{tab:quadrature_orders}.}
  \label{alg:xi_integration}
\end{algorithm}

\subsubsection{Validation of the \texorpdfstring{$\xi$}{xi} Quadrature}

To validate the angular quadrature independently of the energy
integrations, we first consider fixed incident and scattered photon
energies, $\mathcal{E}$ and $\mathcal{E}'$. Using the integral notation
introduced in Eq.~\eqref{eq:I_xi}, we compute the angle-integrated
kernel 
\begin{equation}
2\pi \mathcal{I}_{[-1,1]}[1](\mathcal{E},\mathcal{E}';\tau),
\label{eq:angle_integrated_validation}
\end{equation}
and the Legendre angular moments
\begin{equation}
2\pi \mathcal{I}_{[-1,1]}[P_n(\xi)](\mathcal{E},\mathcal{E}';\tau),
\label{eq:legendre_validation}
\end{equation}
where $P_n(\xi)$ is the Legendre polynomial of degree $n$.

\begin{figure*}[!htbp]
  \centering
  \subfloat[$\mathcal{E}/(k_{\rm B}T) = 0.1$]{%
    \includegraphics[width=0.49\textwidth]{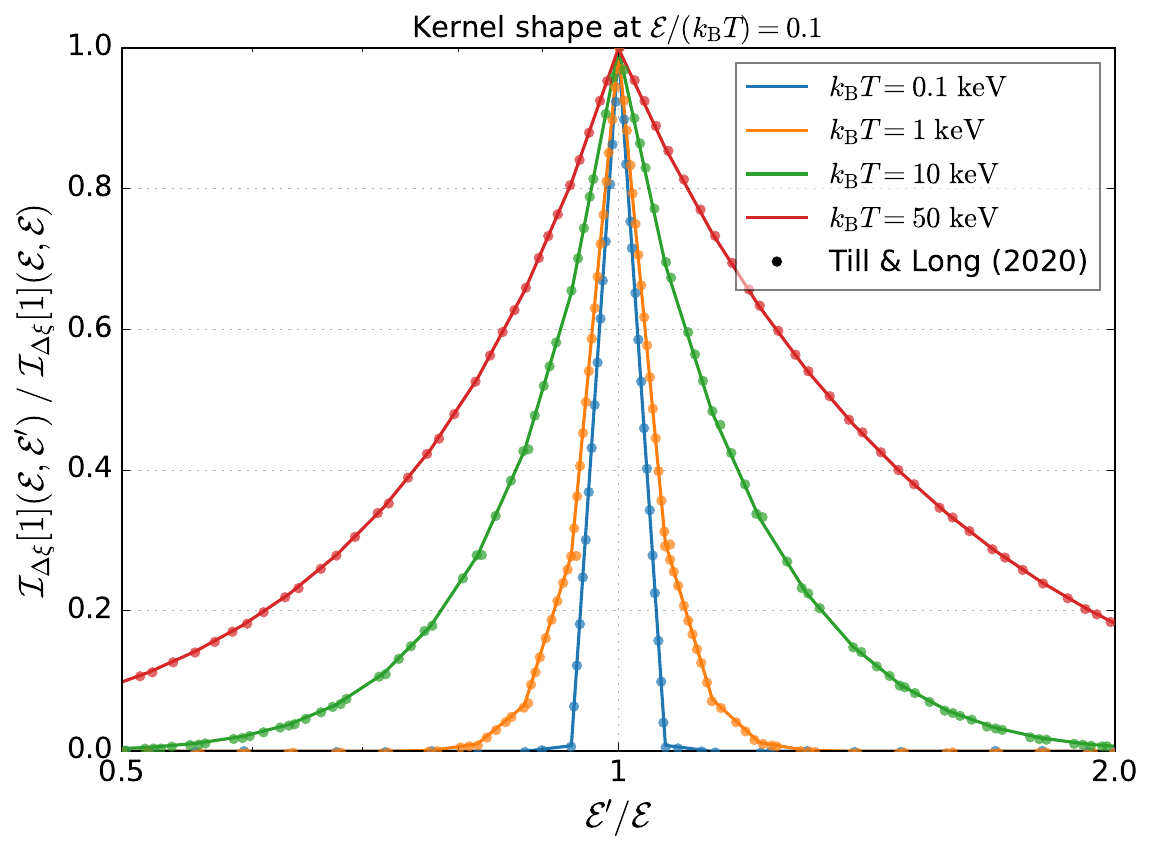}}
  \hfill
  \subfloat[$\mathcal{E}/(k_{\rm B}T) = 5$]{%
    \includegraphics[width=0.49\textwidth]{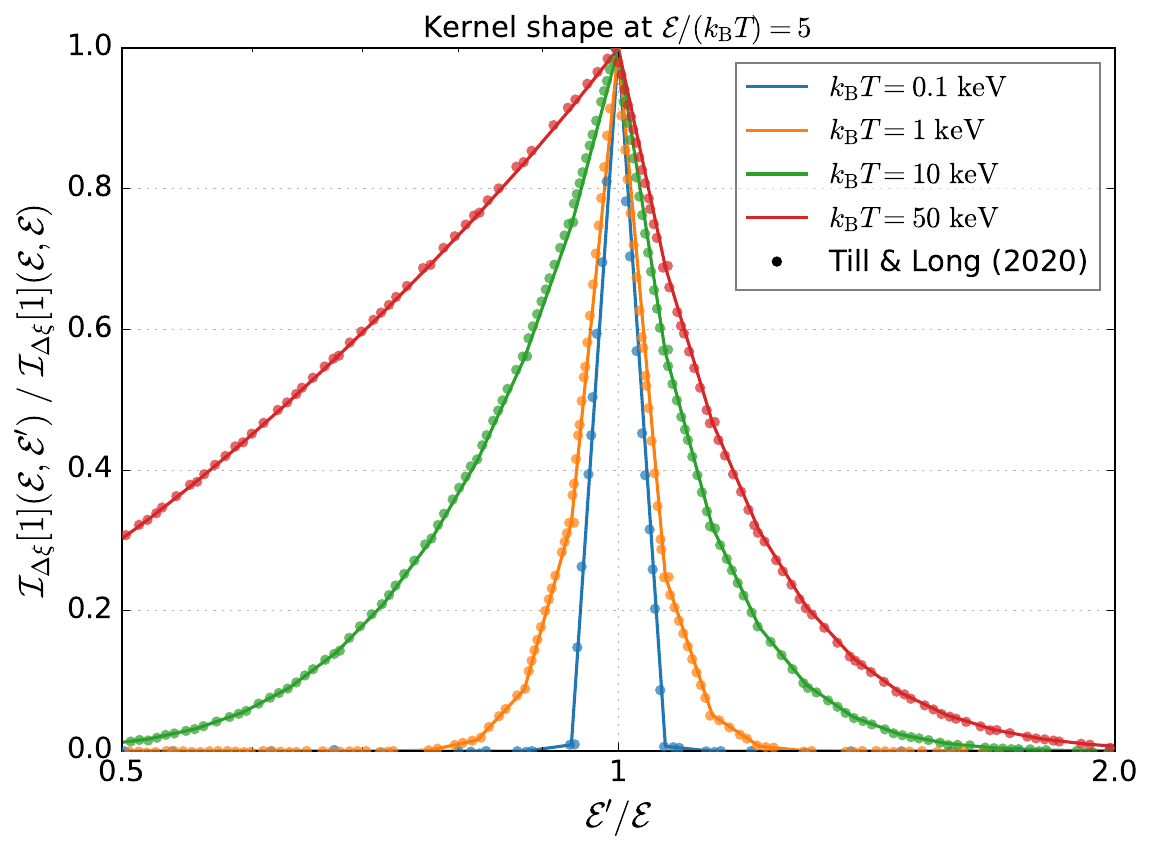}}
  \caption{Angle-integrated kernel
    $2\pi\mathcal{I}_{[-1,1]}[1](\mathcal{E},\mathcal{E}';\tau)$,
    normalized to its value at $\mathcal{E}' = \mathcal{E}$, as a function of
    $\mathcal{E}'/\mathcal{E}$ at fixed incident energy $\mathcal{E}$ at $k_{\rm B}T=0.1$, $1$, $10$, and
    $50\,{\rm keV}$. Solid lines show the present calculation and markers
    show digitized data from \citet{Compton_till_comparison_}.}
  \label{fig:till_kernel_shape}
\end{figure*}

\begin{figure}[!htbp]
  \centering
  \includegraphics[width=\columnwidth]{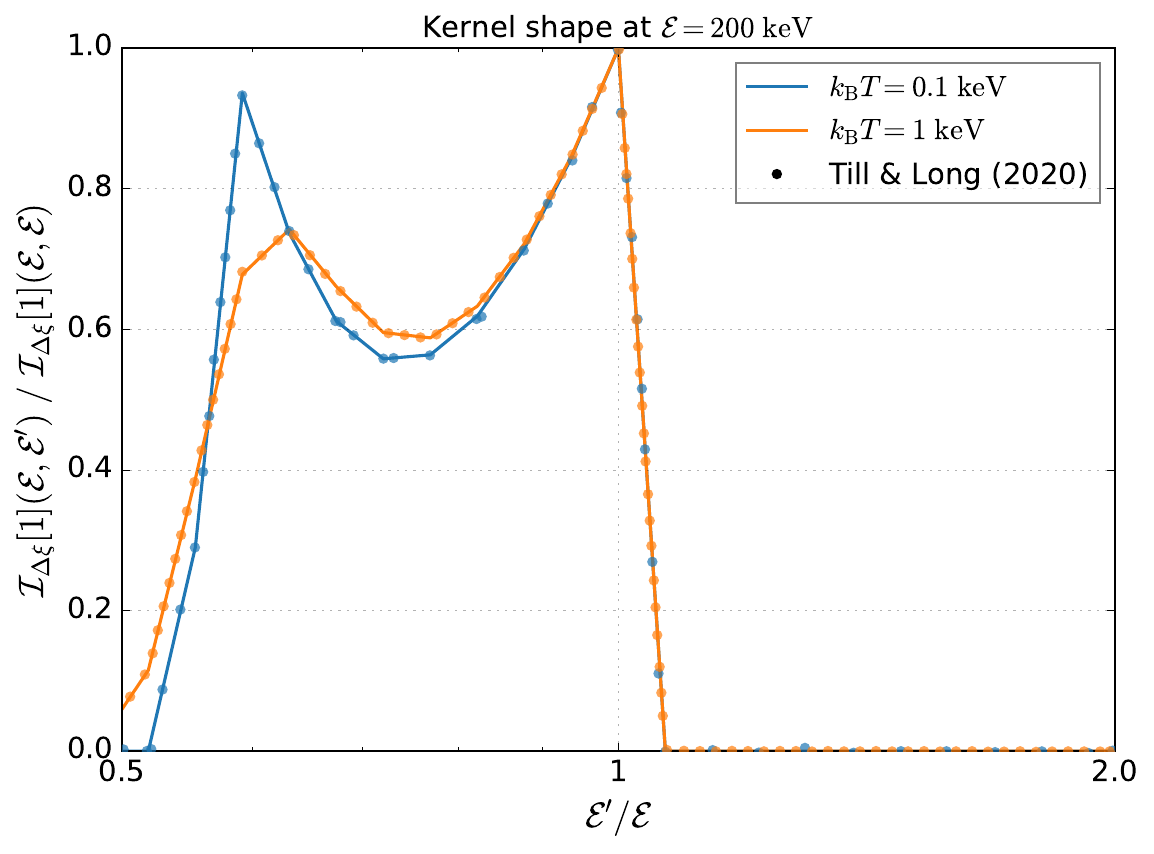}
  \caption{Same as Figure~\ref{fig:till_kernel_shape} but for a
  high-energy incident photon with $\mathcal{E}=200\,{\rm keV}$ at $k_{\rm B}T=0.1$ and $1\,{\rm keV}$.
  Solid lines show the present calculation and markers
    show digitized data from \citet{Compton_till_comparison_}.}
  \label{fig:till_kernel_shape_high_energy}
\end{figure}

Figures~\ref{fig:till_kernel_shape} and~\ref{fig:till_kernel_shape_high_energy} show the angle-integrated
integral $2\pi\mathcal{I}_{[-1,1]}[1]$, normalized to its value at $\mathcal{E'}=\mathcal{E}$,
against digitized results from \citet{Compton_till_comparison_}.

\begin{figure*}[!htbp]
  \centering
  \subfloat[$\mathcal{E} = 40\,{\rm keV}$]{%
    \includegraphics[width=0.49\textwidth]{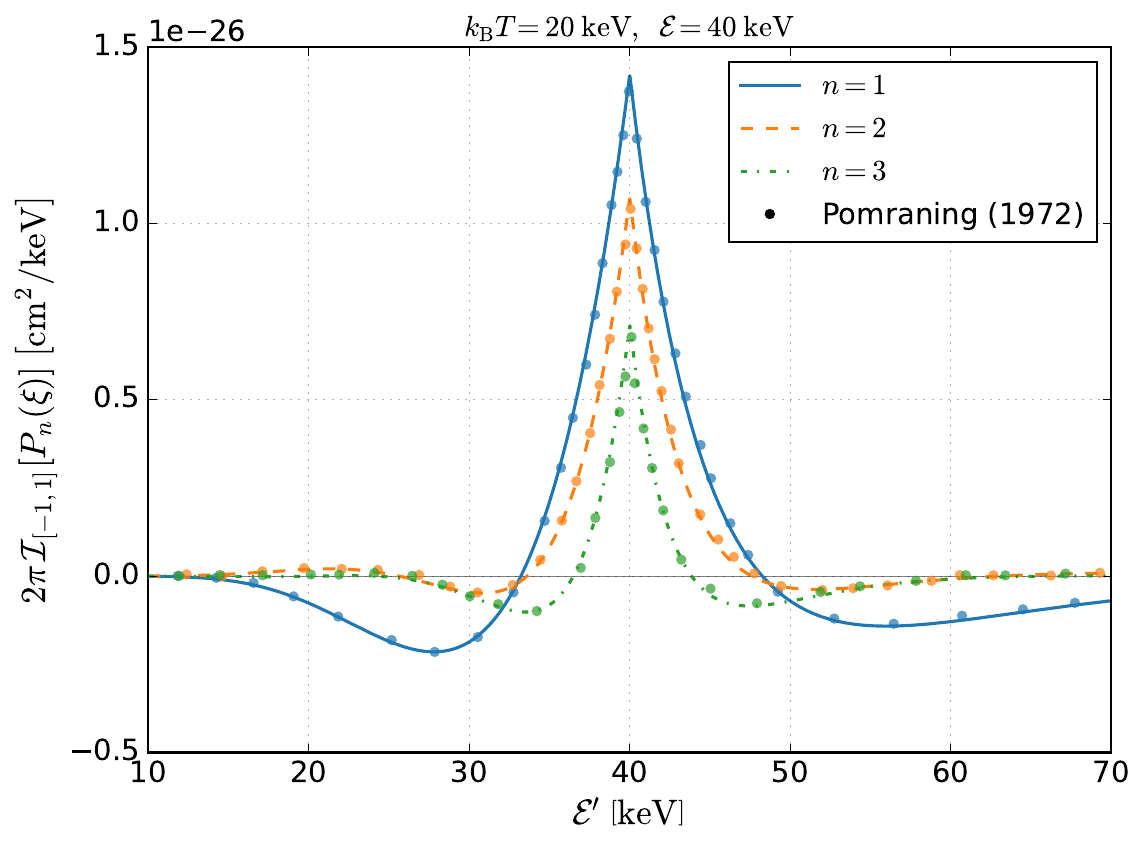}}
  \hfill
  \subfloat[$\mathcal{E} = 100\,{\rm keV}$]{%
    \includegraphics[width=0.49\textwidth]{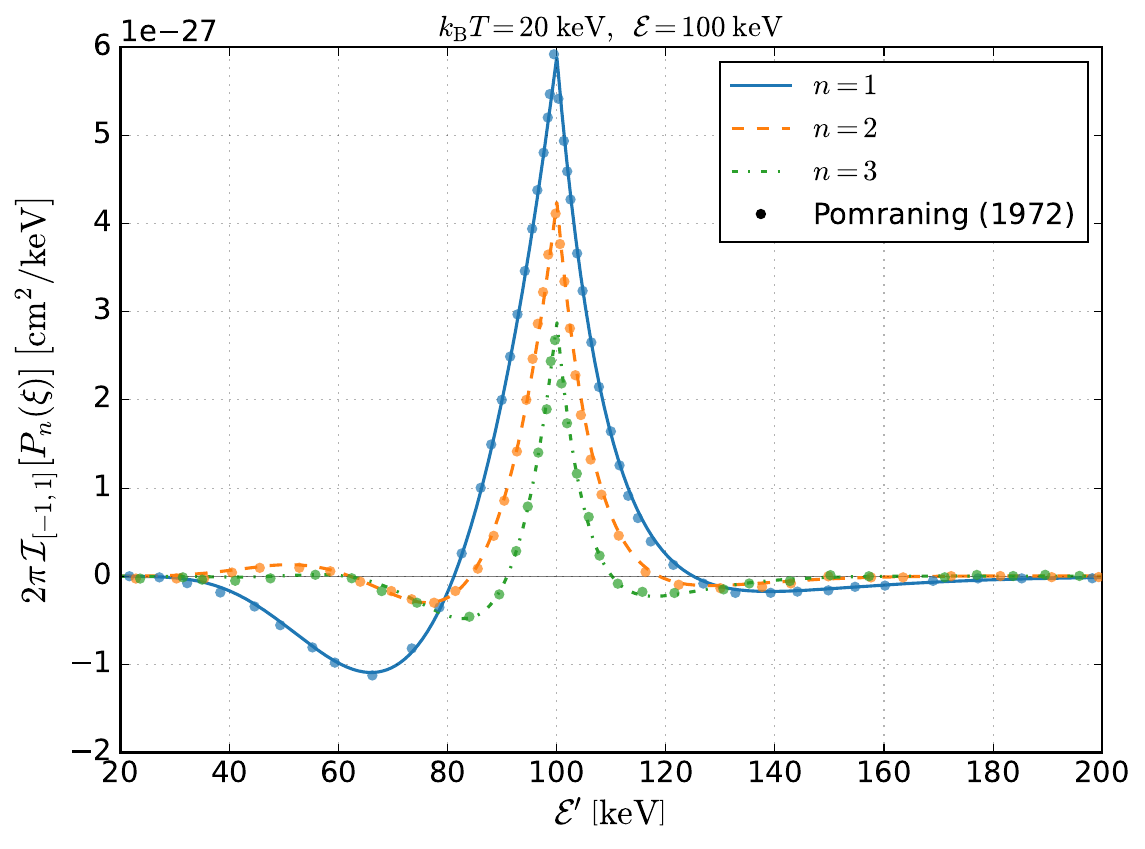}}
  \caption{Legendre-weighted angular integrals
  $2\pi\mathcal{I}_{[-1,1]}[P_n(\xi)](\mathcal{E},\mathcal{E}';\tau)$
  for $n=1,2,3$ at $k_{\rm B}T=20\,{\rm keV}$. Solid lines show the present
  calculation and markers show digitized data from
  \citet{POMRANING19721047,pomraning1973equations}.}
  \label{fig:pomraning_moments}
\end{figure*}

\begin{figure}[!htbp]
  \centering
  \includegraphics[width=\columnwidth]{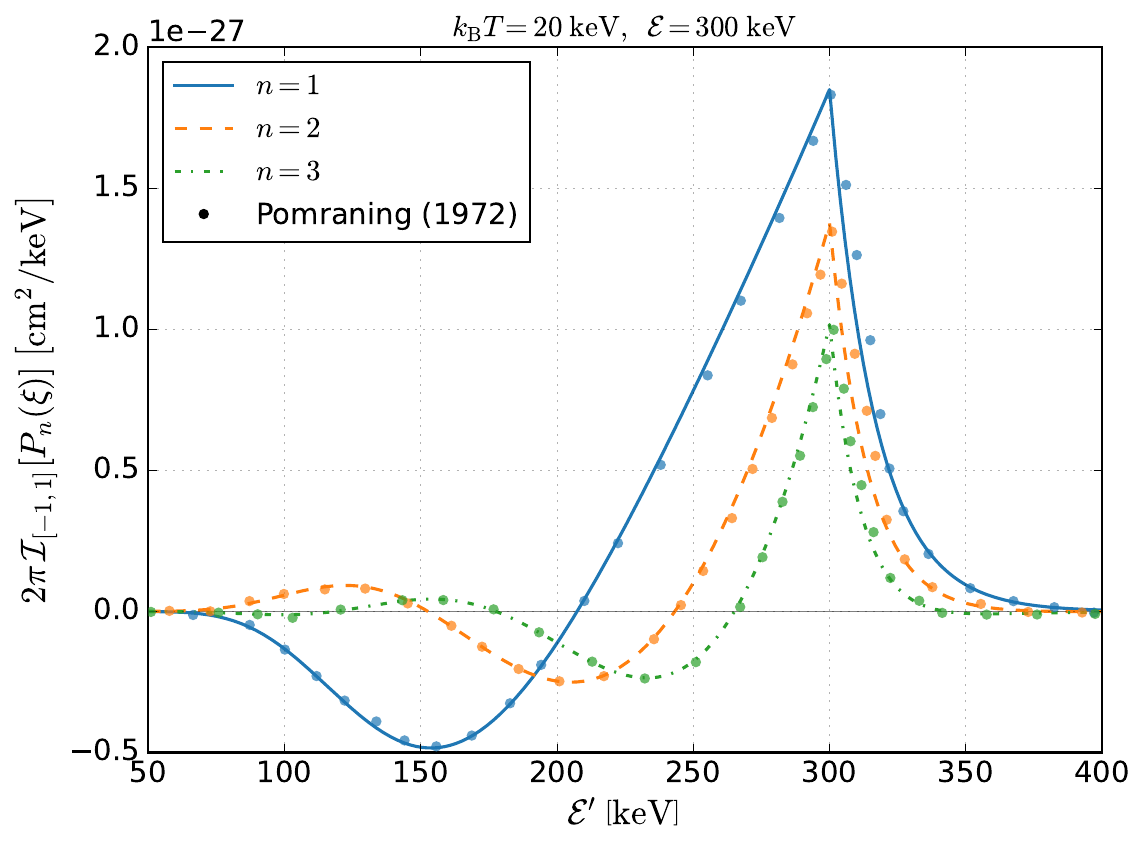}
  \caption{Same as Figure~\ref{fig:pomraning_moments} but for
  $\mathcal{E}=300\,{\rm keV}$. Solid lines show the present
  calculation and markers show digitized data from
  \citet{POMRANING19721047,pomraning1973equations}.}
  \label{fig:pomraning_moments_high_energy}
\end{figure}

Figures~\ref{fig:pomraning_moments} and~\ref{fig:pomraning_moments_high_energy} show the Legendre-weighted
angular integrals $2\pi\mathcal{I}_{[-1,1]}[P_n(\xi)]$ for $n=1,2,3$.
The comparison is made at $k_{\rm B}T=20\,{\rm keV}$ as a function of the
scattered photon energy $\mathcal{E}'$, for incident photon energies
$\mathcal{E}=40$, $100$, and $300\,{\rm keV}$. We compare the results
to digitized data from \citet{POMRANING19721047,pomraning1973equations}.

\subsection{\texorpdfstring{$\mathcal{E}'$}{E'} Integration} \label{sec:Ep_integration}

The scattered-energy quadrature is constructed from the same local
structure argument used for the angular quadrature. For fixed incident
photon energy $\mathcal{E}$ and scattering angle $\xi$, the kernel is
largest near the minimum of $\lambda_+$. In the cold limit this minimum
gives the Compton recoil energy
\begin{equation}
\mathcal{E}'_{\rm pk}(\xi)
=
\frac{\mathcal{E}}{1+\gamma(1-\xi)} ,
\label{eq:Eprime_peak}
\end{equation}
which is the rest-frame Compton relation from
Eq.~\eqref{eq:compton_recoil_rest}. Since $\mathcal{E}'_{\rm pk}$ is an
increasing function of $\xi$, an angular bin
$[\xi_{\rm low},\xi_{\rm high}]$ maps to a cold recoil band
\begin{equation}
\left[
\mathcal{E}'_{\rm pk}(\xi_{\rm low}),
\mathcal{E}'_{\rm pk}(\xi_{\rm high})
\right].
\label{eq:Eprime_recoil_band}
\end{equation}
The scattered-energy quadrature is designed to resolve this band
together with its thermal broadening.

To estimate the thermal broadening of the cold recoil band, we again
expand $\lambda_+$ about its local minimum. At fixed $\mathcal{E}$ and
$\xi$, the second-order expansion about $\mathcal{E}'_{\rm pk}(\xi)$
gives the scattered-energy width
\begin{equation}
\delta\mathcal{E}'(\xi)
=
\frac{\mathcal{E}\sqrt{\tau a(2+2\gamma a+\gamma^2 a)}}{(1+\gamma a)^2},
\label{eq:Eprime_width}
\end{equation}
where $a \equiv 1-\xi$.
This provides an estimate of the $\mathcal{E}'$ range over which the kernel
contributes significantly, obtained by broadening the cold recoil band
by a few local widths.
For an angular bin $[\xi_{\rm low},\xi_{\rm high}]$, we define
\begin{equation}
\delta\mathcal{E}'_{\rm low}
=
\delta\mathcal{E}'(\xi_{\rm low}),
\qquad
\delta\mathcal{E}'_{\rm high}
=
\delta\mathcal{E}'(\xi_{\rm high}) .
\label{eq:Eprime_edge_widths}
\end{equation}
These two widths set the energy intervals used by the
$\mathcal{E}'$-quadrature panels below.

\begin{figure*}[!htbp]
  \centering
  \subfloat[$\mathcal{E}=0.1\,{\rm keV}$]{%
    \includegraphics[width=0.49\textwidth]{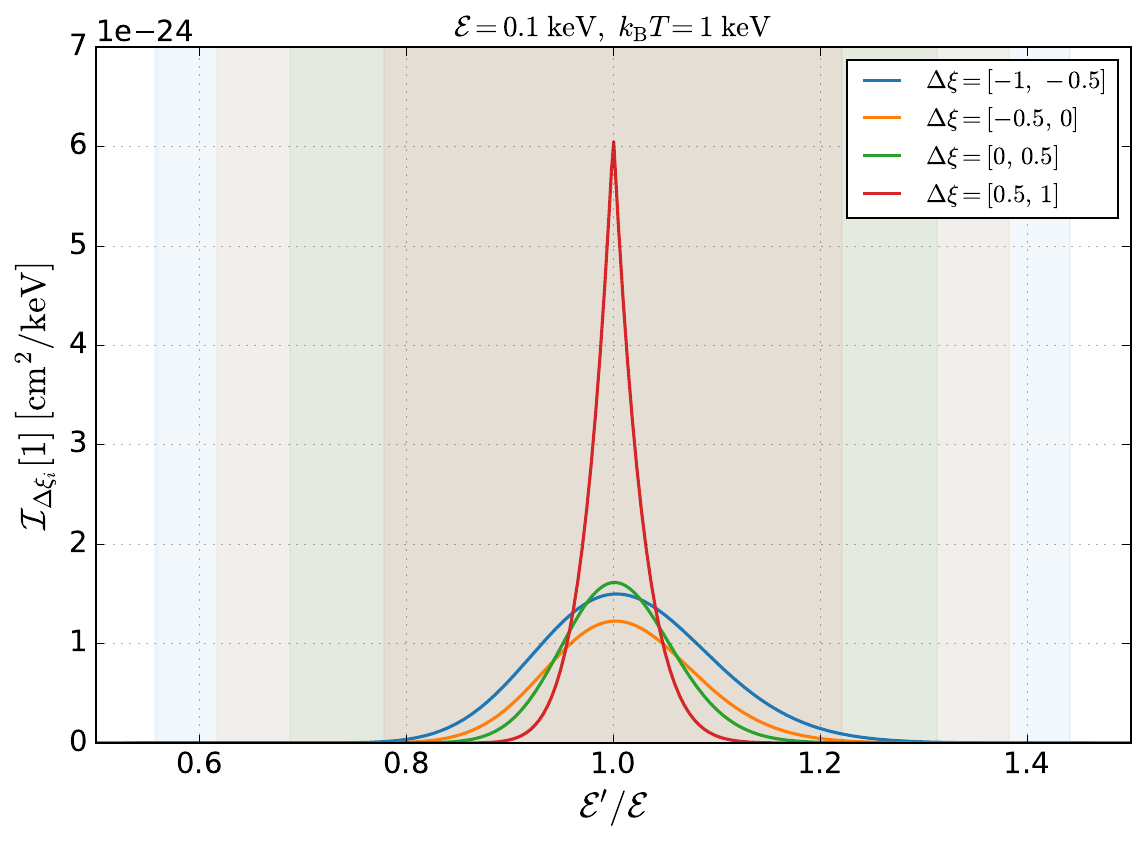}}
  \hfill
  \subfloat[$\mathcal{E}=100\,{\rm keV}$]{%
    \includegraphics[width=0.49\textwidth]{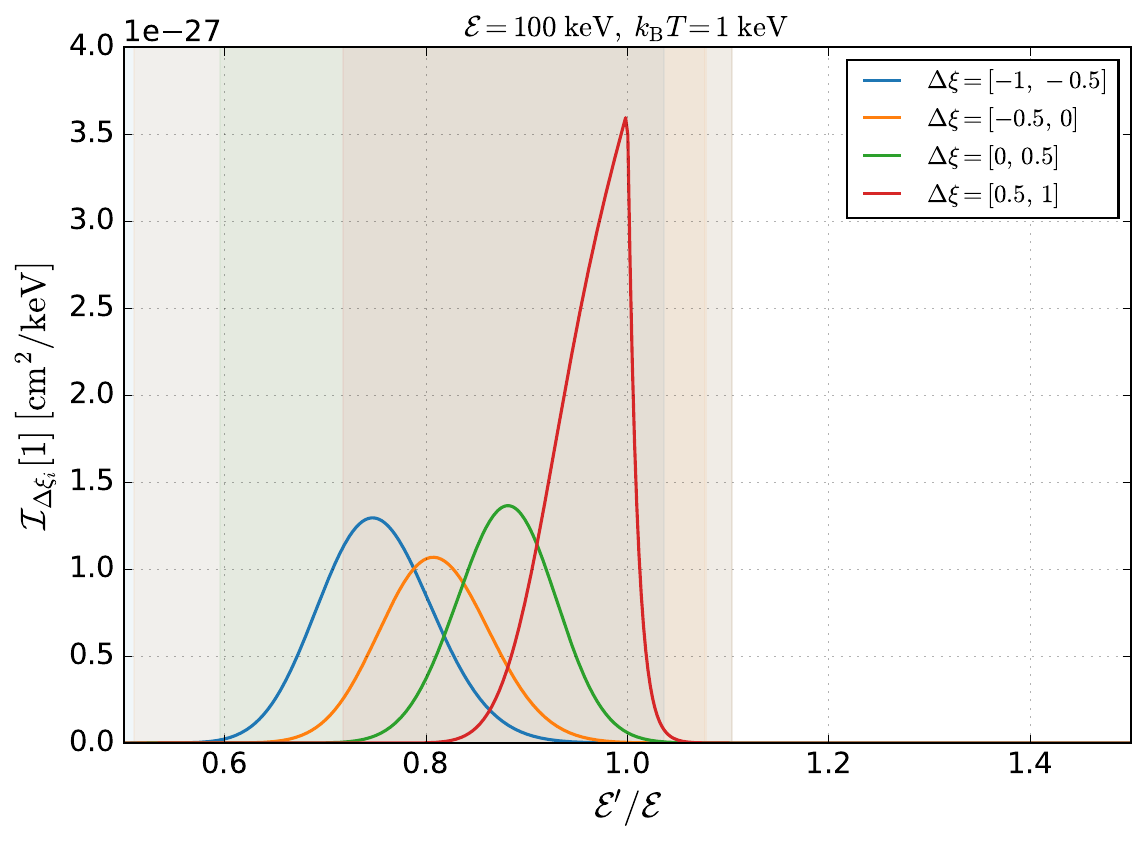}}
  \caption{$\mathcal{I}_{\Delta\xi_i}[1]$ as a function of
  $\mathcal{E}'/\mathcal{E}$ at $k_{\rm B}T=1\,{\rm keV}$, decomposed into four
  uniform angular bins. Shaded bands show the main integration interval
  $[\mathcal{E}'_-,\mathcal{E}'_+]$ for each angular bin.}
  \label{fig:ep_profile_4bins}
\end{figure*}

For a target scattered-energy group
$[\mathcal{E}'_{\rm low},\mathcal{E}'_{\rm high}]$, we define the
retained interval as the overlap between the group and the recoil band
padded by five local thermal widths:
\begin{equation}
\mathcal{E}'_- =
\max\!\left[
\mathcal{E}'_{\rm low},
\mathcal{E}'_{\rm pk}(\xi_{\rm low})
-5\delta\mathcal{E}'_{\rm low}
\right],
\label{eq:Eprime_retained_interval_1}
\end{equation}
\begin{multline}
\mathcal{E}'_+ =
\min\!\Big[
\mathcal{E}'_{\rm high},\\
\max\!\left(
\mathcal{E}'_{\rm pk}(\xi_{\rm low})
+5\delta\mathcal{E}'_{\rm low},
\mathcal{E}'_{\rm pk}(\xi_{\rm high})
+5\delta\mathcal{E}'_{\rm high}
\right)
\Big]
\label{eq:Eprime_retained_interval_2}
\end{multline}
The lower bound is set by the low-$\xi$ recoil peak, which lies at
the lowest $\mathcal{E}'$ in the band. For the upper bound, both
angular endpoints can contribute the highest broadened envelope, so
we take the larger of the two. The segment $[\mathcal{E}'_-,\mathcal{E}'_+]$ dominates the
contribution to the group integral. 
Figure~\ref{fig:ep_profile_4bins} shows the $\xi$-integrated kernel as a function of $\mathcal{E}'/\mathcal{E}$, separated into four angular
bins at $k_{\rm B}T=1\,{\rm keV}$. The shaded intervals show the main
interval $[\mathcal{E}'_-,\mathcal{E}'_+]$ for each angular bin.

The main interval does not truncate the group integral. It only
marks the part of the target group where the recoil ridge is expected to
be resolved directly. Because the kernel decays exponentially away from
the recoil ridge, the side intervals carry little weight. The
low-energy side interval $[\mathcal{E}'_{\rm low},\mathcal{E}'_-]$ is integrated with a
reflected-log rule, which clusters nodes near $\mathcal{E}'_-$. And the high-energy side
interval $[\mathcal{E}'_+,\mathcal{E}'_{\rm high}]$ is integrated with a
logarithmic rule when $\mathcal{E}'_{\rm high}/\mathcal{E}'_+>2$, and
otherwise with ordinary Gauss--Legendre quadrature. Thus the full
scattered-energy group is always covered, while the quadrature rule
changes according to the local structure of the integrand.

Inside the main interval $[\mathcal{E}'_-,\mathcal{E}'_+]$, the $\mathcal{E}'$-integrand can have
steep shoulders near the thermally broadened recoil-band edges, with a
smoother region between them. We therefore define the interior
boundaries
\begin{equation}
  \mathcal{E}'_{{\rm out},{\rm L}}
  =
  \min\!\left[
  \mathcal{E}'_+,
  \mathcal{E}'_{\rm pk}(\xi_{\rm low})
  +2\delta\mathcal{E}'_{\rm low}
  \right],
  \label{eq:Eprime_outer_left_bound}
\end{equation}
\begin{equation}
  \mathcal{E}'_{{\rm int},{\rm L}}
  =
  \max\!\left[
  \mathcal{E}'_-,
  \mathcal{E}'_{\rm pk}(\xi_{\rm low})
  +2\delta\mathcal{E}'_{\rm low}
  \right],
  \label{eq:Eprime_interior_bounds_1}
\end{equation}
\begin{equation}
\mathcal{E}'_{{\rm int},{\rm R}}
=
\min\!\left[
\mathcal{E}'_+,
\mathcal{E}'_{\rm pk}(\xi_{\rm high})
-2\delta\mathcal{E}'_{\rm high}
\right].
\label{eq:Eprime_interior_bounds_2}
\end{equation}
\begin{equation}
\mathcal{E}'_{{\rm out},{\rm R}}
=
\max\!\left[
\mathcal{E}'_-,
\mathcal{E}'_{\rm pk}(\xi_{\rm high})
-2\delta\mathcal{E}'_{\rm high}
\right].
\label{eq:Eprime_outer_right_bound}
\end{equation}

In angular bins that extend close to $\xi=1$, the two edge widths can
be very different because $\delta\mathcal{E}'(\xi)$ collapses in the
forward limit. When
\begin{equation}
\frac{\delta\mathcal{E}'_{\rm low}}
     {\delta\mathcal{E}'_{\rm high}}
> 10 ,
\label{eq:Eprime_double_peak_condition}
\end{equation}
the $\mathcal{E}'$-integrand contains two scales: a broad thermally
broadened recoil feature associated with $\xi_{\rm low}$, and a much
narrower forward-endpoint feature centered near
$\mathcal{E}'_{\rm pk}(\xi_{\rm high})$. We isolate the narrow feature
by introducing an additional panel
\begin{equation}
  \mathcal{E}'_{{\rm f},-}
  =
  \max\!\left[
  \mathcal{E}'_-,
  \mathcal{E}'_{\rm pk}(\xi_{\rm high})
  -5\delta\mathcal{E}'_{\rm high}
  \right],
  \label{eq:Eprime_forward_bounds_1}
\end{equation}
\begin{equation}
\mathcal{E}'_{{\rm f},+}
=
\min\!\left[
\mathcal{E}'_+,
\mathcal{E}'_{\rm pk}(\xi_{\rm high})
+5\delta\mathcal{E}'_{\rm high}
\right].
\label{eq:Eprime_forward_bounds_2}
\end{equation}
This additional split prevents the broad recoil feature from setting the
resolution needed for the much narrower endpoint peak.

Algorithm~\ref{alg:Eprime_integration} summarizes the
$\mathcal{E}'$-quadrature for fixed incident energy $\mathcal{E}$ and
angle bin $[\xi_{\rm low},\xi_{\rm high}]$.
\begin{algorithm}
  \hrule\smallskip
  \begin{algorithmic}[1]
  \Require $\mathcal{E}$,
  $[\mathcal{E}'_{\rm low},\mathcal{E}'_{\rm high}]$,
  and $[\xi_{\rm low},\xi_{\rm high}]$
  \Ensure
  $\mathcal{I}_{\Delta\mathcal{E}'\Delta\xi}
  \approx \int_{\mathcal{E}'_{\rm low}}^{\mathcal{E}'_{\rm high}}
  \mathcal{I}_{\Delta\xi}\,d\mathcal{E}'$
  \State $\mathcal{I}_{\Delta\mathcal{E}'\Delta\xi}\gets 0$
  
  \If{Eq.~\eqref{eq:Eprime_double_peak_condition} holds
  and $\mathcal{E}'_{{\rm f},+}>\mathcal{E}'_{{\rm f},-}$}
    \State $\mathcal{I}_{\Delta\mathcal{E}'\Delta\xi}\mathrel{+}=
    \mathrm{RLogGL}(\mathcal{I}_{\Delta\xi},N_{\rm edge},
    \mathcal{E}'_{\rm low},\mathcal{E}'_-)$
    \Comment{left tail}
  
    \State $\mathcal{I}_{\Delta\mathcal{E}'\Delta\xi}\mathrel{+}=
    \mathrm{GL}(\mathcal{I}_{\Delta\xi},N_{\rm int},
    \mathcal{E}'_-,\mathcal{E}'_{{\rm f},-})$
    \Comment{broad ridge}
  
    \State $\mathcal{I}_{\Delta\mathcal{E}'\Delta\xi}\mathrel{+}=
    \mathrm{GL}(\mathcal{I}_{\Delta\xi},N_{\rm el},
    \mathcal{E}'_{{\rm f},-},
    \mathcal{E}'_{{\rm f},+})$
    \Comment{forward peak}
  
    \State $\mathcal{I}_{\Delta\mathcal{E}'\Delta\xi}\mathrel{+}=
    \mathrm{GL}(\mathcal{I}_{\Delta\xi},N_{\rm int},
    \mathcal{E}'_{{\rm f},+},\mathcal{E}'_+)$
    \Comment{broad ridge}
  
    \State $\mathcal{I}_{\Delta\mathcal{E}'\Delta\xi}\mathrel{+}=
    \mathrm{TailGL}(\mathcal{I}_{\Delta\xi},N_{\rm edge},
    \mathcal{E}'_+,\mathcal{E}'_{\rm high})$
    \Comment{right tail}
    \State \Return $\mathcal{I}_{\Delta\mathcal{E}'\Delta\xi}$
  \EndIf
  
  \If{$\mathcal{E}'_-\geq\mathcal{E}'_+$}
  \Comment{ridge outside group}
    \If{$\mathcal{E}'_{\rm high}\leq
    \mathcal{E}'_{\rm pk}(\xi_{\rm low})$}
    \Comment{group below ridge}
      \State $\mathcal{I}_{\Delta\mathcal{E}'\Delta\xi}\mathrel{+}=
      \mathrm{RLogGL}(\mathcal{I}_{\Delta\xi},N_{\rm edge},
      \mathcal{E}'_{\rm low},\mathcal{E}'_{\rm high})$
      \State \Return $\mathcal{I}_{\Delta\mathcal{E}'\Delta\xi}$
    \Else
    \Comment{group above ridge}
      \State $\mathcal{I}_{\Delta\mathcal{E}'\Delta\xi}\mathrel{+}=
      \mathrm{TailGL}(\mathcal{I}_{\Delta\xi},N_{\rm edge},
      \mathcal{E}'_{\rm low},\mathcal{E}'_{\rm high})$
      \State \Return $\mathcal{I}_{\Delta\mathcal{E}'\Delta\xi}$
    \EndIf
  \EndIf
  
  \State $\mathcal{I}_{\Delta\mathcal{E}'\Delta\xi}\mathrel{+}=
  \mathrm{RLogGL}(\mathcal{I}_{\Delta\xi},N_{\rm edge},
  \mathcal{E}'_{\rm low},\mathcal{E}'_-)$
  \Comment{left tail}
  
  \If{$\mathcal{E}'_{\rm pk}(\xi_{\rm low})
  +2\delta\mathcal{E}'_{\rm low}
  \geq
  \mathcal{E}'_{\rm pk}(\xi_{\rm high})
  -2\delta\mathcal{E}'_{\rm high}$}
    \State $\mathcal{I}_{\Delta\mathcal{E}'\Delta\xi}\mathrel{+}=
    \mathrm{GL}(\mathcal{I}_{\Delta\xi},N_{\rm int},
    \mathcal{E}'_-,\mathcal{E}'_+)$
    \Comment{single ridge panel}
  \Else
    \State $\mathcal{I}_{\Delta\mathcal{E}'\Delta\xi}\mathrel{+}=
    \mathrm{GL}(\mathcal{I}_{\Delta\xi},N_{\rm edge},
    \mathcal{E}'_-,\mathcal{E}'_{{\rm out},{\rm L}})$
    \Comment{left edge}
  
    \State $\mathcal{I}_{\Delta\mathcal{E}'\Delta\xi}\mathrel{+}=
    \mathrm{GL}(\mathcal{I}_{\Delta\xi},N_{\rm int},
    \mathcal{E}'_{{\rm int},{\rm L}},
    \mathcal{E}'_{{\rm int},{\rm R}})$
    \Comment{interior}
  
    \State $\mathcal{I}_{\Delta\mathcal{E}'\Delta\xi}\mathrel{+}=
    \mathrm{GL}(\mathcal{I}_{\Delta\xi},N_{\rm edge},
    \mathcal{E}'_{{\rm out},{\rm R}},\mathcal{E}'_+)$
    \Comment{right edge}
  \EndIf
  
  \State $\mathcal{I}_{\Delta\mathcal{E}'\Delta\xi}\mathrel{+}=
  \mathrm{TailGL}(\mathcal{I}_{\Delta\xi},N_{\rm edge},
  \mathcal{E}'_+,\mathcal{E}'_{\rm high})$
  \Comment{right tail}
  \State \Return $\mathcal{I}_{\Delta\mathcal{E}'\Delta\xi}$
  \end{algorithmic}
  \smallskip\hrule
  \caption{Structure-aware $\mathcal{E}'$-integration for fixed incident energy
  $\mathcal{E}$ and angle bin
  $[\xi_{\rm low},\xi_{\rm high}]$.
  The quadrature rules are defined in
  Appendix~\ref{sec:quadrature_rules_appendix}. The retained, interior,
  and forward-core bounds are defined in
  Eqs.~\eqref{eq:Eprime_retained_interval_1}--%
  \eqref{eq:Eprime_forward_bounds_2}.}
  \label{alg:Eprime_integration}
\end{algorithm}

\subsubsection{Validation of the \texorpdfstring{$\mathcal{E}'+\xi$}{E'+xi} Quadrature}

\begin{figure*}[!htbp]
  \centering
  \subfloat[First moment ($m=1$)]{%
    \includegraphics[width=0.49\textwidth]{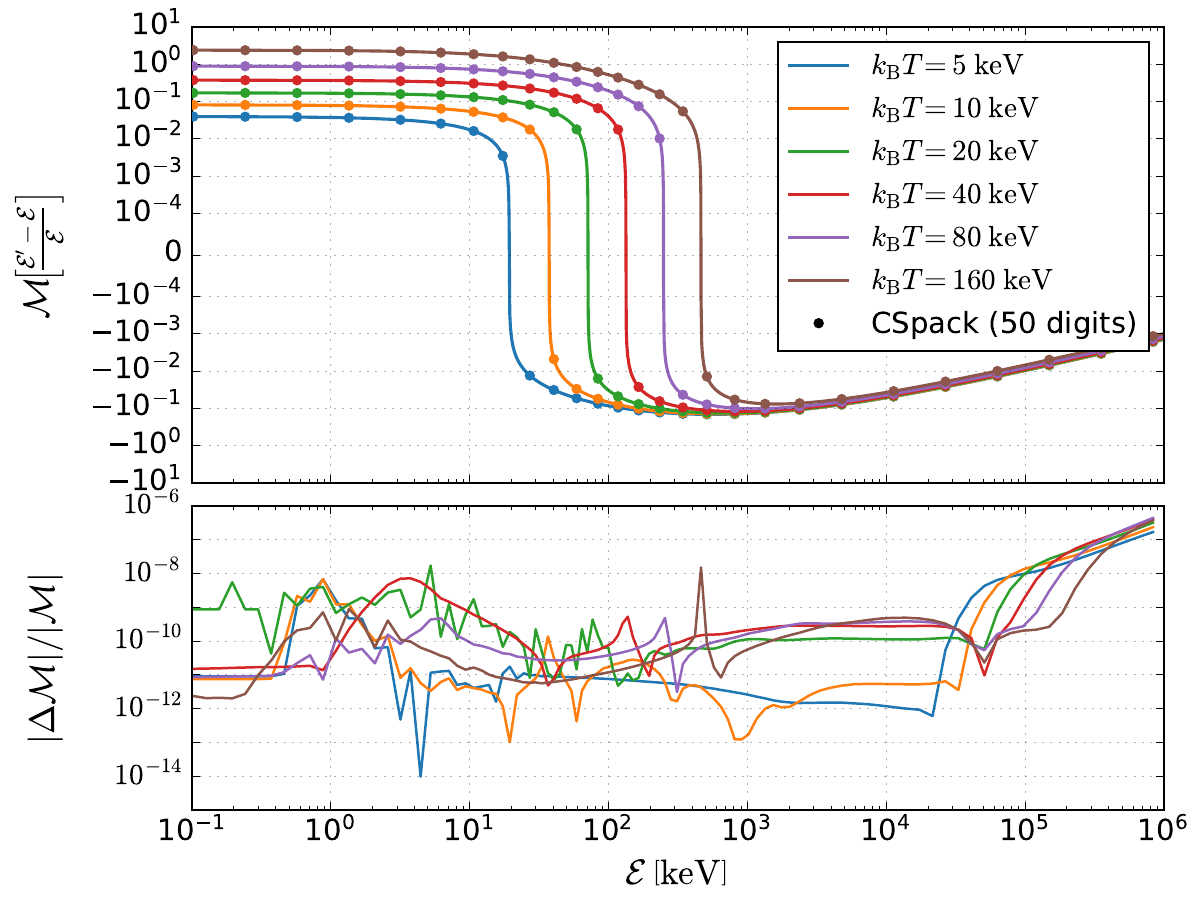}}
  \hfill
  \subfloat[Second moment ($m=2$)]{%
    \includegraphics[width=0.49\textwidth]{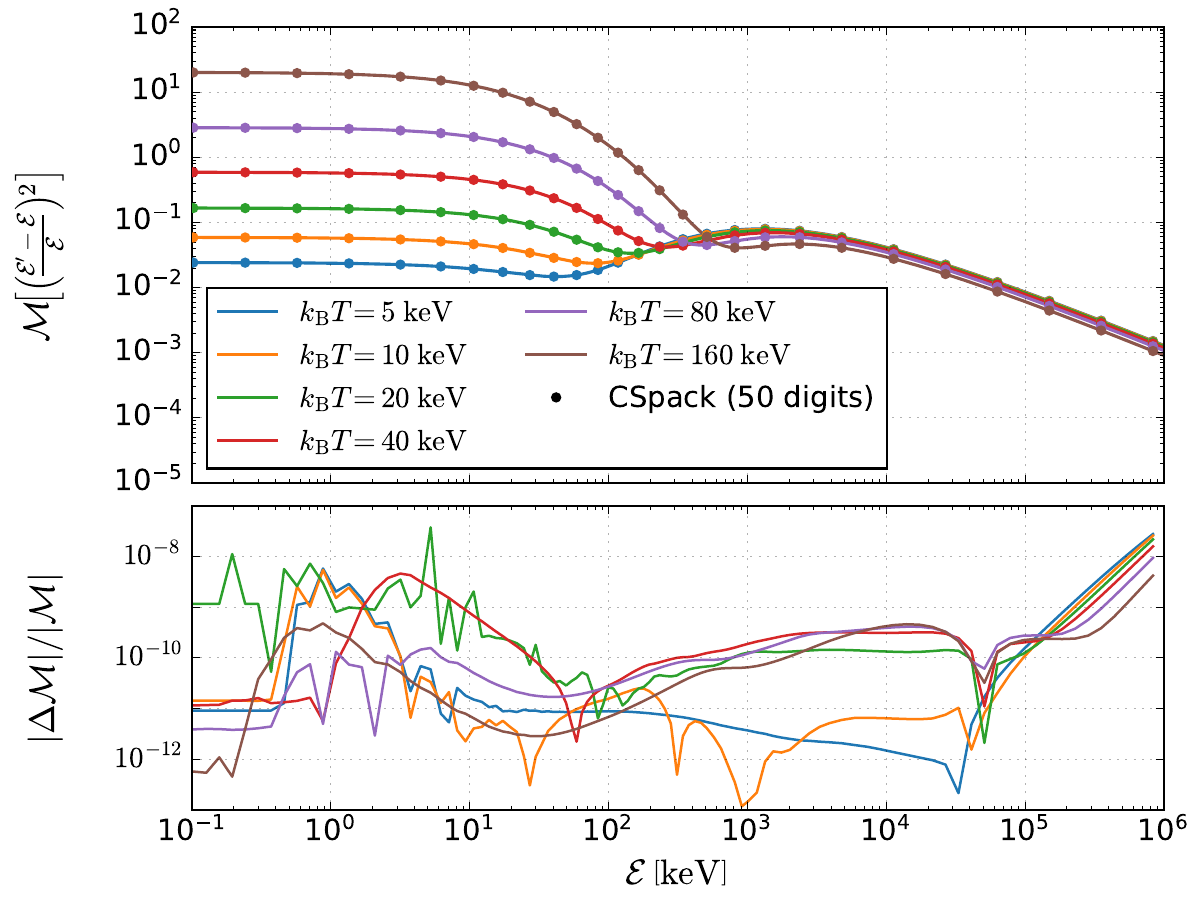}}
  \caption{Energy-transfer moments
  $\mathcal{M}\!\left[((\mathcal{E}'-\mathcal{E})/\mathcal{E})^m\right]$
  for $m=1,2$, shown as a function of the incident photon energy
  $\mathcal{E}$ at $k_{\rm B}T=5$--$160\,{\rm keV}$. Solid lines show the
  present calculation and markers show the closed-form moments implemented in \textsc{CSpack} \citep{Sarkar2019,CSpack}, evaluated with 50 significant digits and averaged over the Maxwell--J\"uttner distribution at the same precision. The lower panels show the relative deviation $|\Delta\mathcal{M}|/|\mathcal{M}|$ of the present calculation from these moments.}
  \label{fig:sarkar_moments}
\end{figure*}

\begin{figure*}[!htbp]
  \centering
  \subfloat[Compton mean free path]{%
    \includegraphics[width=0.49\textwidth]{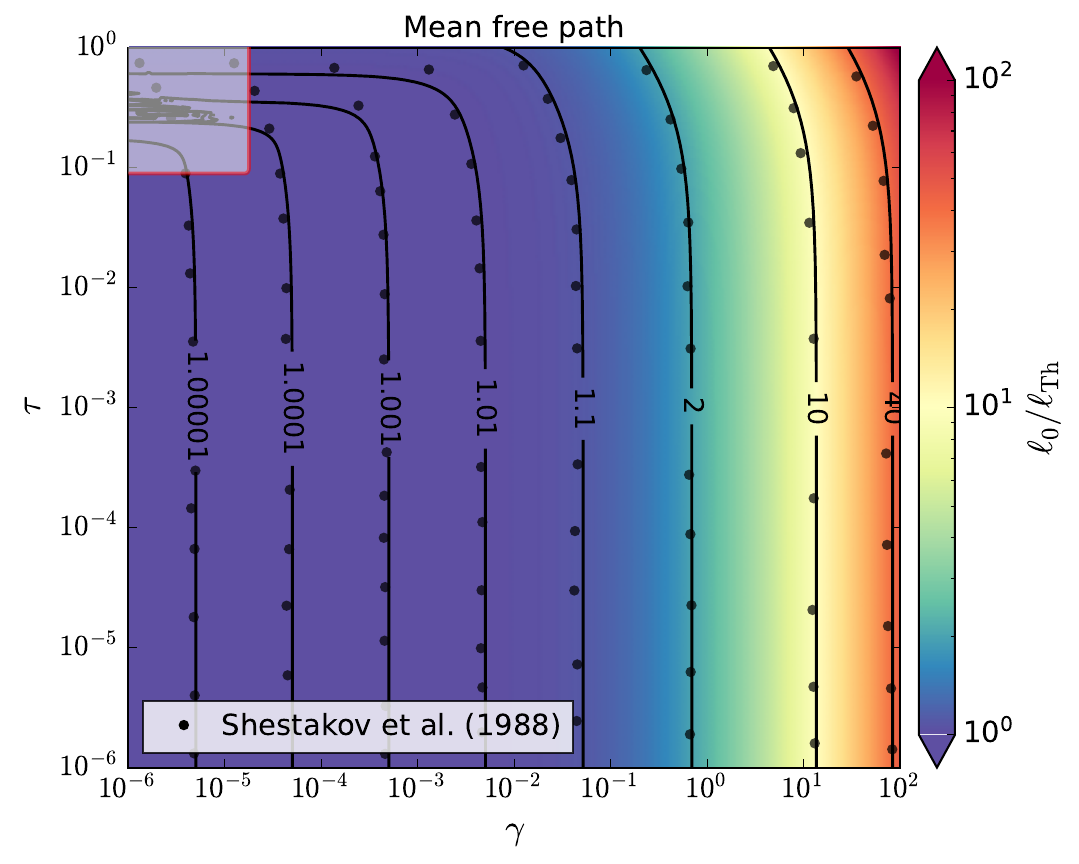}}
  \hfill
  \subfloat[Transport mean free path]{%
    \includegraphics[width=0.49\textwidth]{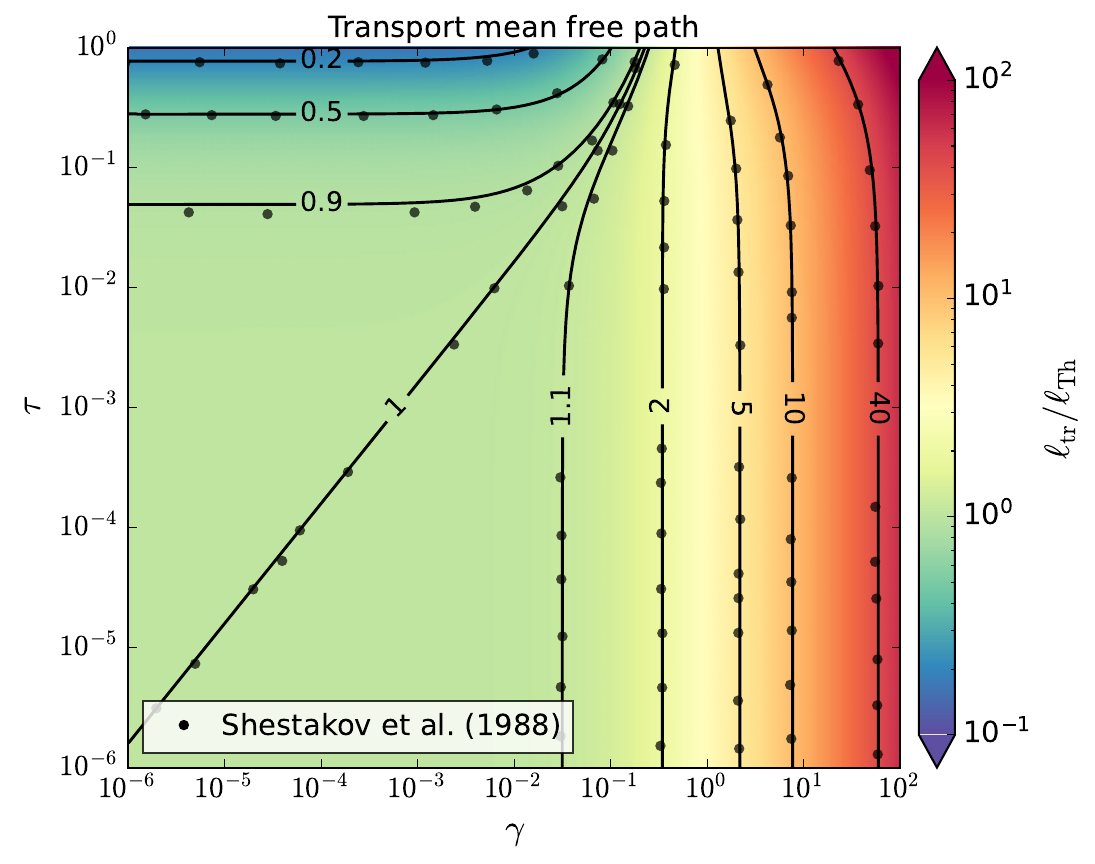}}
  \caption{Contours of the Compton mean free path
  $\ell_0/\ell_{\rm Th}$ and transport mean free path
  $\ell_{\rm tr}/\ell_{\rm Th}$ in the $(\gamma,\tau)$ plane. Solid
  contours show the present calculation and dashed contours show
  digitized data from \citet{SHESTAKOV1988577}. The red box marks the
  region where cancellation in the pointwise kernel evaluation limits
  the attainable accuracy, as discussed in
  Section~\ref{sec:cancellation_control}.}
  \label{fig:shestakov_mfp}
\end{figure*}

To validate the $\mathcal{E}'+\xi$ quadrature, we compare
energy-integrated quantities computed with the present scheme against
published results. For brevity we define the "moments"
\begin{equation}
\mathcal{M}[X]
=
\frac{2\pi}{\sigma_{\rm Th}}
\mathcal{I}_{[0,\infty]\,[-1,1]}[X].
\label{eq:moment_functional}
\end{equation}
The energy-transfer "moments" are
\begin{equation}
\mathcal{M}\!\left[
\left(
\frac{\mathcal{E}'-\mathcal{E}}{\mathcal{E}}
\right)^m
\right],
\qquad
m=0,1,2.
\label{eq:energy_moment}
\end{equation}
We also define a mixed angular-energy moment
\begin{equation}
\mathcal{M}\!\left[
\frac{\mathcal{E}'}{\mathcal{E}}\xi
\right].
\label{eq:angular_energy_moment}
\end{equation}
Define the Compton mean free path and transport mean free path,
normalized by the Thomson value $\ell_{\rm Th}=1/(N_e\sigma_{\rm Th})$, where $N_e$ cancels and the ratios depend only on the microscopic cross section \citep{SHESTAKOV1988577}
\begin{equation}
\frac{\ell_0}{\ell_{\rm Th}}
=
\frac{1}{\mathcal{M}[1]},
\qquad
\frac{\ell_{\rm tr}}{\ell_{\rm Th}}
=
\frac{1}{\mathcal{M}[1]
-
\mathcal{M}\!\left[\frac{\mathcal{E}'}{\mathcal{E}}\xi\right]}.
\label{eq:mean_free_paths}
\end{equation}
Figure~\ref{fig:sarkar_moments} compares the two energy-transfer
moments
\begin{equation}
\mathcal{M}\!\left[
\left(
\frac{\mathcal{E}'-\mathcal{E}}{\mathcal{E}}
\right)^m
\right],
\qquad m=1,2,
\label{eq:sarkar_energy_moments}
\end{equation}
against a high-precision evaluation of the closed-form moments implemented in \textsc{CSpack} \citep{CSpack}.
These closed forms give the moments for a single electron momentum \citep{Sarkar2019}. We evaluate them with 50 significant digits and average them over the Maxwell--J\"uttner distribution at the same precision. The extended precision makes the reference far more accurate than the $10^{-8}$ level tested here at all photon energies and temperatures. For applications, \textsc{CSpack} also provides expressions for low photon energies that are evaluated efficiently in double precision \citep{Sarkar2019}. The present scheme agrees with the 50-digit average to within $10^{-6}$ at all temperatures shown, and to $\sim10^{-8}$ for $\mathcal{E}\le10$\,MeV.
The moment functional in Eq.~\eqref{eq:moment_functional} requires
integrating over $\mathcal{E}'\in[0,\infty)$. Because the kernel
decays exponentially beyond the thermally broadened recoil band
(Section~\ref{sec:Ep_integration}), the semi-infinite domain is
replaced by a finite interval extending $50\,\delta\mathcal{E}'$
Eq.~\eqref{eq:Eprime_width} beyond the cold recoil band on each side.
The interval is subdivided into three log-spaced core panels and two
tail panels, each integrated with Gauss--Legendre quadrature of order
256 ($N_{\rm edge}=N_{\rm int}=256$).
For Figure~\ref{fig:sarkar_moments}, which extends to $\mathcal{E}=10^6$\,keV and $k_{\rm B}T=160$\,keV, the tails are instead extended to the scattered energies kinematically allowed for electrons with Lorentz factors up to $1+40\tau$, and the interval is clipped to these limits. It is divided into geometric panels no wider than a factor of two, each integrated with Algorithm~\ref{alg:Eprime_integration} at the same orders, and the angular orders are raised to $N_{\rm peak}=1536$ and $N_{\rm tail}=512$ to resolve the forward-peaked kernel at $\mathcal{E}\gg m_ec^2$.
Figure~\ref{fig:shestakov_mfp} compares the Compton mean free path
$\ell_0/\ell_{\rm Th}$ and transport mean free path
$\ell_{\rm tr}/\ell_{\rm Th}$, computed from
Eq.~\eqref{eq:mean_free_paths}, against digitized contours from
\citet{SHESTAKOV1988577}. The two panels show the results in the dimensionless
$(\gamma,\tau)$ plane. The red box marks the low-$\gamma$, high-$\tau$
region where the cancellation discussed in
Section~\ref{sec:cancellation_control} limits the attainable accuracy (unless higher than double-double precision arithmetic is used).
The $\mathcal{E}'$ integration uses the same paneling and quadrature
order as for the energy-transfer moments
($N_{\rm edge}=N_{\rm int}=256$). The angular quadrature
order is elevated to $N_{\rm peak}=768$ and $N_{\rm tail}=256$ to
resolve the low-$\gamma$ regime, where the total cross section
approaches the Thomson value and the quantity of interest,
$\ell_0/\ell_{\rm Th} - 1$, is a small residual.

\begin{figure}[!htbp]
  \centering
  \includegraphics[width=\columnwidth]{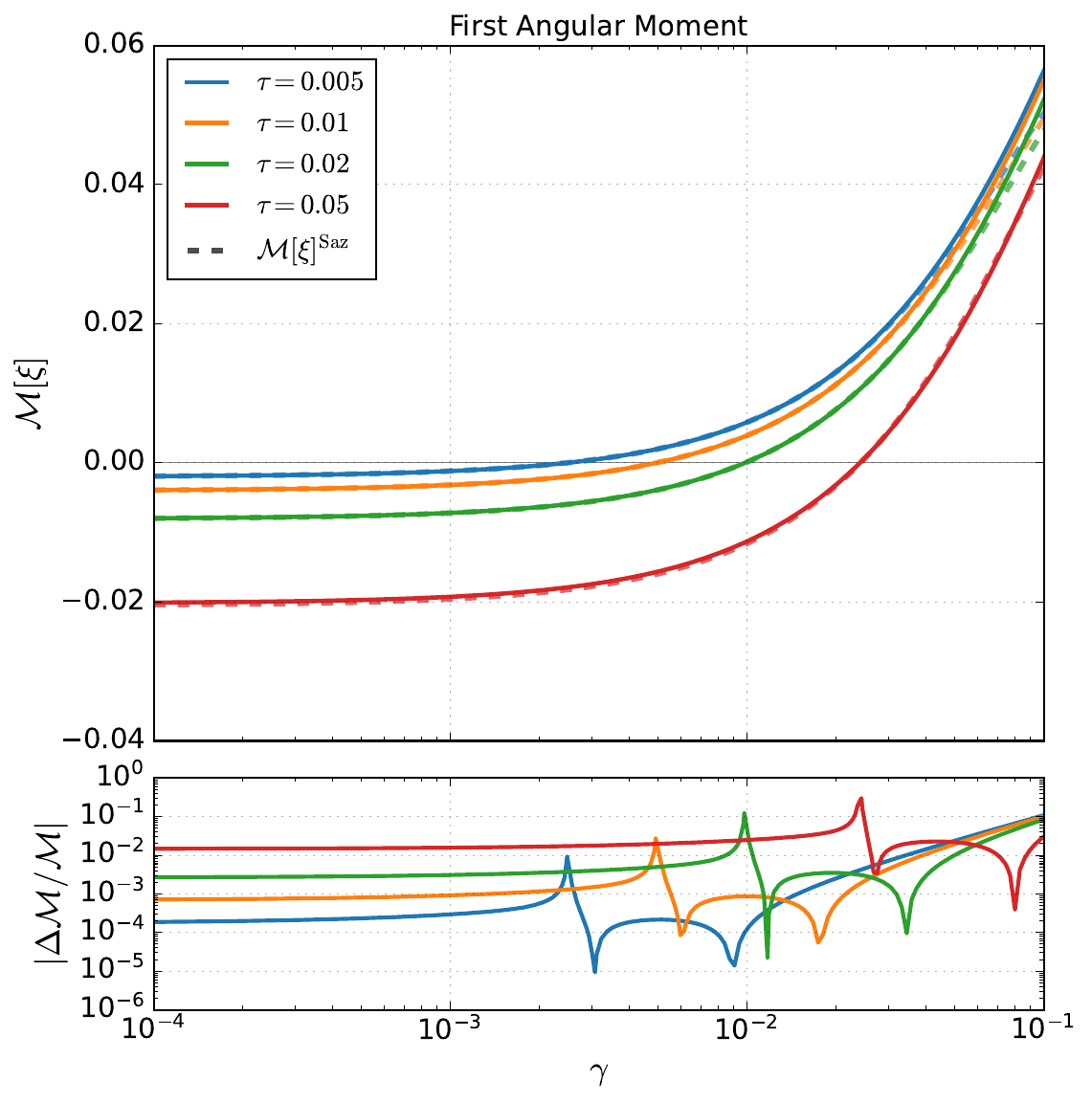}
  \caption{First angular moment $\mathcal{M}[\xi]$ as a function of
  $\gamma$ at several values of $\tau$. Solid lines show the present
  numerical calculation and dashed lines show the second-order
  approximation Eq.~\eqref{eq:sazonov_first_moment}. The lower panel shows the absolute relative
  error $|\Delta\mathcal{M}/\mathcal{M}|$ between the two on a
  logarithmic scale.}
  \label{fig:sazonov_first_moment}
\end{figure}

As a further angular-moment validation, we compare the first Legendre
moment of the angular scattering function against the mildly
relativistic expansion of \citet{Sazonov2000}. The first angular
moment is
\begin{equation}
\mathcal{M}[\xi]
=
\frac{2\pi}{\sigma_{\rm Th}}
\int_0^\infty d\mathcal{E}'
\int_{-1}^{1} d\xi\;
\xi\,\mathcal{K}(\mathcal{E}\to\mathcal{E}',\xi;\tau).
\label{eq:first_angular_moment_def}
\end{equation}
Multiplying equation~(15) of \citet{Sazonov2000} by $\xi$ and
integrating over all scattering angles yields a second-order
approximation valid for $\gamma\ll 1$ and $\tau\ll 1$:
\begin{multline}
\mathcal{M}[\xi]^{\rm Saz}
=
\frac{4}{5}\gamma
- \frac{2}{5}\tau
- \frac{29}{10}\gamma^2 \\
+ \frac{12}{5}\gamma\tau
- \frac{1}{5}\tau^2
+ O\!\left(\gamma^3,\gamma^2\tau,\gamma\tau^2,\tau^3\right).
\label{eq:sazonov_first_moment}
\end{multline}
Figure~\ref{fig:sazonov_first_moment} compares the numerically
computed $\mathcal{M}[\xi]$ against
Eq.~\eqref{eq:sazonov_first_moment} over a range of $\gamma$ at
several temperatures.

\subsection{\texorpdfstring{$\mathcal{E}$}{E} Integration} \label{sec:E_integration}

The incident-energy quadrature integrates over the source group
$\Delta\mathcal{E}_g$. The inner integral
$\mathcal{I}_{\Delta\mathcal{E}_{g'}\Delta\xi_i}[X]
(\mathcal{E};\tau)$ is weighted by a prescribed incident spectrum
$w(\mathcal{E},T)$. The integral is then normalized by
\begin{equation}
W_g(T)
=
\int_{\mathcal{E}_{g-1/2}}^{\mathcal{E}_{g+1/2}}
w(\mathcal{E},T)\,d\mathcal{E}.
\label{eq:group_weight_normalization}
\end{equation}
The outscattering cross section for a transition from source group $g$
to target group $g'$ in angular bin $\Delta\xi_i$
(Eq.~\eqref{eq:angle_bin_definition}) is
\begin{multline}
\sigma_{\mathrm{out}}(g\rightarrow g',i; T)
=\\
\frac{2\pi}{W_g(T)}
\int_{\Delta\mathcal{E}_g}
w(\mathcal{E},T)\,
\mathcal{I}_{\Delta\mathcal{E}_{g'}\Delta\xi_i}[1]
(\mathcal{E};\tau)\,
d\mathcal{E}
\label{eq:sigma_out_angle_resolved}
\end{multline}
The factor $2\pi$ accounts for azimuthal symmetry.
The corresponding angle-integrated form is
\begin{multline}
\sigma_{\mathrm{out}}(g\rightarrow g'; T)
=\\
\frac{2\pi}{W_g(T)}
\int_{\Delta\mathcal{E}_g}
w(\mathcal{E},T)\,
\mathcal{I}_{\Delta\mathcal{E}_{g'}[-1,1]}[1]
(\mathcal{E};\tau)\,
d\mathcal{E}
\label{eq:sigma_out}
\end{multline}
Replacing the unit integrand with an arbitrary function
$X$ and defining 
\begin{multline}
\langle X\rangle(g, g', i; T)
=\frac{2\pi}{\sigma_{\mathrm{out}}(g\rightarrow g',i; T)\,W_g(T)}\times\\
\int_{\Delta\mathcal{E}_g}w(\mathcal{E},T)\,\mathcal{I}_{\Delta\mathcal{E}_{g'}\Delta\xi_i}[X](\mathcal{E};\tau)\,d\mathcal{E}
\label{eq:multigroup_weighted_average}
\end{multline}
When the angular domain spans the full range $\Delta\xi_i=[-1,1]$, the
angular index $i$ is omitted from the notation.

At low temperatures the Compton kernel becomes very narrow,
and scattering transfers photons predominantly into adjacent groups.
To capture this group-to-adjacent-group coupling, we
identify boundary layers near the edges of the source group
using the same thermal width from Eq.~\eqref{eq:Eprime_width},
evaluated at the backscattering angle $\xi=-1$. Defining
\begin{equation}
\delta\mathcal{E}_{g\pm1/2}
\equiv
\delta\mathcal{E}'(\mathcal{E}_{g\pm1/2},\xi=-1,T),
\label{eq:boundary_thermal_width}
\end{equation}
the inner edges of the two boundary layers are
\begin{equation}
\mathcal{E}_{-}
=
\mathcal{E}_{g-1/2}
+
5\delta\mathcal{E}_{g-1/2},
\label{eq:incident_energy_left_boundary}
\end{equation}
and
\begin{equation}
\mathcal{E}_{+}
=
\mathcal{E}_{g+1/2}
-
5\delta\mathcal{E}_{g+1/2}.
\label{eq:incident_energy_right_boundary}
\end{equation}
The source group is divided into a left boundary layer
$[\mathcal{E}_{g-1/2},\mathcal{E}_{-}]$, a central region
$[\mathcal{E}_{-},\mathcal{E}_{+}]$, and a right boundary layer
$[\mathcal{E}_{+},\mathcal{E}_{g+1/2}]$. For a peaked weighting function, such as the Planck or Wien spectrum,
the central region is further divided at the spectral peak
$\mathcal{E}_{\mathrm{pk}}(T)$ when the peak lies within the source
group. Algorithm~\ref{alg:E_integration} summarizes the complete
incident-energy quadrature.
\begin{algorithm}
  \hrule\smallskip
  \begin{algorithmic}[1]
  \Require $w(\mathcal{E},T)$,
  $[\mathcal{E}_{\rm low},\mathcal{E}_{\rm high}]$,
  $[\mathcal{E}'_{\rm low},\mathcal{E}'_{\rm high}]$,
  $[\xi_{\rm low},\xi_{\rm high}]$,
  \Ensure
  $I \approx
  \int_{\mathcal{E}_{\rm low}}^{\mathcal{E}_{\rm high}}
  w\,\mathcal{I}_{\Delta\mathcal{E}_{g'}\Delta\xi_i}[X]\,d\mathcal{E}$
  \State $f(\mathcal{E})\gets
  w(\mathcal{E},T)\,
  \mathcal{I}_{\Delta\mathcal{E}_{g'}\Delta\xi_i}[X]
  (\mathcal{E};\tau)$

  \If{$\mathcal{E}_{-}\geq\mathcal{E}_{+}$}
  \Comment{boundary layers overlap}
    \State \Return $\mathrm{Middle}(f,N_{\mathcal{E}},\mathcal{E}_{\rm low},\mathcal{E}_{\rm high})$
  \EndIf

  \State $I\mathrel \gets
  \mathrm{MapGL}(f,N_{\mathcal{E}},\mathcal{E}_{\rm low},\mathcal{E}_{-})$
  \Comment{left boundary}

  \State $I\mathrel{+}=
  \mathrm{Middle}(f,N_{\mathcal{E}},\mathcal{E}_{-},\mathcal{E}_{+})$
  \Comment{central region}

  \State $I\mathrel{+}=
  \mathrm{MapGL}(f,N_{\mathcal{E}},\mathcal{E}_{+},\mathcal{E}_{\rm high})$
  \Comment{right boundary}

  \State \Return $I$
  \end{algorithmic}
  \smallskip\hrule
  \caption{Structure-aware $\mathcal{E}$-integration for fixed target group
  $\Delta\mathcal{E}_{g'}$ and angular bin $\Delta\xi_i$.
  The quadrature rules $\mathrm{MapGL}$ and $\mathrm{Middle}$ are
  defined in Appendix~\ref{sec:quadrature_rules_appendix}.
  The boundary-layer edges $\mathcal{E}_{\pm}$ are defined in
  Eqs.~\eqref{eq:incident_energy_left_boundary}--%
  \eqref{eq:incident_energy_right_boundary}.}
  \label{alg:E_integration}
\end{algorithm}

\subsubsection{Validation of the
\texorpdfstring{$\mathcal{E}+\mathcal{E}'+\xi$}{E+E'+xi} Quadrature}
\label{sec:E_Ep_xi_validation}

We validate the full multigroup-multiangle quadrature
against previously published data.

Using the outscattering cross section
(Eq.~\eqref{eq:sigma_out}), the outgoing-group
distribution is
\begin{equation}
P(g\rightarrow g'; T)
\equiv
\frac{\sigma_{\mathrm{out}}(g\rightarrow g'; T)}
{\displaystyle\sum_h
 \sigma_{\mathrm{out}}(g\rightarrow h; T)}.
\label{eq:outgoing_group_pdf}
\end{equation}
For a specified group transition, we similarly define the
angular distribution
\begin{equation}
P(g\rightarrow g',i; T)
\equiv
\frac{\sigma_{\mathrm{out}}(g\rightarrow g',i; T)}
{\displaystyle\sum_j
 \sigma_{\mathrm{out}}(g\rightarrow g',j; T)}.
\label{eq:angular_pdf}
\end{equation}

\begin{figure}[!htbp]
  \centering
  \includegraphics[width=\linewidth]
    {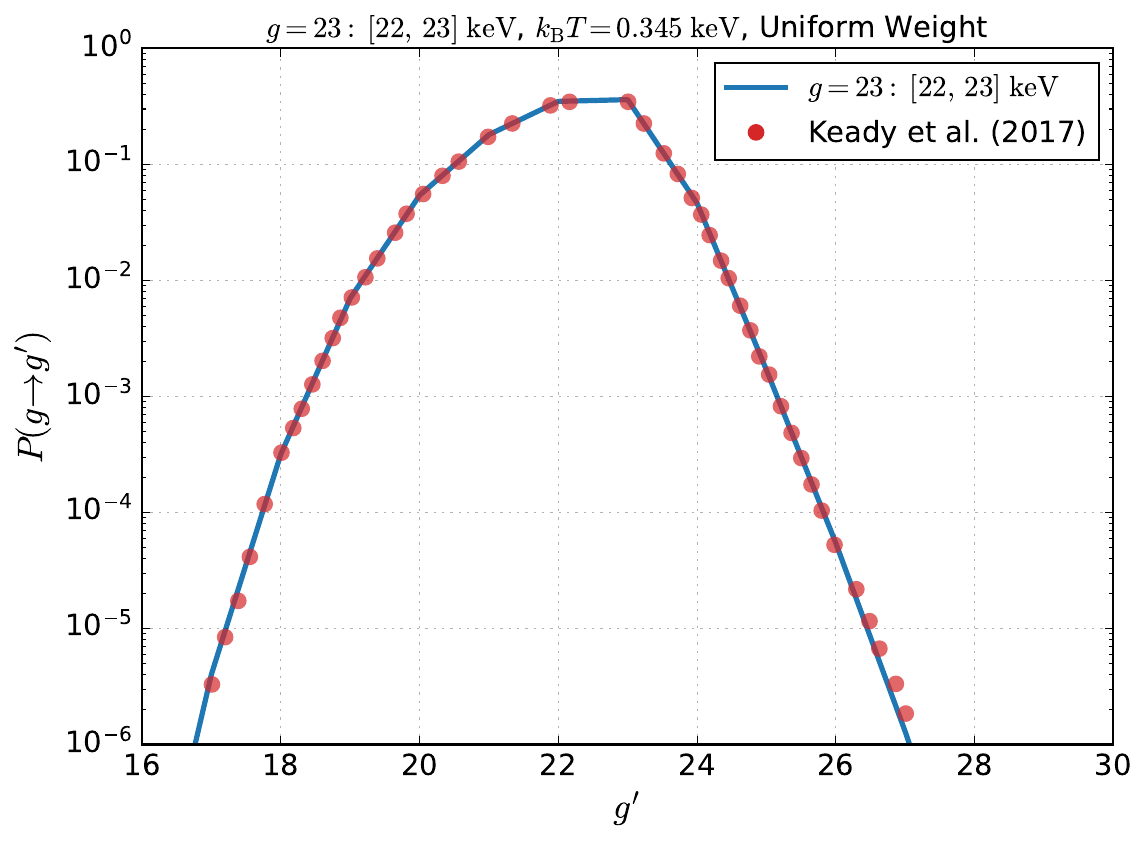}
  \caption{
  Normalized outgoing-group distribution $P(g\rightarrow g')$ for
  source group $g=23$ with uniform weighting at
  $k_{\rm B}T=0.345~\mathrm{keV}$. The solid curve shows the present
  calculation and markers show digitized results from
  \citet{Keady2017}.
  The calculation uses a 30-group with
  boundaries at $0.75$, $1$, $2$, $\ldots$, $30~\mathrm{keV}$
  (angle-integrated).}
  \label{fig:keady_outgoing_group_pdf}
\end{figure}

We validate these distributions against the multigroup calculations of
\citet{Keady2017} at $k_{\rm B}T=0.345~\mathrm{keV}$. The first set of
comparisons uses the 30-group grid uniform
incident-energy weighting. The grid boundaries lie at $0.75$, $1$, $2$,
$\ldots$, $30~\mathrm{keV}$, so that group~1 spans
$0.75$--$1.0~\mathrm{keV}$ and groups~2--30 each span
$1~\mathrm{keV}$. Figure~\ref{fig:keady_outgoing_group_pdf} compares
the outgoing-group distribution $P(g\rightarrow g')$ for
source group $g=23$  with the digitized
results.

\begin{figure*}[!htbp]
  \centering
  \subfloat[$g'=18,\,19$ (downscatter)]{%
    \includegraphics[width=0.49\textwidth]
      {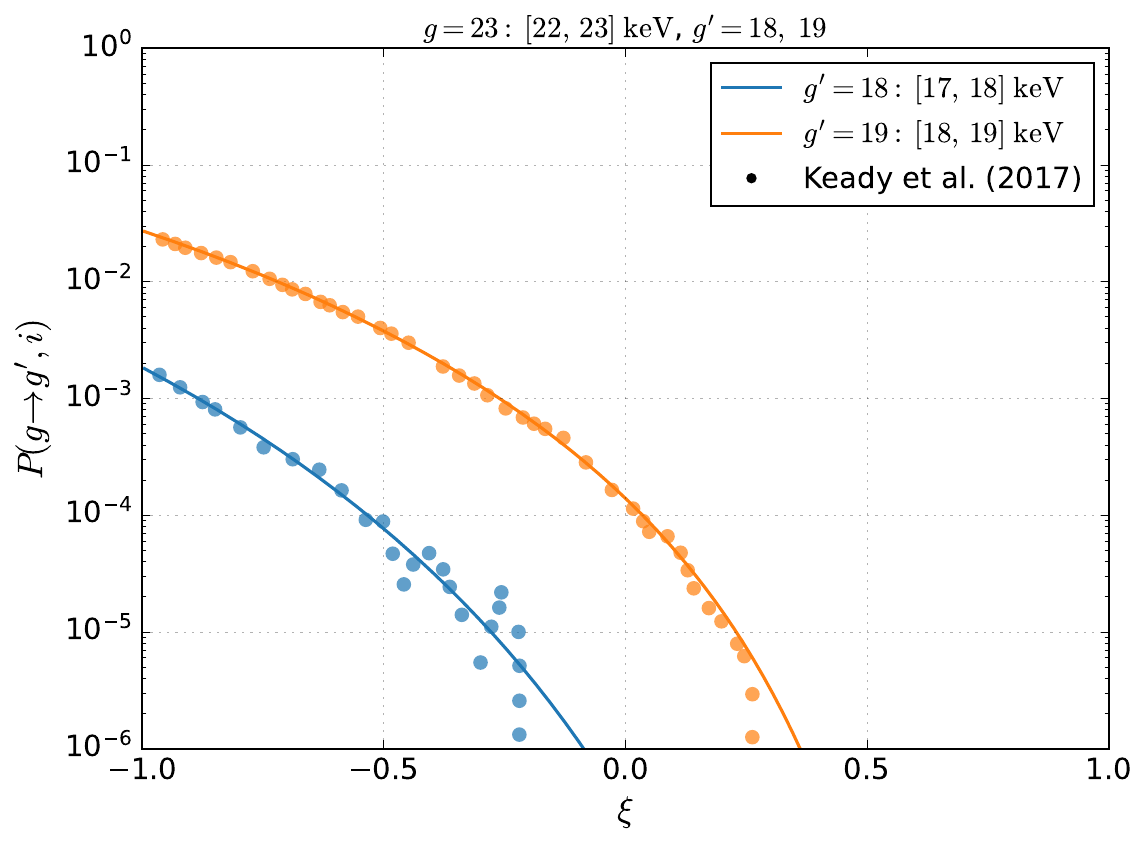}}
  \hfill
  \subfloat[$g'=20,\,21$ (downscatter)]{%
    \includegraphics[width=0.49\textwidth]
      {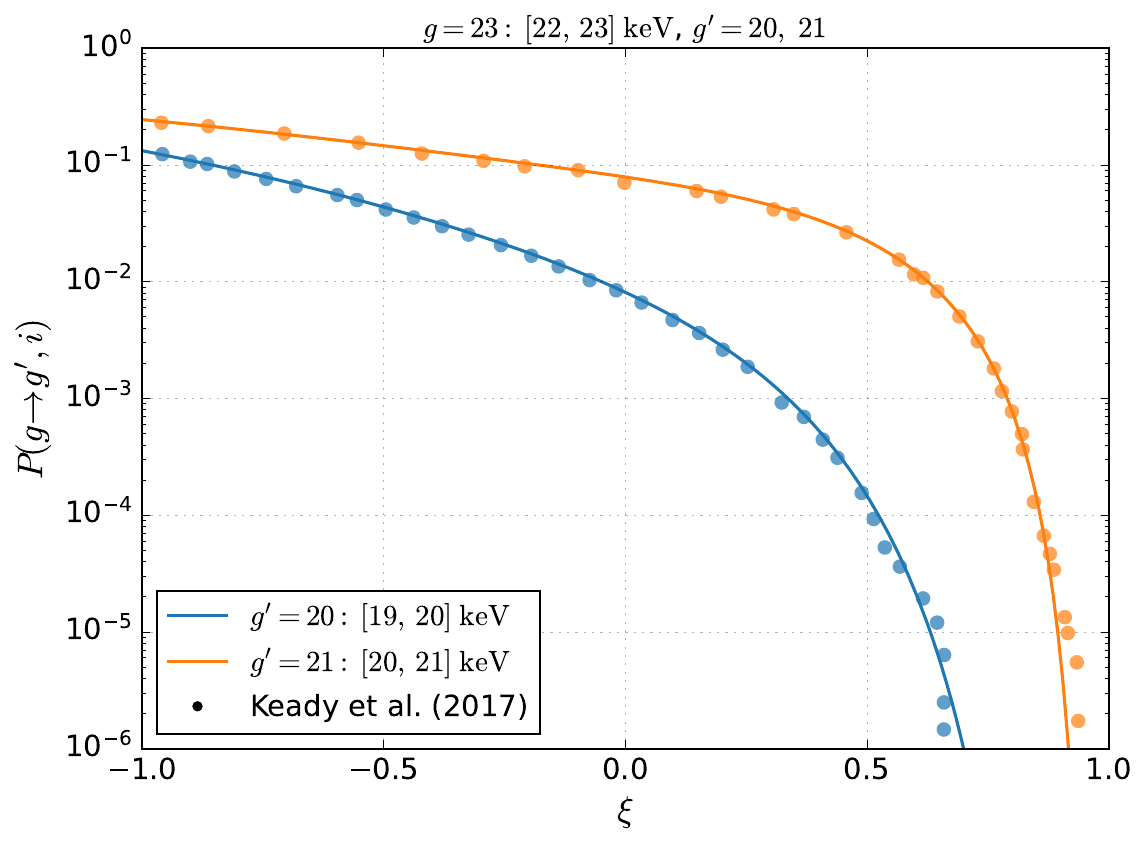}}\\[6pt]
  \subfloat[$g'=22,\,23$]{%
    \includegraphics[width=0.49\textwidth]
      {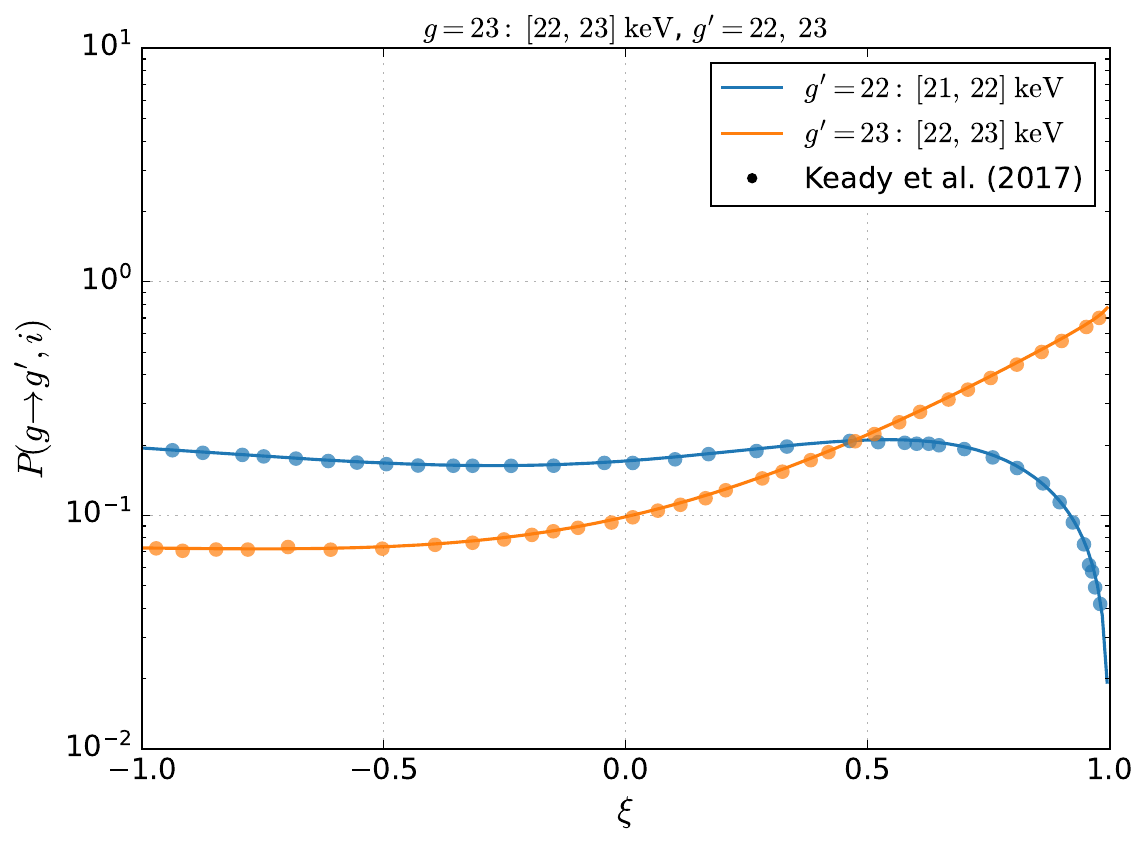}}
  \hfill
  \subfloat[$g'=24,\,25$ (upscatter)]{%
    \includegraphics[width=0.49\textwidth]
      {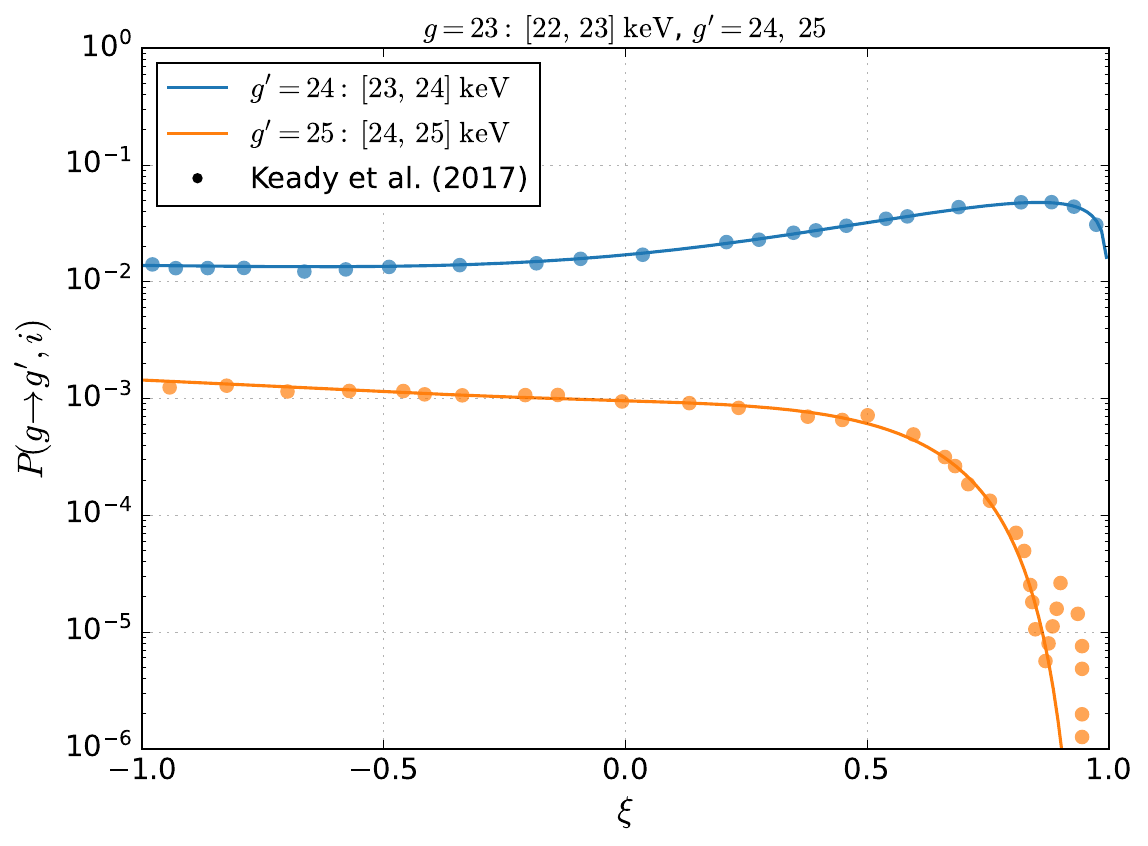}}
  \caption{
  Conditional angular distributions $P(g\rightarrow g',i)$ for
  source group $g=23$ with uniform weighting at
  $k_{\rm B}T=0.345~\mathrm{keV}$. Each panel shows two target groups,
  spanning downscattering, inscattering, and upscattering.
  Solid lines show the present calculation and
  markers show the digitized results of \citet{Keady2017}.
  The calculation uses the same 30-group energy grid as
  Figure~\ref{fig:keady_outgoing_group_pdf} with 200 uniform angular bins.}
  \label{fig:keady_angular_pdf}
\end{figure*}

\begin{figure}[!htbp]
  \centering
  \includegraphics[width=\linewidth]
    {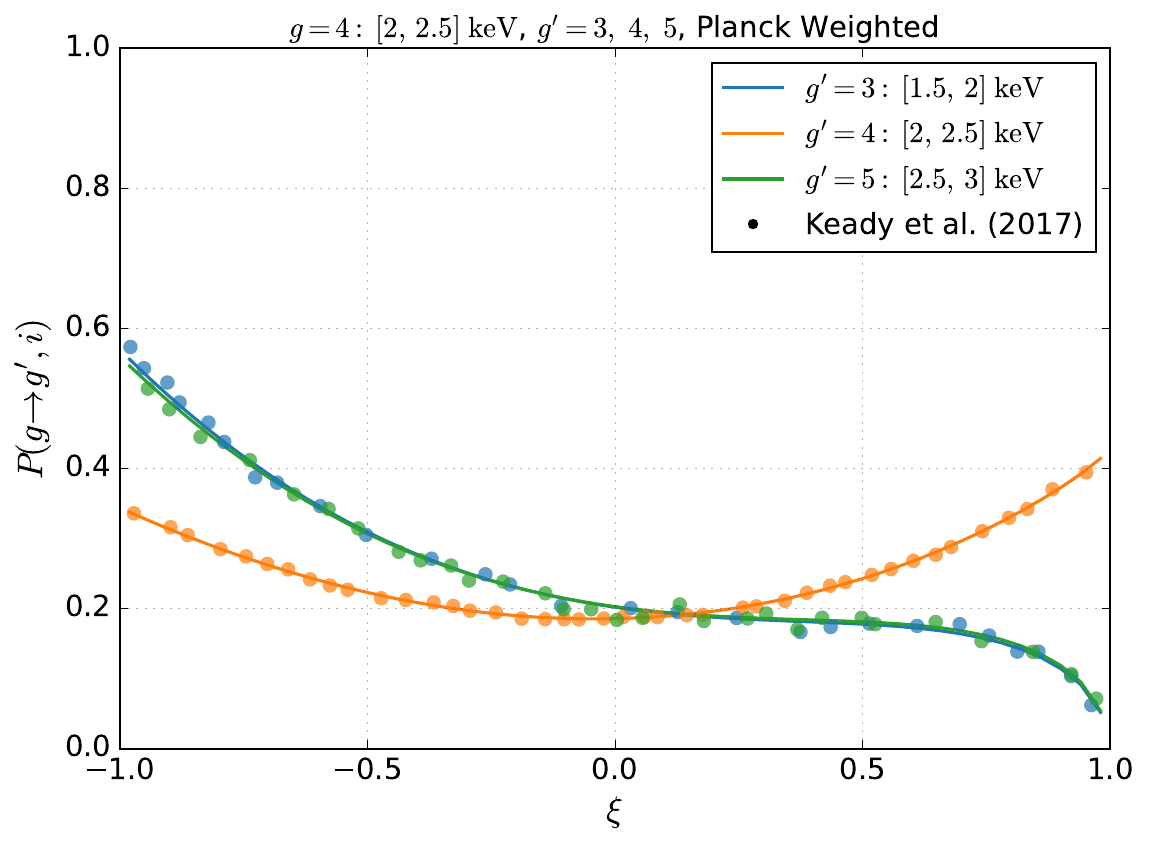}
  \caption{
  Conditional angular distribution $P(g\rightarrow g',i)$ for source
  group $g=4$ and target groups
  $g'=3,4,5$ with Planck weighting at
  $k_{\rm B}T=0.345~\mathrm{keV}$. The solid curves show the present calculation
  and markers show digitized results from \citet{Keady2017}.
  The calculation uses a 20-group grid with boundaries at $0.55$,
  $1.0$, $1.5$, $\ldots$, $10.5~\mathrm{keV}$ and 50 uniform angular bins.}
  \label{fig:keady_planck_angular_pdf}
\end{figure}

Figure~\ref{fig:keady_angular_pdf} extends the comparison to the
angular distributions $P(g\rightarrow g',i)$ for source
group $g=23$ on the same grid, resolved into 200 uniform angular bins.
The remaining comparisons with \citet{Keady2017} use a 20-group
Planck-weighted grid with
elevated quadrature orders ($N_{\rm peak}=96$, $N_{\rm tail}=32$,
$N_{\rm edge}=N_{\rm int}=48$, $N_{\mathcal{E}}=24$).
The weight function is the capped Planck spectral energy density
(Appendix~\ref{sec:group_denominators}),
\begin{equation}
w(\mathcal{E},T)
=
\begin{cases}
\dfrac{\mathcal{E}^3/(k_{\rm B}T)^3}
{\exp[\mathcal{E}/(k_{\rm B}T)]-1},
& \mathcal{E}/(k_{\rm B}T) < x_{\rm cap},\\[6pt]
\dfrac{x_{\rm cap}^3}{e^{x_{\rm cap}}-1},
& \mathcal{E}/(k_{\rm B}T) \ge x_{\rm cap}.
\end{cases}
\label{eq:capped_planck_weight}
\end{equation}
The cap threshold $x_{\rm cap}$ prevents the weight function from
underflowing at high $\mathcal{E}/(k_{\mathrm B}T)$.
For the Keady comparison, $x_{\rm cap}=50$.
The boundaries lie at $0.55$, $1.0$, $1.5$,
$\ldots$, $10.5~\mathrm{keV}$, so that group~1 spans
$0.55$--$1.0~\mathrm{keV}$ and groups~2--20 each span
$0.5~\mathrm{keV}$. Figure~\ref{fig:keady_planck_angular_pdf} shows
the angular distribution, resolved into 50 uniform angular bins.

\begin{figure*}[!htbp]
  \centering
  \subfloat[$g=6\rightarrow g'=5$]{%
    \includegraphics[width=0.49\textwidth]
      {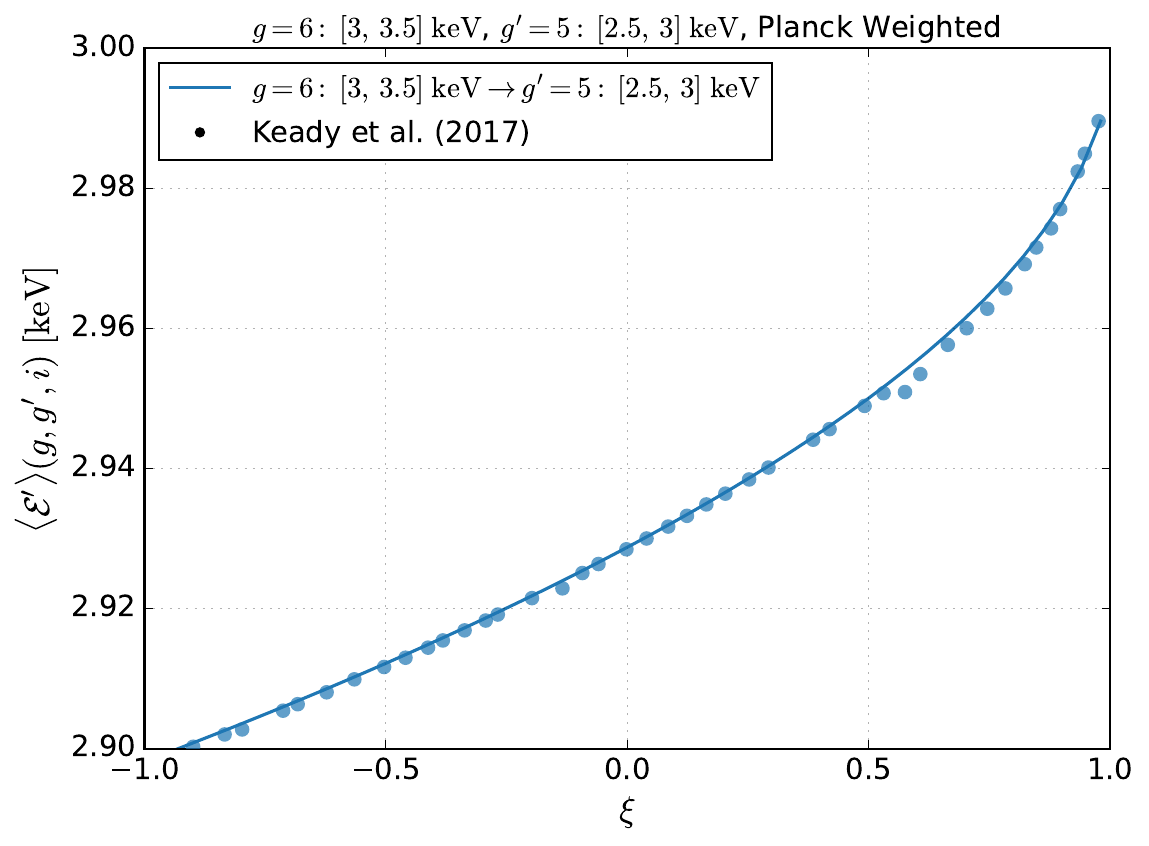}}
  \hfill
  \subfloat[$g=4\rightarrow g'=3$]{%
    \includegraphics[width=0.49\textwidth]
      {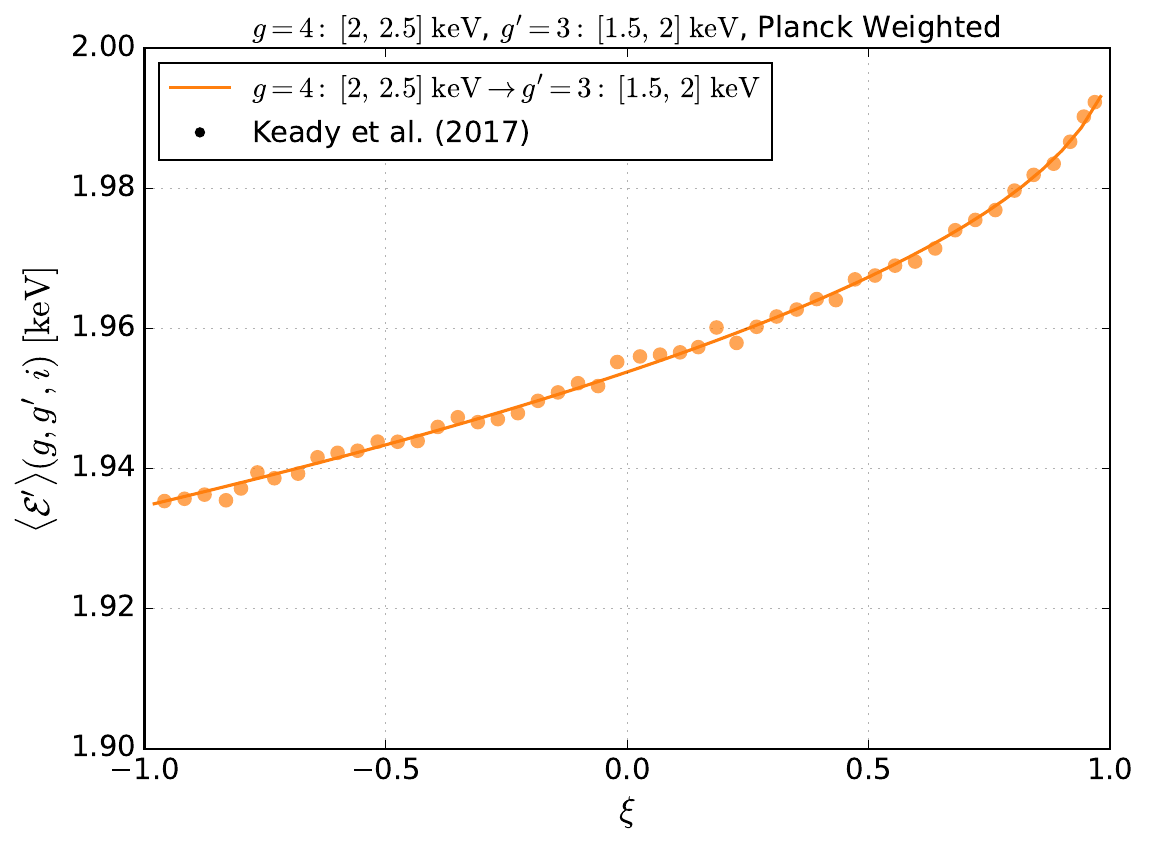}}
  \caption{
  Conditional mean outgoing energy
  $\left\langle \mathcal{E}'\right\rangle(g, g', i; T)$ as a
  function of angular bin with Planck weighting at
  $k_{\rm B}T=0.345~\mathrm{keV}$. Solid lines show the present calculation,
  and markers show digitized results from \citet{Keady2017}.
  The calculation uses the same 20-group grid and 50 angular bins as
  Figure~\ref{fig:keady_planck_angular_pdf}.}
  \label{fig:keady_avg_freq}
\end{figure*}

On the same grid, we compare the conditional mean outgoing energy
$\left\langle \mathcal{E}' \right\rangle(g, g', i; T)$
in each angular bin.
Figure~\ref{fig:keady_avg_freq} compares
$\left\langle \mathcal{E}'\right\rangle(g, g', i; T)$
with the results of \citet{Keady2017} for two transitions.

We next consider quantities formed from energy-weighted moments of the
multigroup kernel. The angle-integrated outscattering cross section is
defined in Eq.~\eqref{eq:sigma_out}.
The corresponding inscattering coefficient is
\begin{equation}
\sigma_{\mathrm{in}}(g\rightarrow g'; T)
=
\sigma_{\mathrm{out}}(g\rightarrow g'; T)\,
\left\langle \frac{\mathcal{E}'}{\mathcal{E}}\right\rangle(g, g'; T).
\label{eq:sigma_in}
\end{equation}
Following \citet{TILL2020_COMPTON_DATA_COMPARE}, we combine the outscattering and
inscattering coefficients into four groupwise diagnostic factors. For a
fixed group $g$, the mean amplification factor is
\begin{equation}
a_g(T)
=
\frac{
\displaystyle\sum_{g''}
\sigma_{\mathrm{in}}(g\rightarrow g''; T)
}{
\displaystyle\sum_{g''}
\sigma_{\mathrm{out}}(g\rightarrow g''; T)
}.
\label{eq:mean_amplification_factor}
\end{equation}
Separating the sums according to the direction of energy redistribution
gives the downscattering and upscattering factors,
\begin{equation}
d_g(T)
=
\frac{
\displaystyle\sum_{g''<g}
\sigma_{\mathrm{in}}(g\rightarrow g''; T)
}{
\displaystyle\sum_{g''<g}
\sigma_{\mathrm{out}}(g\rightarrow g''; T)
},
\label{eq:mean_downscattering_factor}
\end{equation}
and
\begin{equation}
u_g(T)
=
\frac{
\displaystyle\sum_{g''>g}
\sigma_{\mathrm{in}}(g\rightarrow g''; T)
}{
\displaystyle\sum_{g''>g}
\sigma_{\mathrm{out}}(g\rightarrow g''; T)
}.
\label{eq:mean_upscattering_factor}
\end{equation}
The corresponding within-group ratio is
\begin{equation}
c_g(T)
=
\frac{
\sigma_{\mathrm{in}}(g\rightarrow g; T)
}{
\displaystyle\sum_{g''}\sigma_{\mathrm{out}}(g\rightarrow g''; T)
}.
\label{eq:within_group_ratio}
\end{equation}
Here the ordering of the group indices follows the energy ordering of
the grid, so $g''<g$ denotes transitions toward lower photon energies
and $g''>g$ denotes transitions toward higher photon energies.

\begin{figure*}[!htbp]
  \centering
  \subfloat[Mean amplification factor $a_g$]{%
    \includegraphics[width=0.49\textwidth]
      {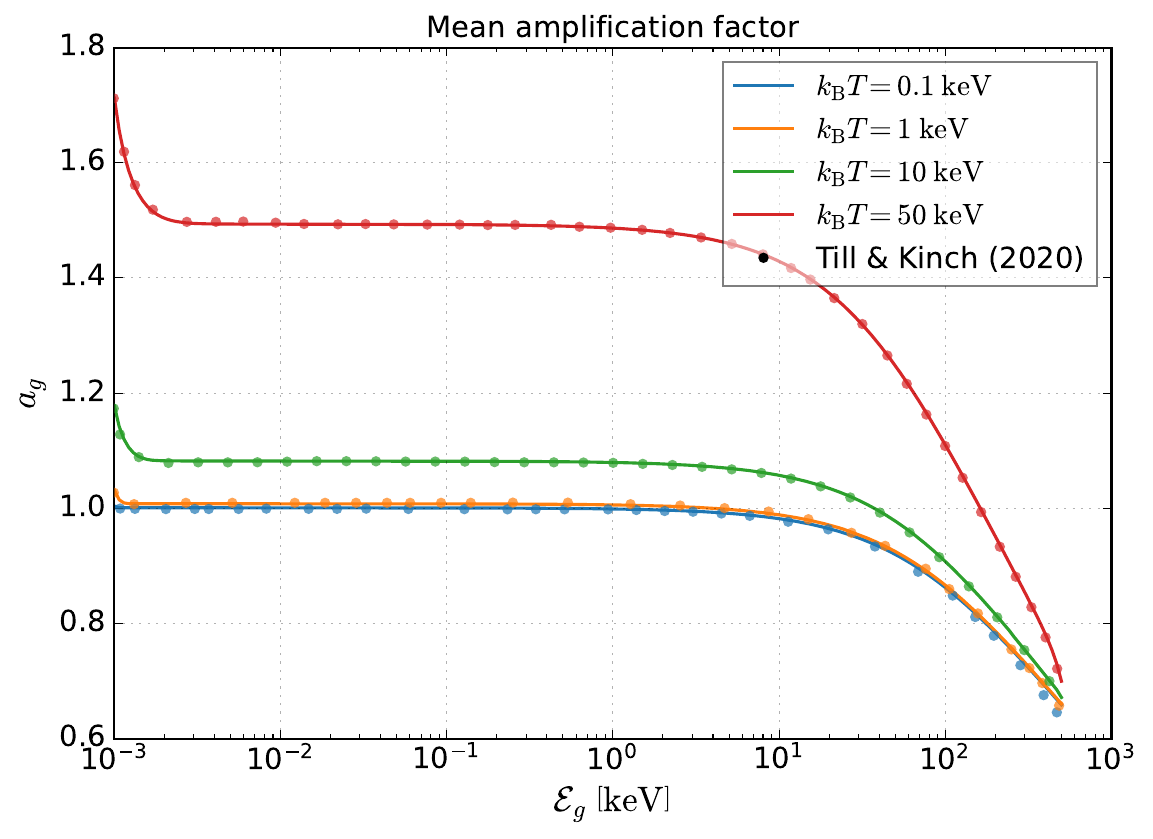}}
  \hfill
  \subfloat[Mean downscattering factor $d_g$]{%
    \includegraphics[width=0.49\textwidth]
      {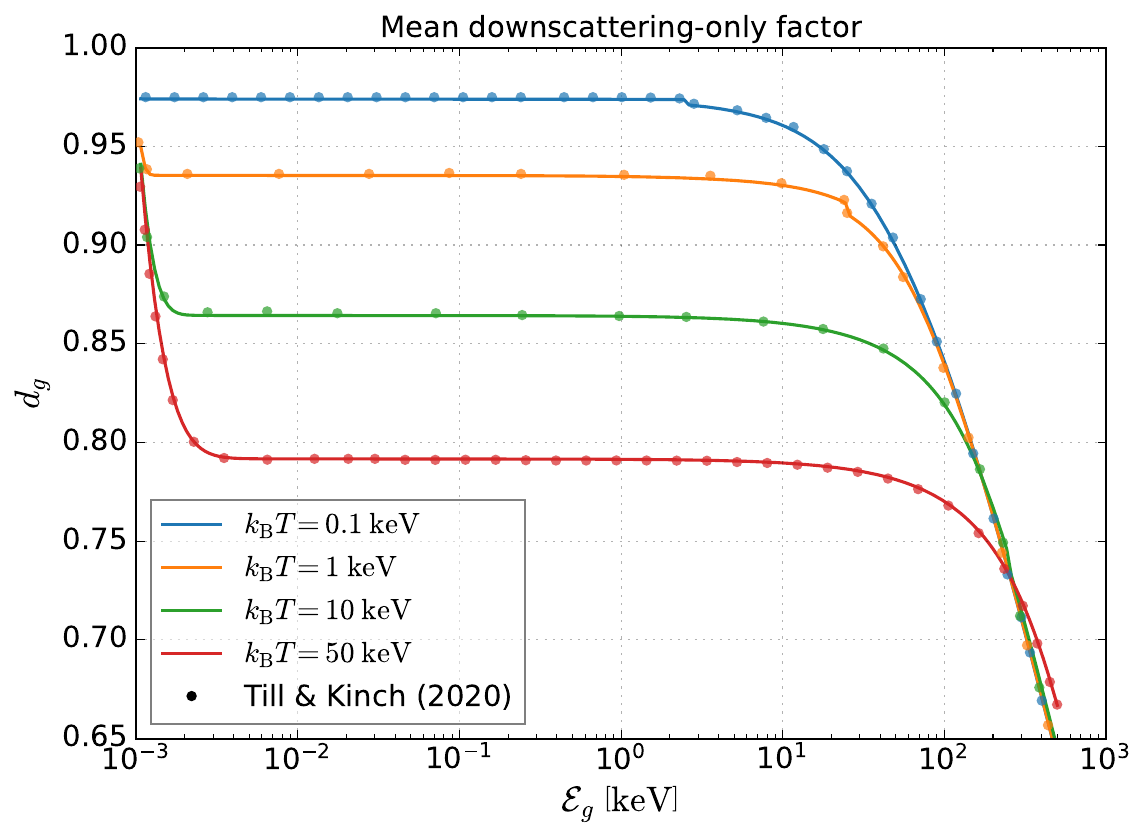}}\\[6pt]
  \subfloat[Mean upscattering factor $u_g$]{%
    \includegraphics[width=0.49\textwidth]
      {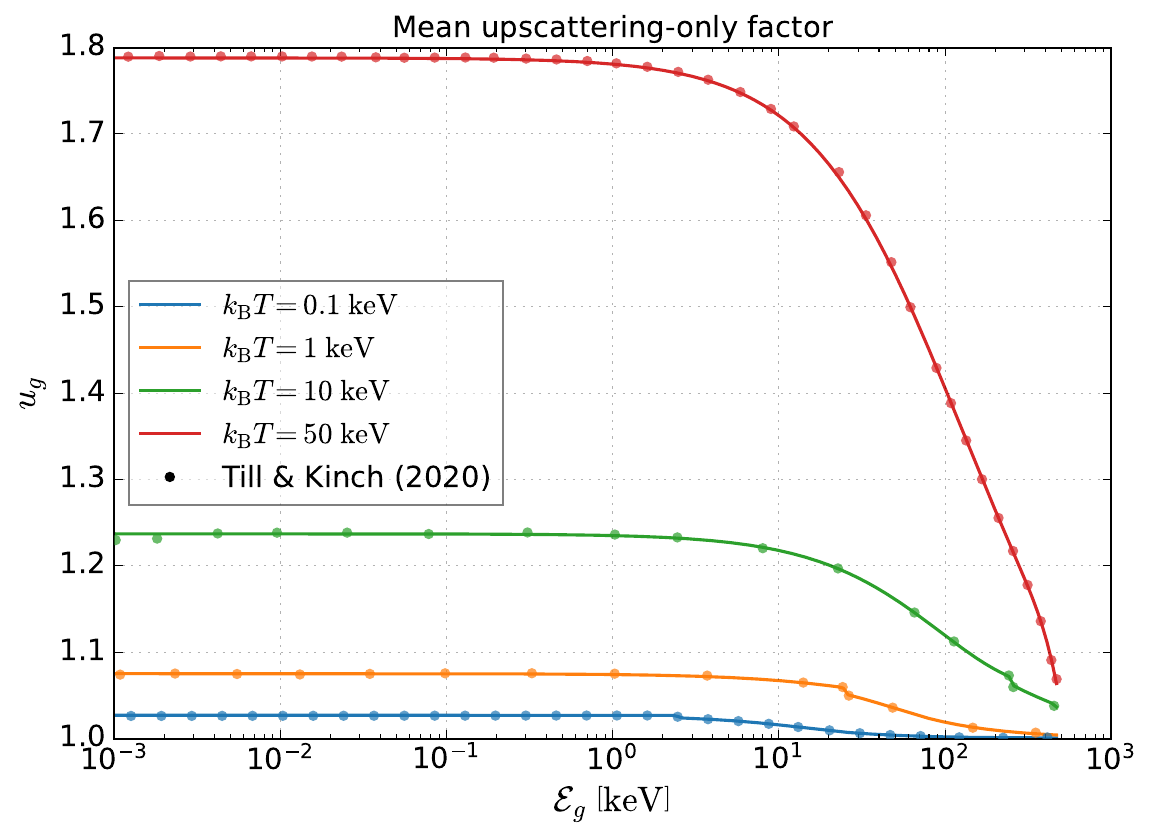}}
  \hfill
  \subfloat[Within-group scattering ratio $c_g$]{%
    \includegraphics[width=0.49\textwidth]
      {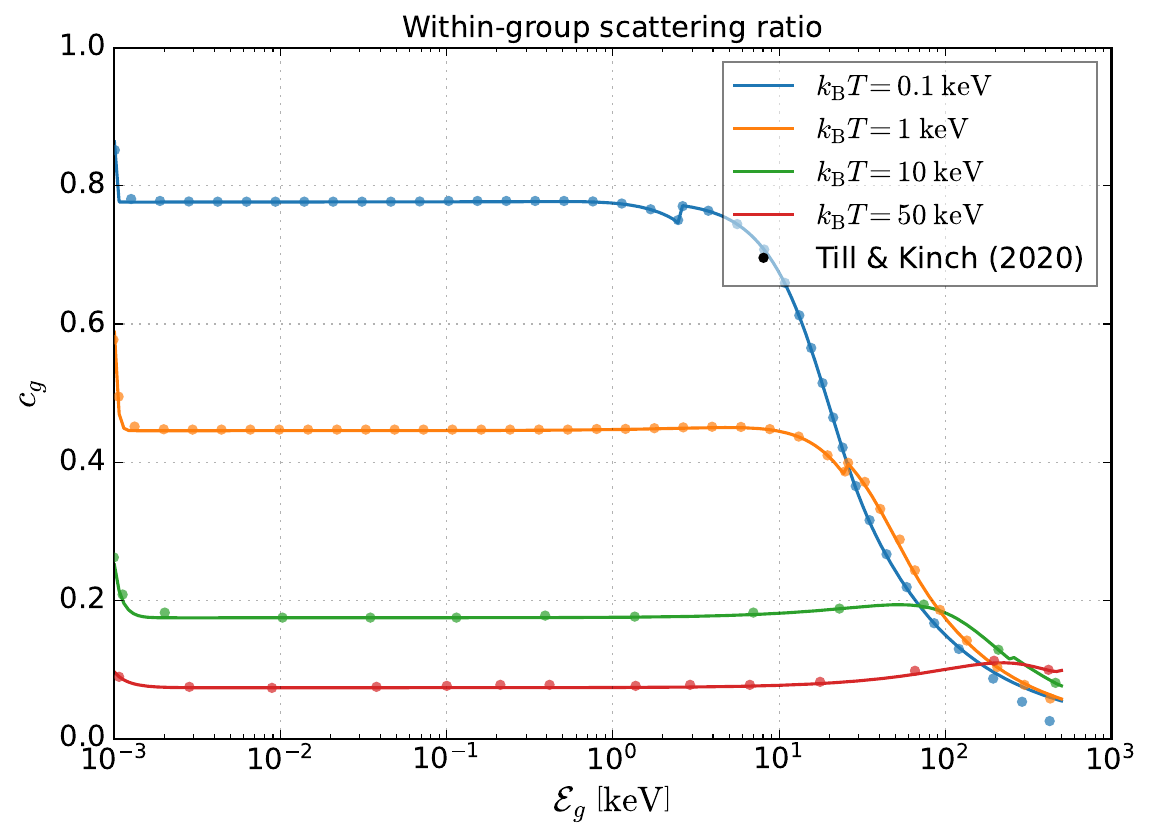}}
  \caption{
  Multigroup diagnostic factors with the capped Planck weighting of
  Eq.~\eqref{eq:capped_planck_weight} at $k_{\rm B}T=0.1$, $1$, $10$, and
  $50~\mathrm{keV}$.
  Solid lines show the present calculation, and markers show digitized
  data from \citet{TILL2020_COMPTON_DATA_COMPARE}.
  The calculation uses a 201-group geometric grid with boundaries
  spanning $9.7\times10^{-4}$--$517~\mathrm{keV}$
  (angle-integrated).}
  \label{fig:till_mg_factors}
\end{figure*}

Figure~\ref{fig:till_mg_factors} compares these four diagnostics with
the multigroup results of \citet{TILL2020_COMPTON_DATA_COMPARE} on a
201-group logarithmic grid with boundaries extending from approximately
$0.97~\mathrm{eV}$ to $517~\mathrm{keV}$. The incident-energy weight
function is the capped Planck spectral energy density of
Eq.~\eqref{eq:capped_planck_weight} with $x_{\rm cap}=25$.

\subsection{Full Temperature Derivative of the Multigroup Cross Section}
\label{sec:full_temperature_derivative}

The kernel-derivative integrals $\mathcal{I}^{\partial T}$
(Eq.~\eqref{eq:mgma_kernel_derivative_integral_objects}) capture the
temperature dependence of the microscopic kernel $\mathcal{K}$ alone.
When the incident-energy weight $w(\mathcal{E},T)$ depends on
temperature, as for the Planck weighting of
Eq.~\eqref{eq:capped_planck_weight}, the multigroup cross section
(Eq.~\eqref{eq:sigma_out_angle_resolved}) acquires additional
temperature dependence through the numerator weighting and the group
normalization $W_g(T)$. Differentiating $\sigma_{\mathrm{out}}(g\rightarrow g', i; T)$ gives
\begin{multline}
\frac{d\sigma_{\mathrm{out}}(g\rightarrow g',i; T)}{dT}
=
\frac{2\pi}{W_g(T)}
\mathcal{I}^{\partial T}_{\Delta\mathcal{E}_g
\Delta\mathcal{E}_{g'}\Delta\xi_i}[w]
\\
+\;
\frac{2\pi}{W_g(T)}
\mathcal{I}_{\Delta\mathcal{E}_g
\Delta\mathcal{E}_{g'}\Delta\xi_i}
\!\left[w\frac{\partial\ln w}{\partial T}\right]
\\\;-\;
\frac{\sigma_{\mathrm{out}}(g\rightarrow g',i; T)}{W_g(T)}
\frac{dW_g}{dT}.
\label{eq:full_dsigma_dT}
\end{multline}
All three terms can be evaluated by the same quadrature machinery
described in Sections~\ref{sec:xi_integration}--\ref{sec:E_Ep_xi_validation}.

\section{Results} \label{sec:results}

\subsection{Multigroup Matrices and Monte Carlo Comparison}
\label{sec:monte_carlo_comparison}

We now present the complete multigroup matrices and the corresponding
kernel-derivative contributions, then compare representative energy and angular profiles
with an independent Monte Carlo calculation
(Appendix~\ref{sec:monte_carlo_derivation_appendix}). The angle-integrated
outscattering cross section is defined in Eq.~\eqref{eq:sigma_out}.
The corresponding kernel-derivative contribution is
\begin{equation}
\sigma^{\partial T}_{\mathrm{out}}(g\rightarrow g'; T)
=
\sigma_{\mathrm{out}}(g\rightarrow g'; T)\,
\left\langle 1\right\rangle^{\partial T}(g, g'; T)
\label{eq:results_sigma_out_derivative}
\end{equation}
For the independent comparison, the Monte Carlo calculation does not
evaluate the closed-form kernel and integrate it numerically. Instead,
it samples the underlying physical scattering process: an
incident photon energy and electron velocity are drawn from their
prescribed distributions, the scattering is performed in the electron
rest frame using the Klein--Nishina cross section, and the outgoing
photon is transformed back to the lab frame and tallied in the
corresponding target-energy group and angular bin. The estimator,
including its normalization and the reweighting used for the kernel-derivative
contribution, is derived in Appendix~\ref{sec:monte_carlo_derivation_appendix}. Because the
deterministic and Monte Carlo calculations follow different numerical
routes, their comparison provides an independent test of the complete
three-dimensional integration.

\begin{figure*}[!htbp]
  \centering
  \subfloat[$k_{\rm B}T=1~\mathrm{keV}$]{%
    \includegraphics[width=0.49\textwidth]
      {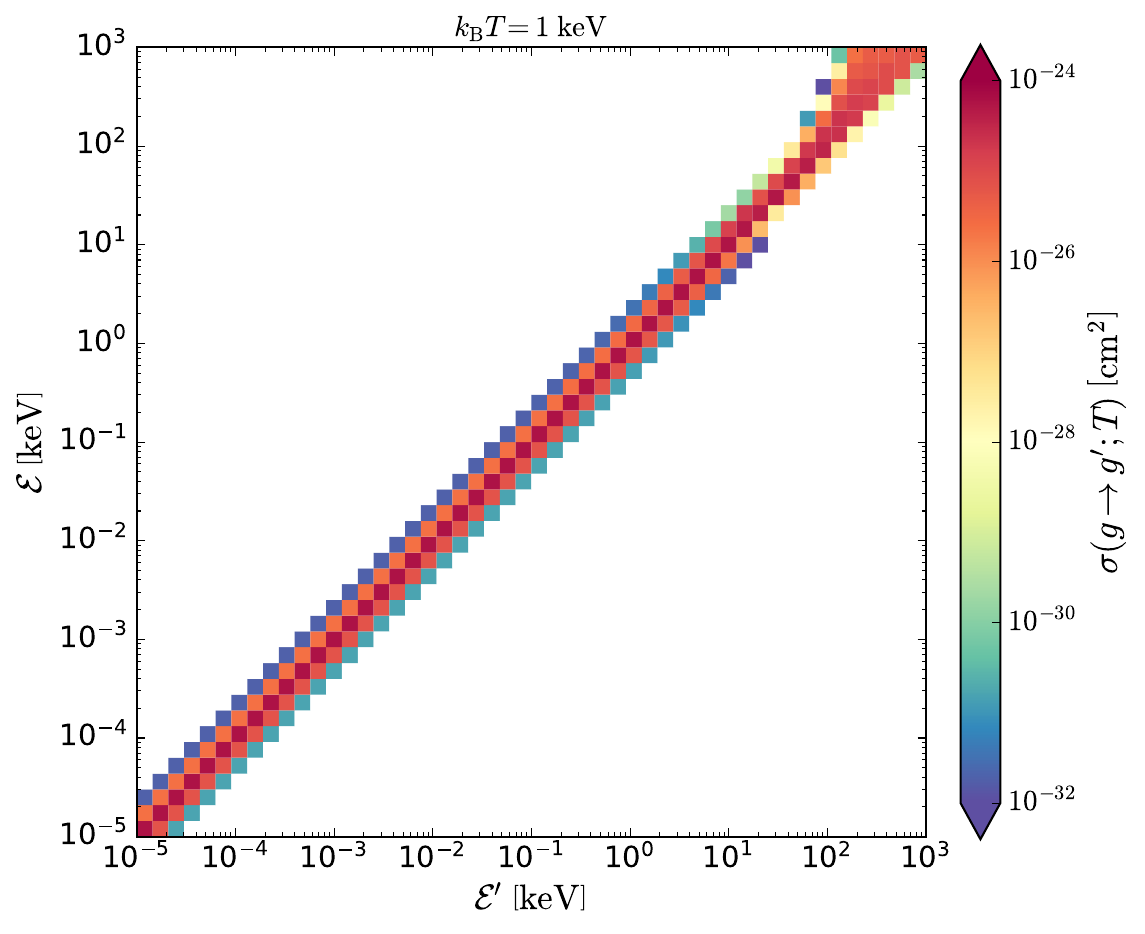}}
  \hfill
  \subfloat[$k_{\rm B}T=10~\mathrm{keV}$]{%
    \includegraphics[width=0.49\textwidth]
      {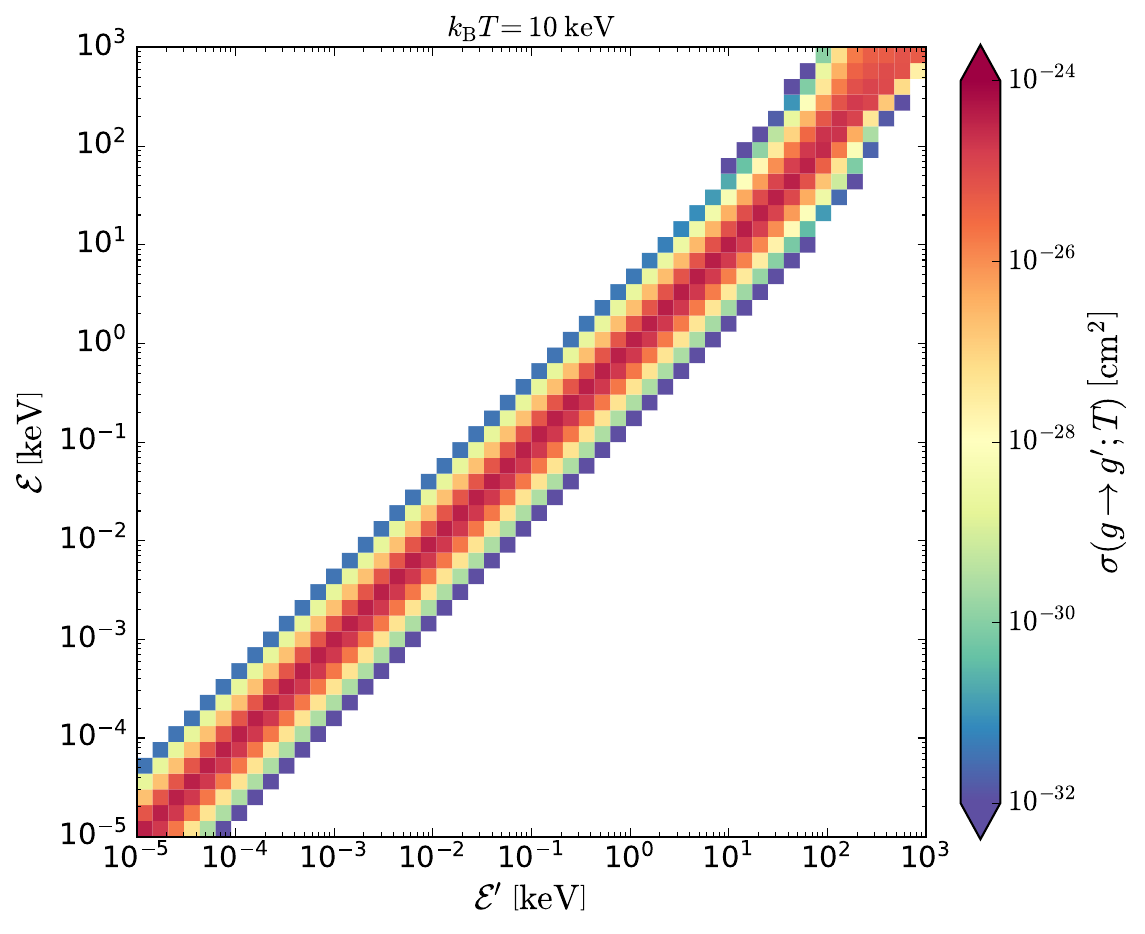}}\\[6pt]
  \subfloat[$k_{\rm B}T=50~\mathrm{keV}$]{%
    \includegraphics[width=0.49\textwidth]
      {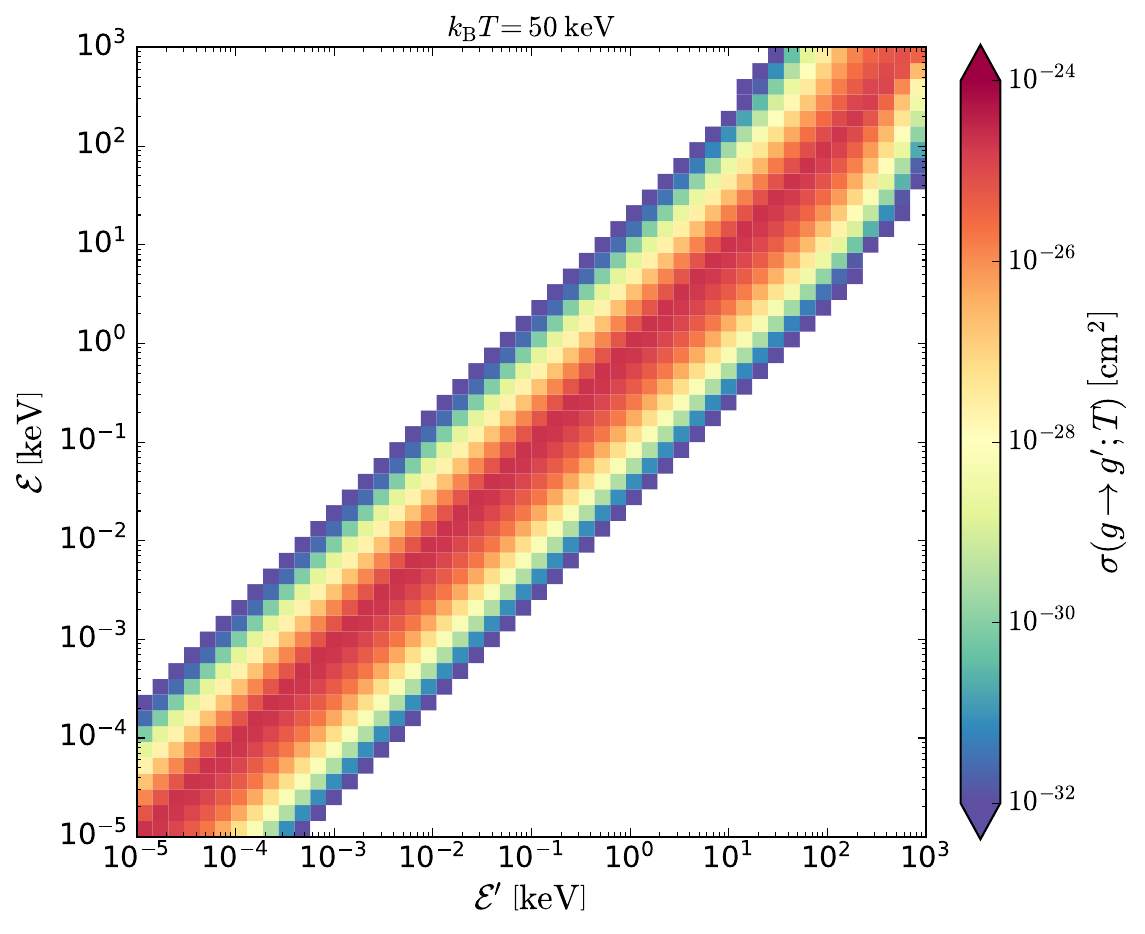}}
  \hfill
  \subfloat[$k_{\rm B}T=100~\mathrm{keV}$]{%
    \includegraphics[width=0.49\textwidth]
      {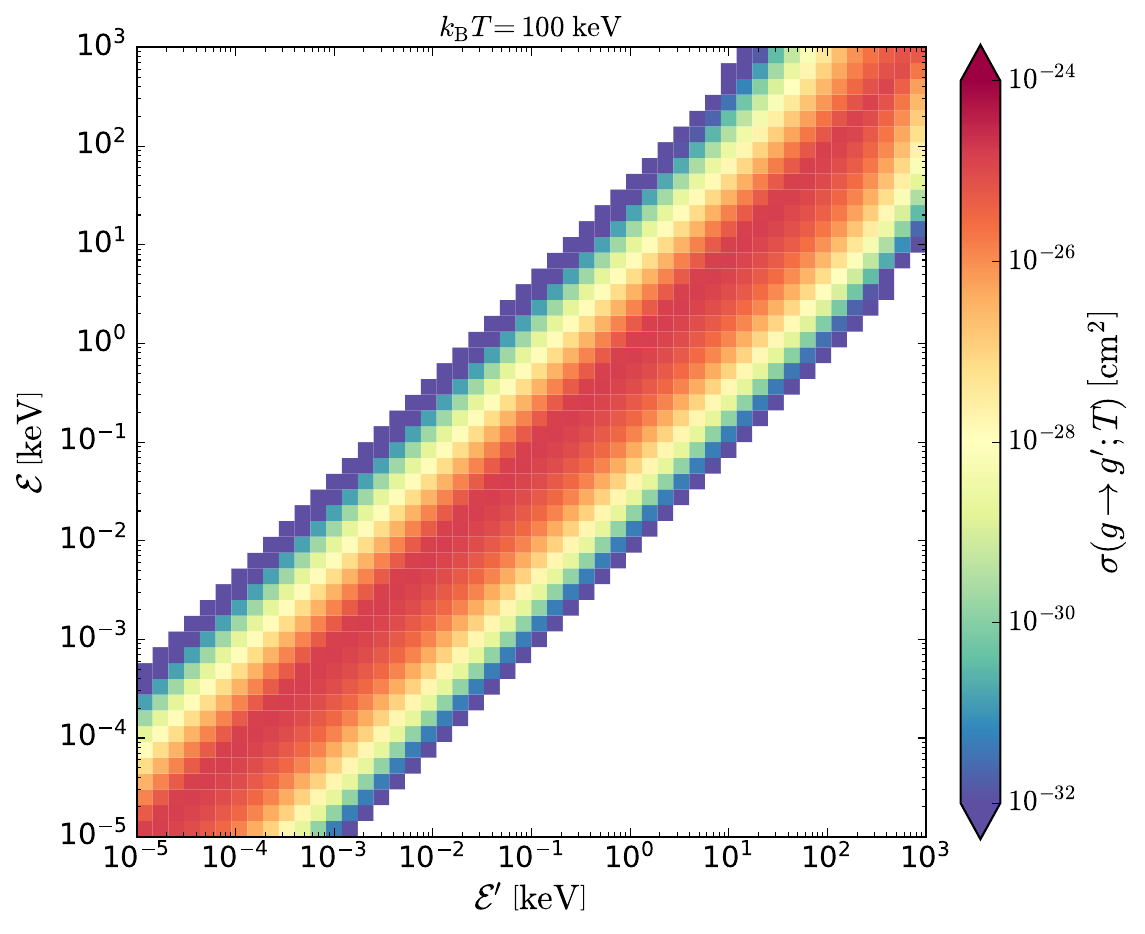}}
  \caption{
  Angle-integrated multigroup outscattering cross section
  $\sigma_{\mathrm{out}}(g\rightarrow g';T)$ on the 50-group geometric grid
  of Section~\ref{sec:monte_carlo_comparison}, spanning
  $10^{-5}$--$10^{3}~\mathrm{keV}$, with the capped Planck weighting of
  Eq.~\eqref{eq:capped_planck_weight}. White cells denote group pairs
  where the cross section is negligible relative to the matrix maximum.}
  \label{fig:sigma_matrix}
\end{figure*}

\begin{figure*}[!htbp]
  \centering
  \subfloat[$k_{\rm B}T=1~\mathrm{keV}$]{%
    \includegraphics[width=0.49\textwidth]
      {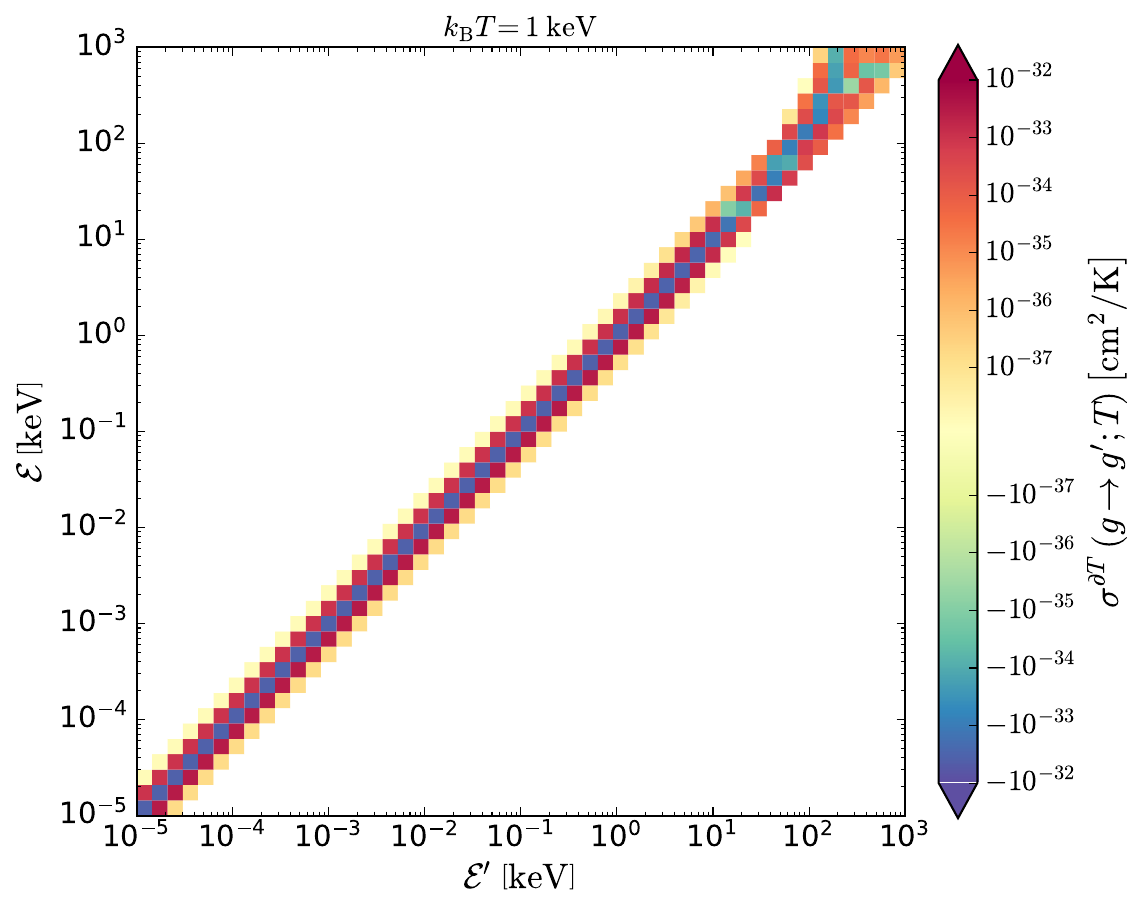}}
  \hfill
  \subfloat[$k_{\rm B}T=10~\mathrm{keV}$]{%
    \includegraphics[width=0.49\textwidth]
      {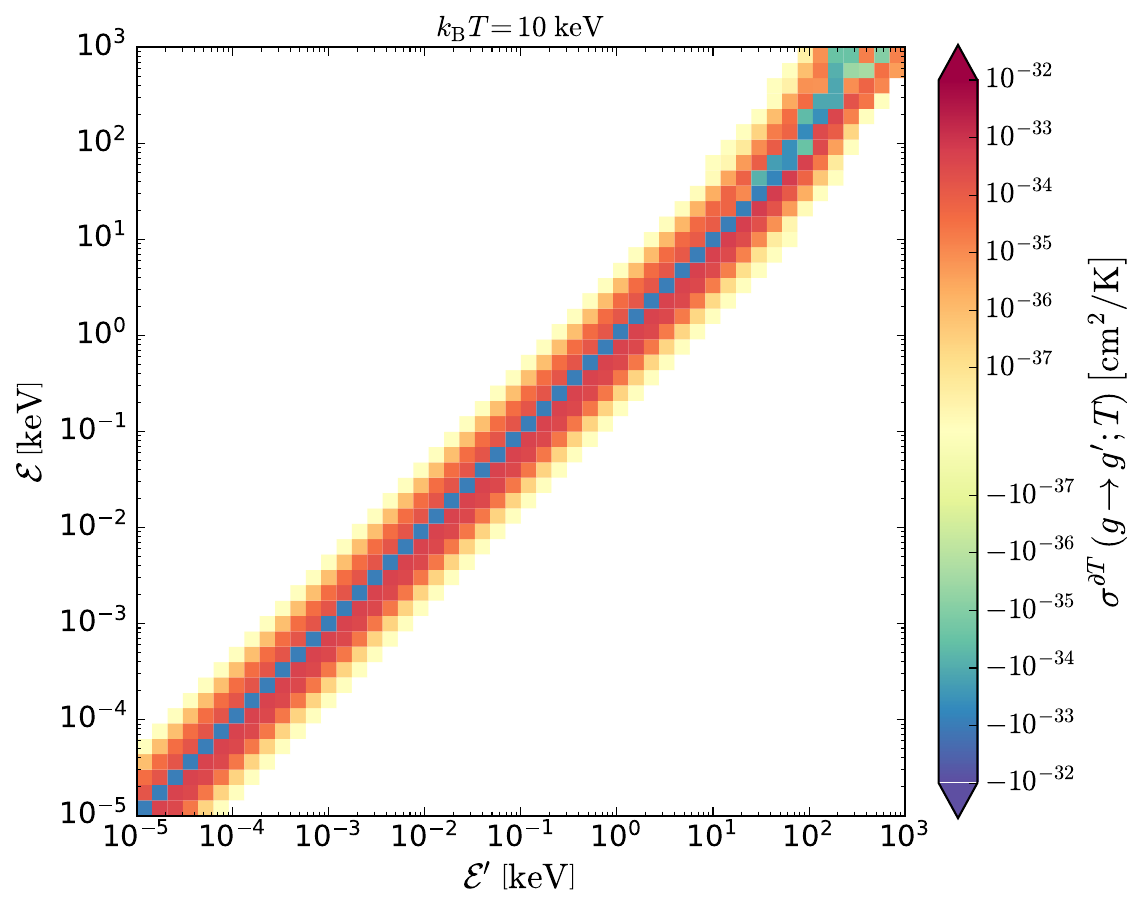}}\\[6pt]
  \subfloat[$k_{\rm B}T=50~\mathrm{keV}$]{%
    \includegraphics[width=0.49\textwidth]
      {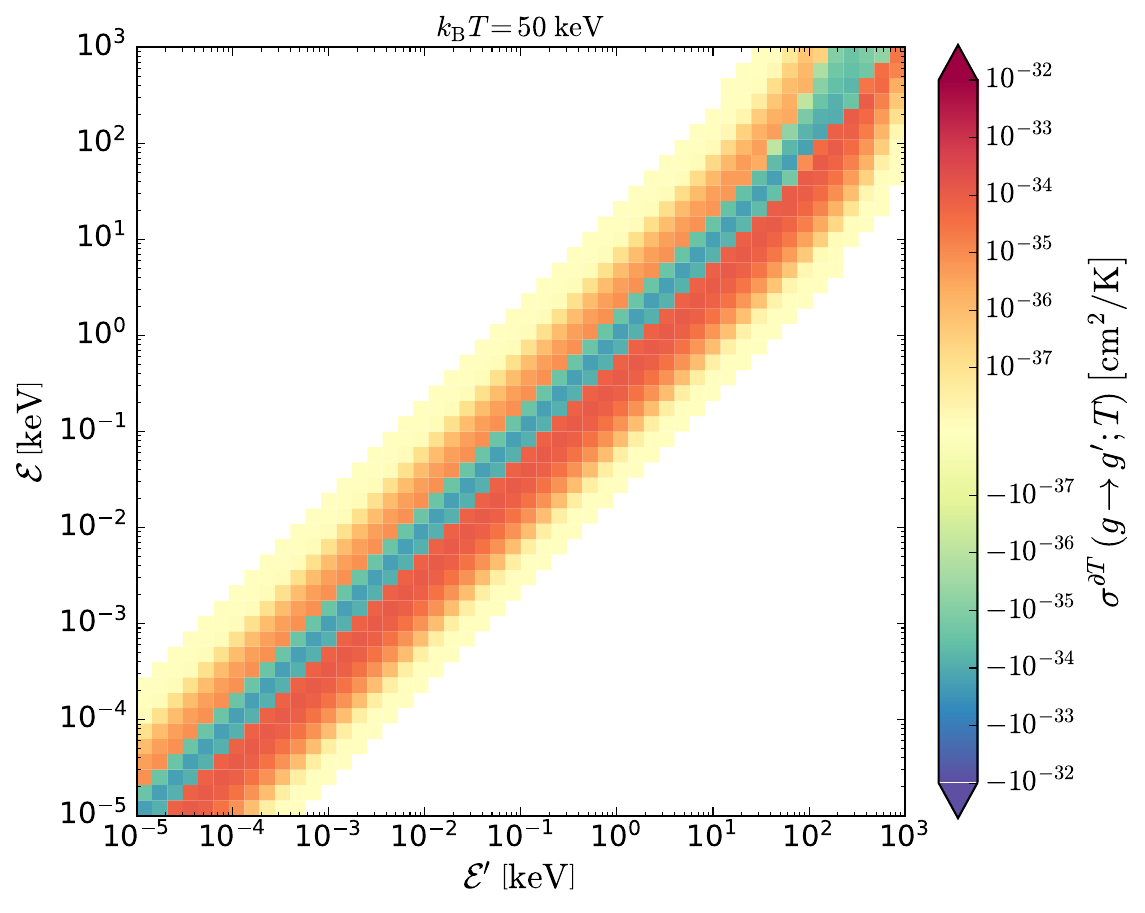}}
  \hfill
  \subfloat[$k_{\rm B}T=100~\mathrm{keV}$]{%
    \includegraphics[width=0.49\textwidth]
      {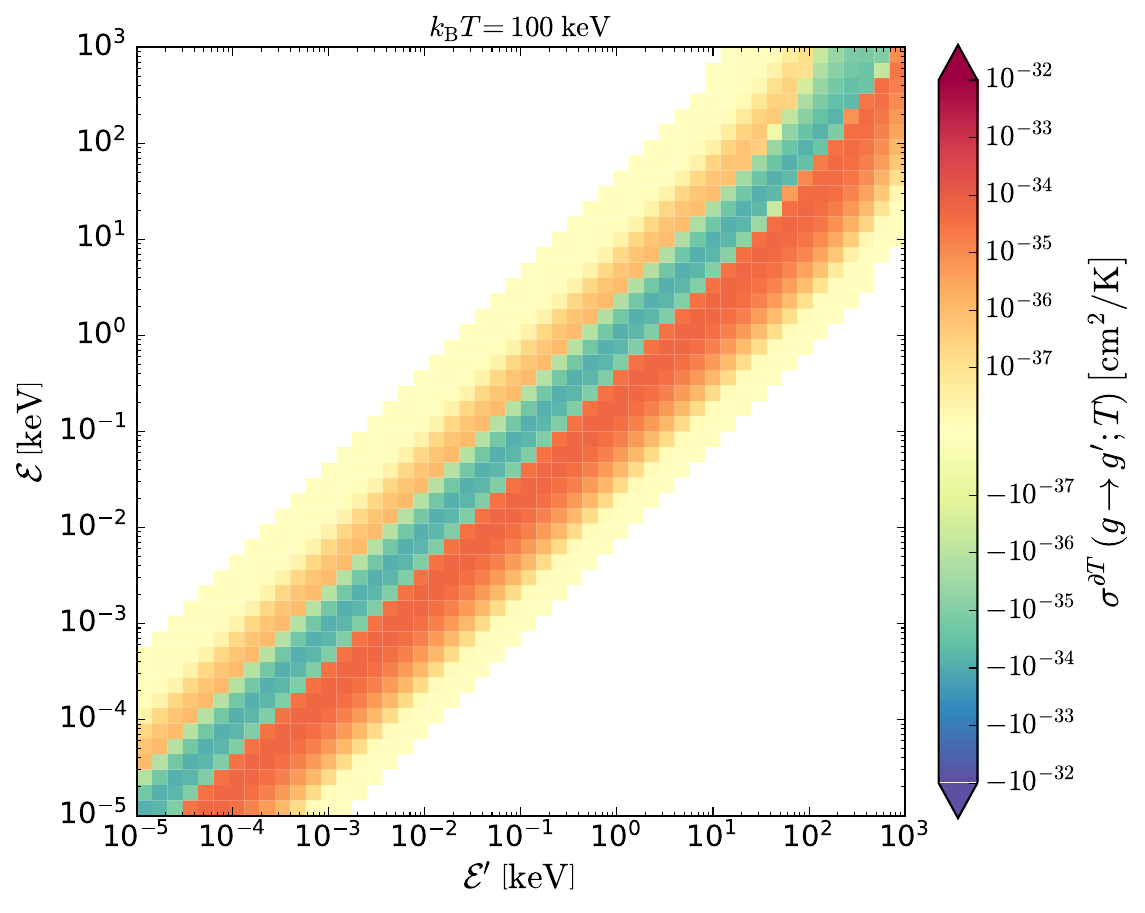}}
  \caption{
  Kernel-derivative contribution
  $\sigma^{\partial T}_{\mathrm{out}}(g\rightarrow g';T)$ on the same
  50-group geometric grid as Figure~\ref{fig:sigma_matrix}, with the
  capped Planck weighting of Eq.~\eqref{eq:capped_planck_weight}. White
  cells denote group pairs where the cross section, not the derivative,
  is negligible relative to the matrix maximum.}
  \label{fig:dsigma_dT_matrix}
\end{figure*}

We first examine the complete angle-integrated matrices
for the outscattering cross section and its kernel-derivative contribution,
defined in Eqs.~\eqref{eq:sigma_out} and
\eqref{eq:results_sigma_out_derivative}.
Figures~\ref{fig:sigma_matrix} and~\ref{fig:dsigma_dT_matrix} show
these quantities at $k_{\rm B}T=1$, $10$, $50$, and $100~\mathrm{keV}$. The
matrices are evaluated on a 50-group geometric grid spanning
$10^{-5}$--$10^{3}~\mathrm{keV}$, using the Planck weight of
Eq.~\eqref{eq:capped_planck_weight} with $x_{\rm cap}=25$.

\begin{figure*}[!htbp]
  \centering
  \subfloat[{$g=18:\;\mathcal{E}\in[7.6\times10^{-3},\,0.011]~\mathrm{keV}$}]{%
    \includegraphics[width=0.49\textwidth]
      {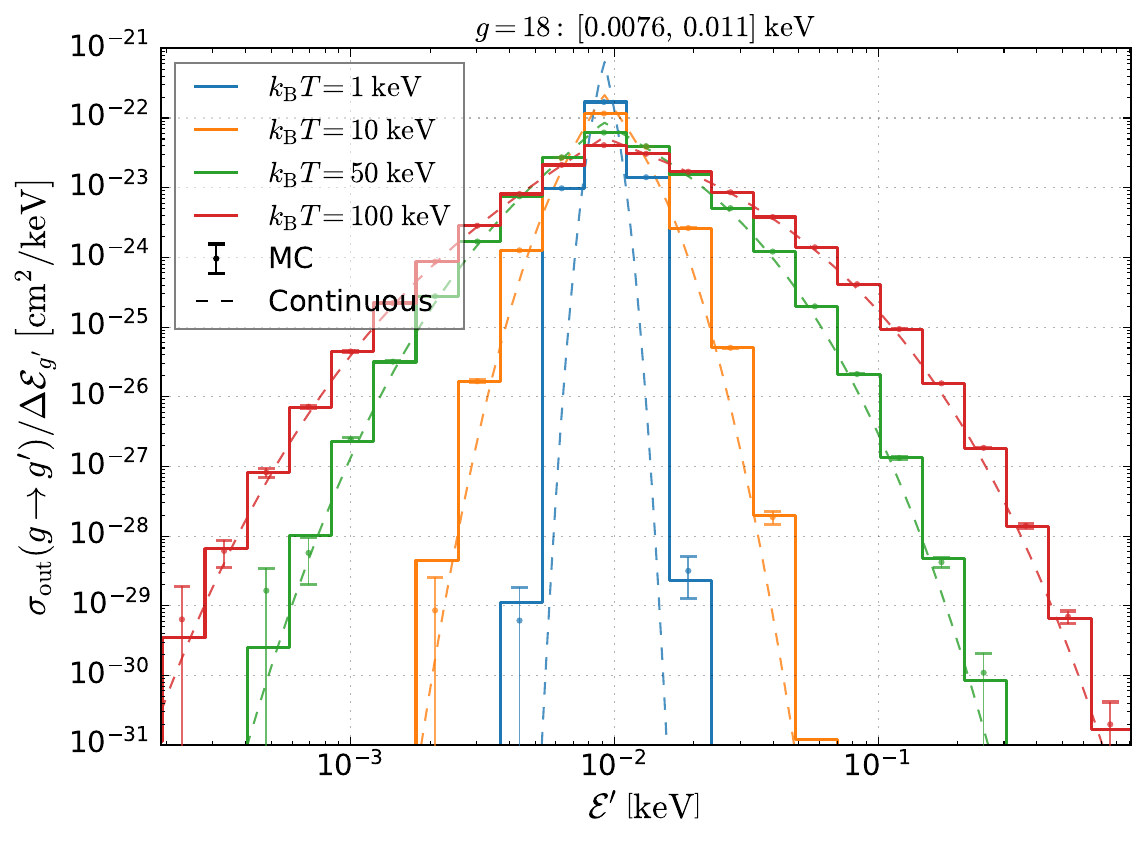}}
  \hfill
  \subfloat[{$g=31:\;\mathcal{E}\in[0.91,\,1.3]~\mathrm{keV}$}]{%
    \includegraphics[width=0.49\textwidth]
      {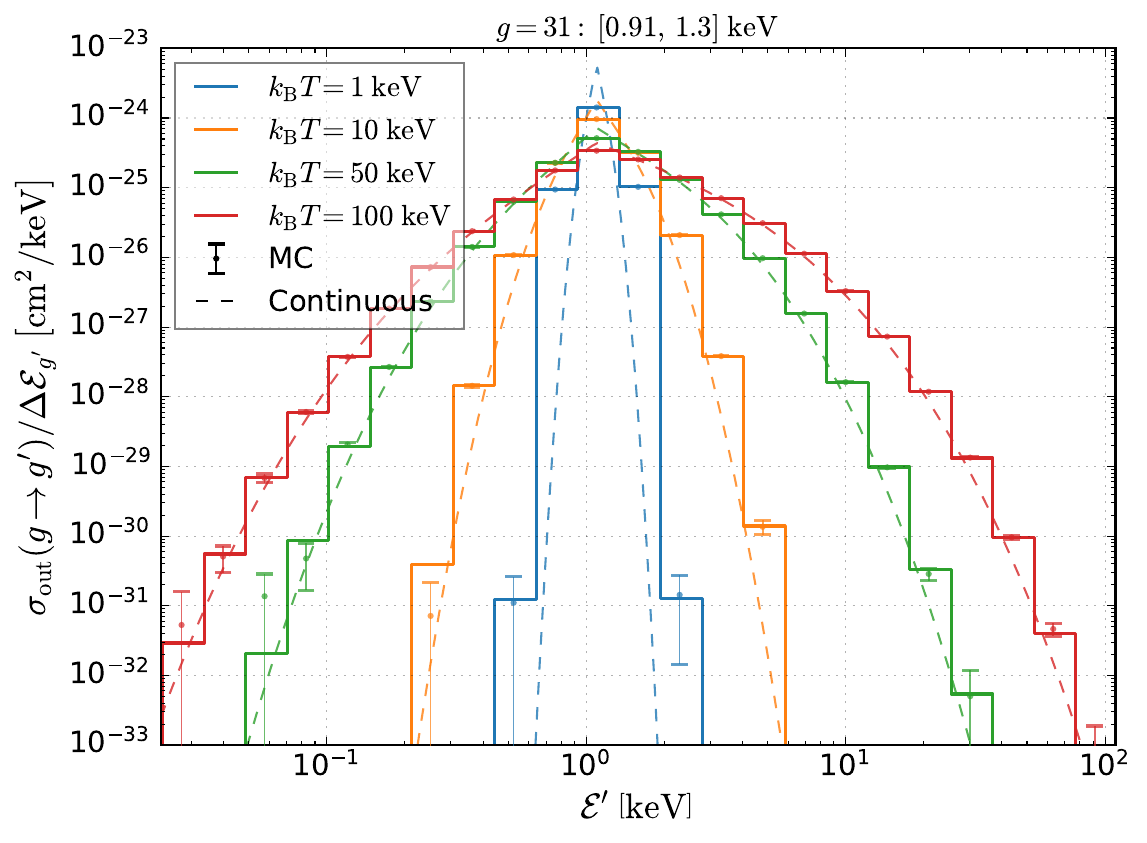}}\\[6pt]
  \subfloat[{$g=37:\;\mathcal{E}\in[8.3,\,12]~\mathrm{keV}$}]{%
    \includegraphics[width=0.49\textwidth]
      {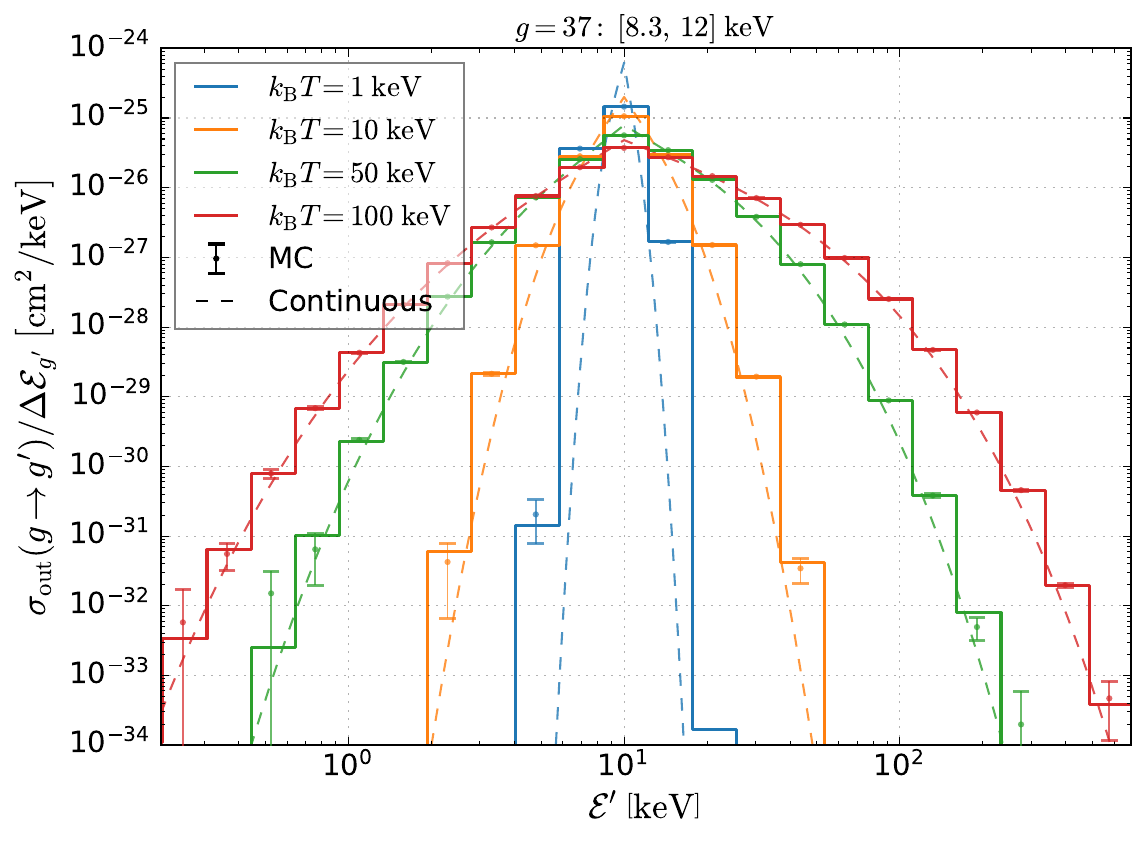}}
  \hfill
  \subfloat[{$g=43:\;\mathcal{E}\in[76,\,110]~\mathrm{keV}$}]{%
    \includegraphics[width=0.49\textwidth]
      {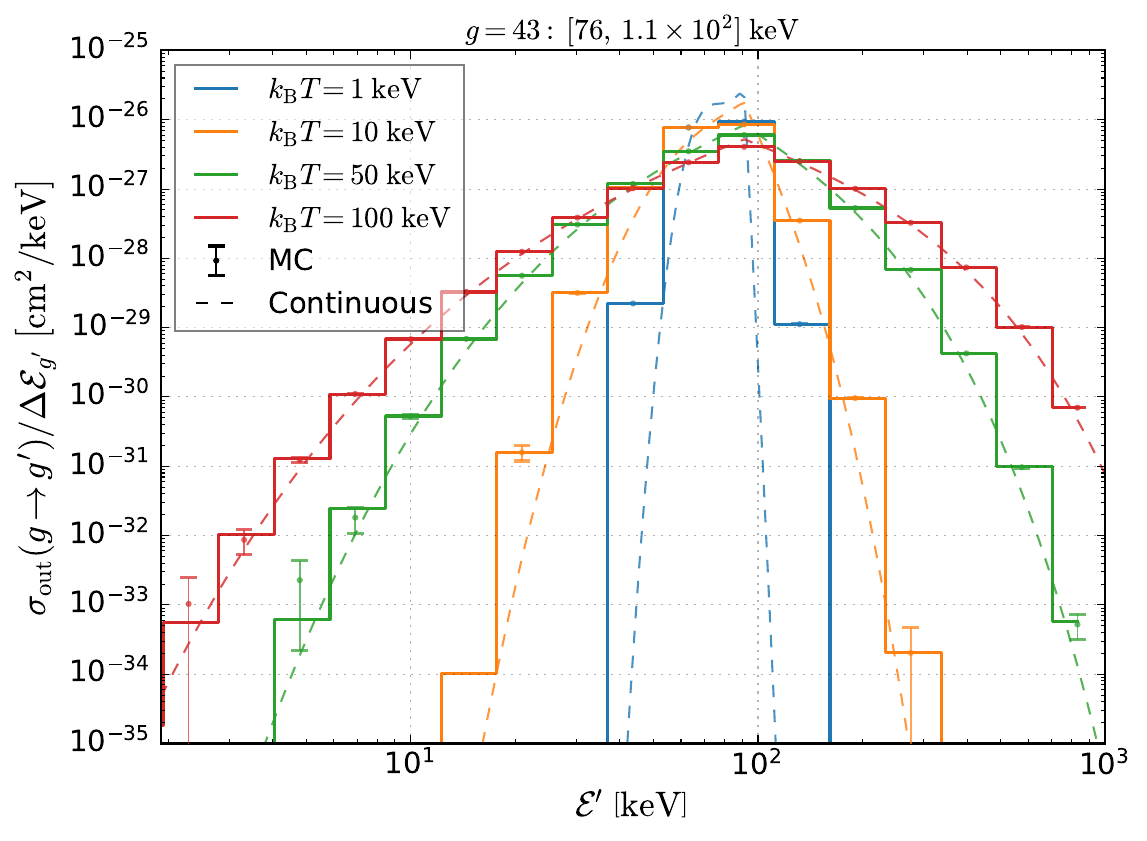}}
  \caption{
  Outgoing-energy profiles of the angle-integrated outscattering cross
  section density for four source groups at $k_{\rm B}T=1$, $10$, $50$, and
  $100~\mathrm{keV}$. Multigroup values are divided by the target
  group width to obtain the density. Solid step lines show the
  deterministic result. Markers show the Monte Carlo mean over 16
  independent realizations of $10^7$ samples each, with error bars
  corresponding to twice the standard error of the mean. Dashed lines
  show the continuous curve,
  $2\pi\,\mathcal{I}_{[-1,1]}[1](\mathcal{E}_g,\mathcal{E}';\tau)$,
  the pointwise kernel density evaluated at the geometric center
  $\mathcal{E}_g = \sqrt{\mathcal{E}_{g-1/2}\,\mathcal{E}_{g+1/2}}$
  of the source group.
  The incident-spectrum weighting is the capped
  Planck weight of Eq.~\eqref{eq:capped_planck_weight}.
  The multigroup grid is the 50-group geometric grid spanning
  $10^{-5}$--$10^{3}~\mathrm{keV}$.}
  \label{fig:sigma_profiles}
\end{figure*}

\begin{figure*}[!htbp]
  \centering
  \subfloat[{$g=18:\;\mathcal{E}\in[7.6\times10^{-3},\,0.011]~\mathrm{keV}$}]{%
    \includegraphics[width=0.49\textwidth]
      {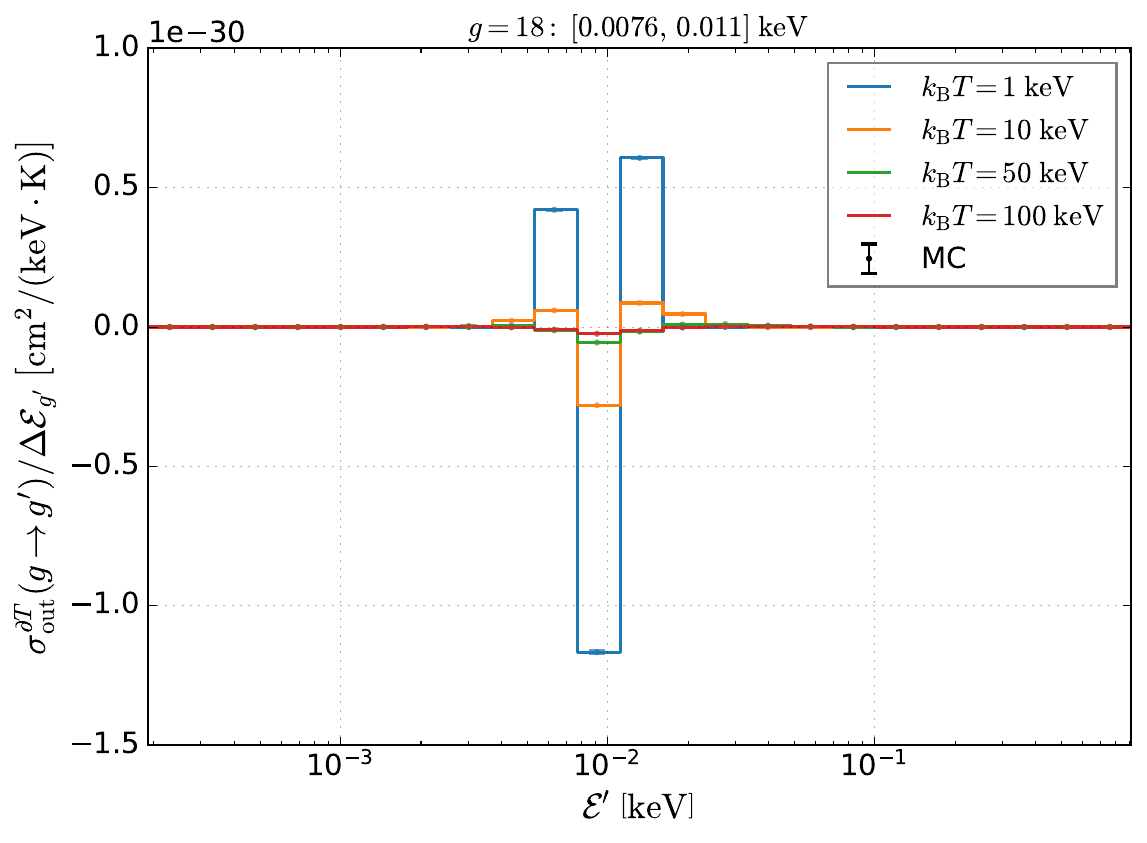}}
  \hfill
  \subfloat[{$g=31:\;\mathcal{E}\in[0.91,\,1.3]~\mathrm{keV}$}]{%
    \includegraphics[width=0.49\textwidth]
      {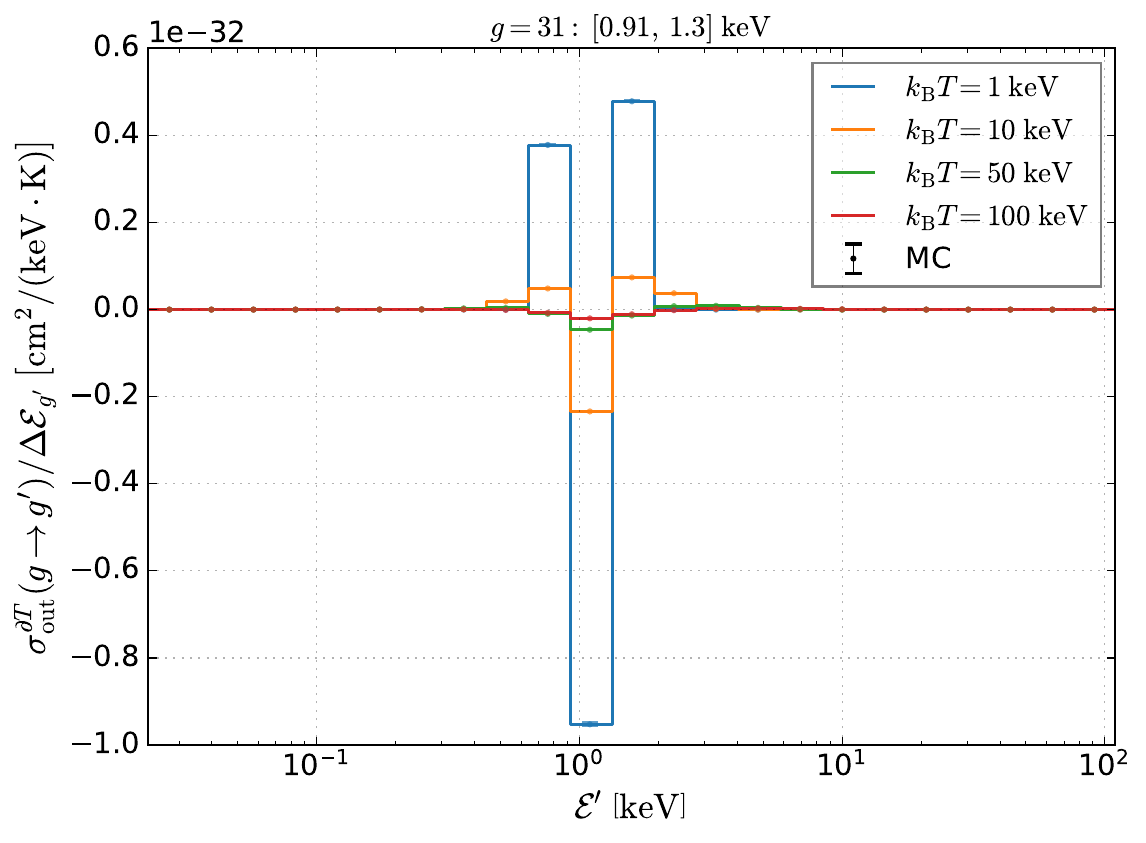}}\\[6pt]
  \subfloat[{$g=37:\;\mathcal{E}\in[8.3,\,12]~\mathrm{keV}$}]{%
    \includegraphics[width=0.49\textwidth]
      {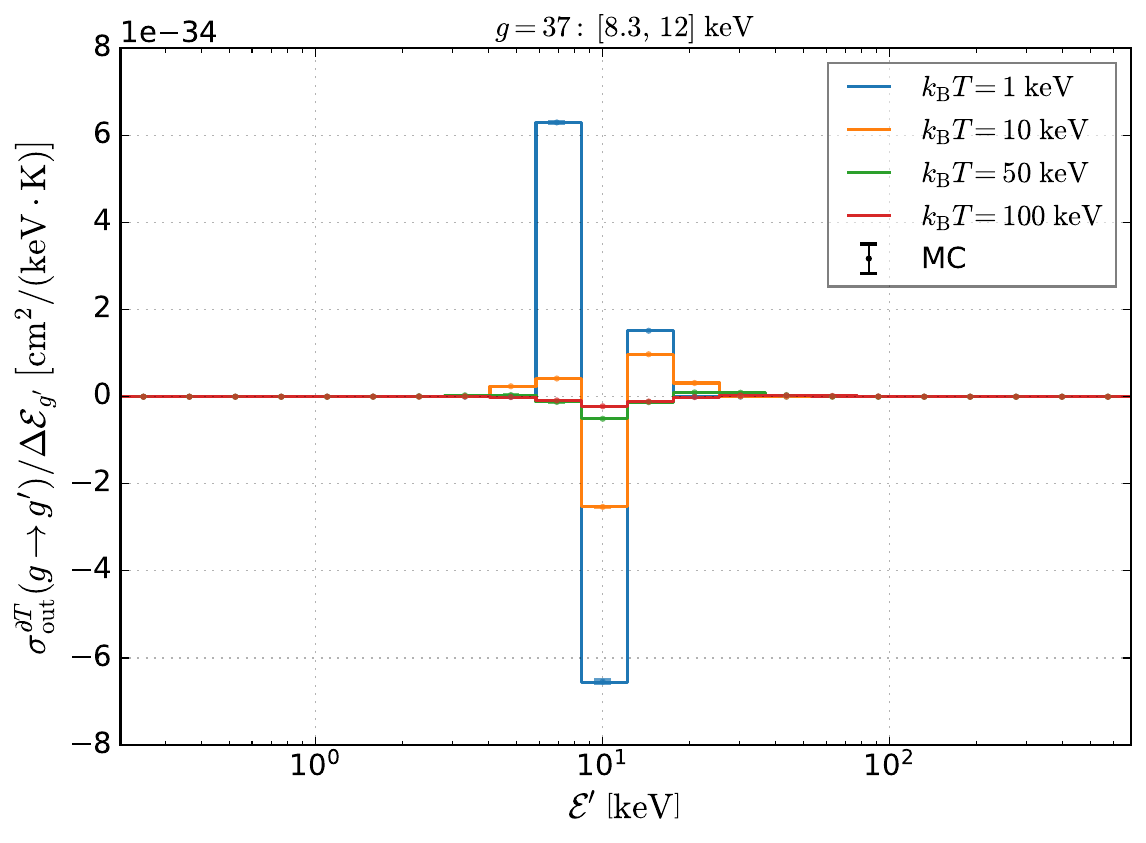}}
  \hfill
  \subfloat[{$g=43:\;\mathcal{E}\in[76,\,110]~\mathrm{keV}$}]{%
    \includegraphics[width=0.49\textwidth]
      {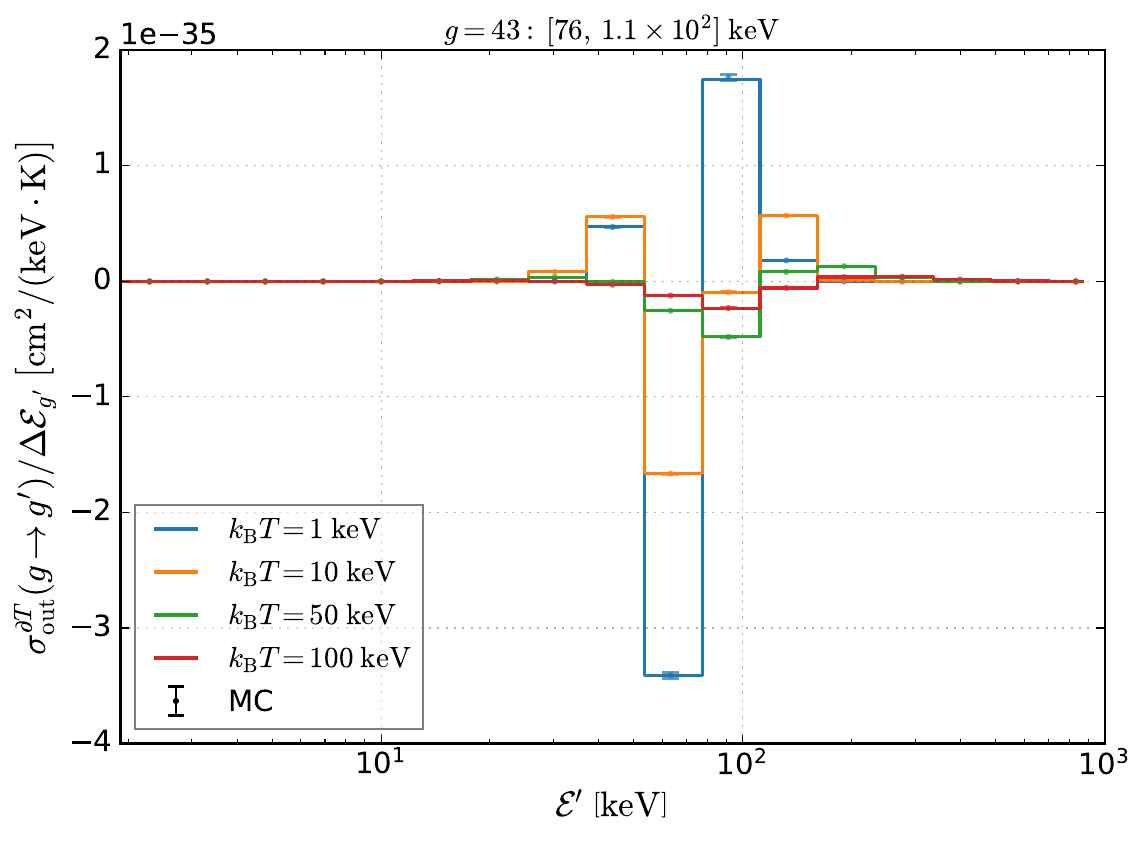}}
  \caption{
  Outgoing-energy profiles of the kernel-derivative contribution to the
  angle-integrated outscattering cross section density for the same
  source groups, temperatures, weighting, and Monte Carlo realizations
  as Figure~\ref{fig:sigma_profiles}. Solid step lines show the
  deterministic result, and markers show the Monte Carlo mean with
  error bars corresponding to twice the standard error of the mean.}
  \label{fig:dsigma_dT_profiles}
\end{figure*}

Figures~\ref{fig:sigma_profiles} and
\ref{fig:dsigma_dT_profiles} compare outgoing-energy profiles of the
angle-integrated outscattering cross section and its kernel-derivative
contribution for four representative source groups centered near
$0.01$, $1$, $10$, and $100~\mathrm{keV}$. The deterministic profiles
are evaluated on the same 50-group grid and with the same Planck
weighting as the matrices above. Each Monte Carlo point
is the mean of 16 independent realizations, each containing $10^7$
samples, and the error bars indicate twice the standard error of the
mean. In Figure~\ref{fig:sigma_profiles}, the continuous curve shows the pointwise kernel density
$2\pi\,\mathcal{I}_{[-1,1]}[1](\mathcal{E}_g,\mathcal{E}';\tau)$
evaluated at the geometric center of the source group.

\begin{figure*}[!htbp]
  \centering
  \subfloat[$k_{\rm B}T=1~\mathrm{keV}$]{%
    \includegraphics[width=0.49\textwidth]
      {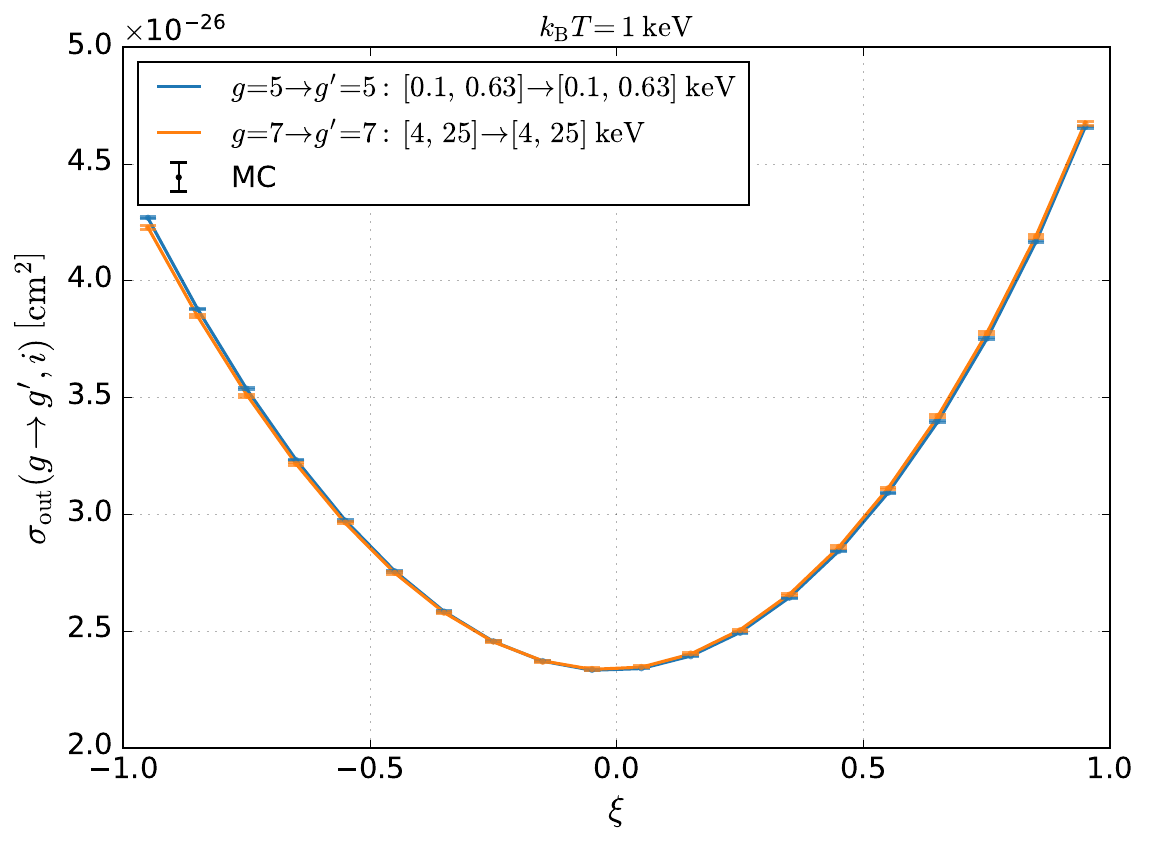}}
  \hfill
  \subfloat[$k_{\rm B}T=10~\mathrm{keV}$]{%
    \includegraphics[width=0.49\textwidth]
      {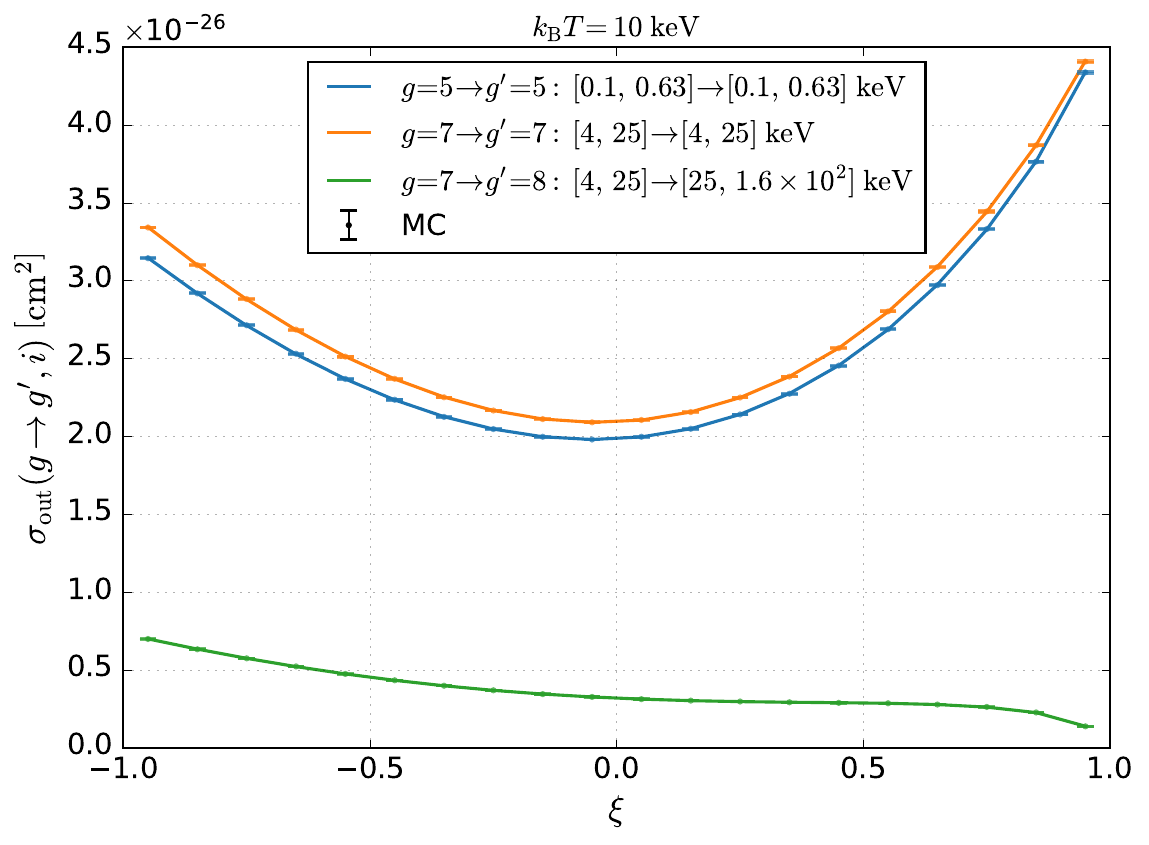}}\\[6pt]
  \subfloat[$k_{\rm B}T=50~\mathrm{keV}$]{%
    \includegraphics[width=0.49\textwidth]
      {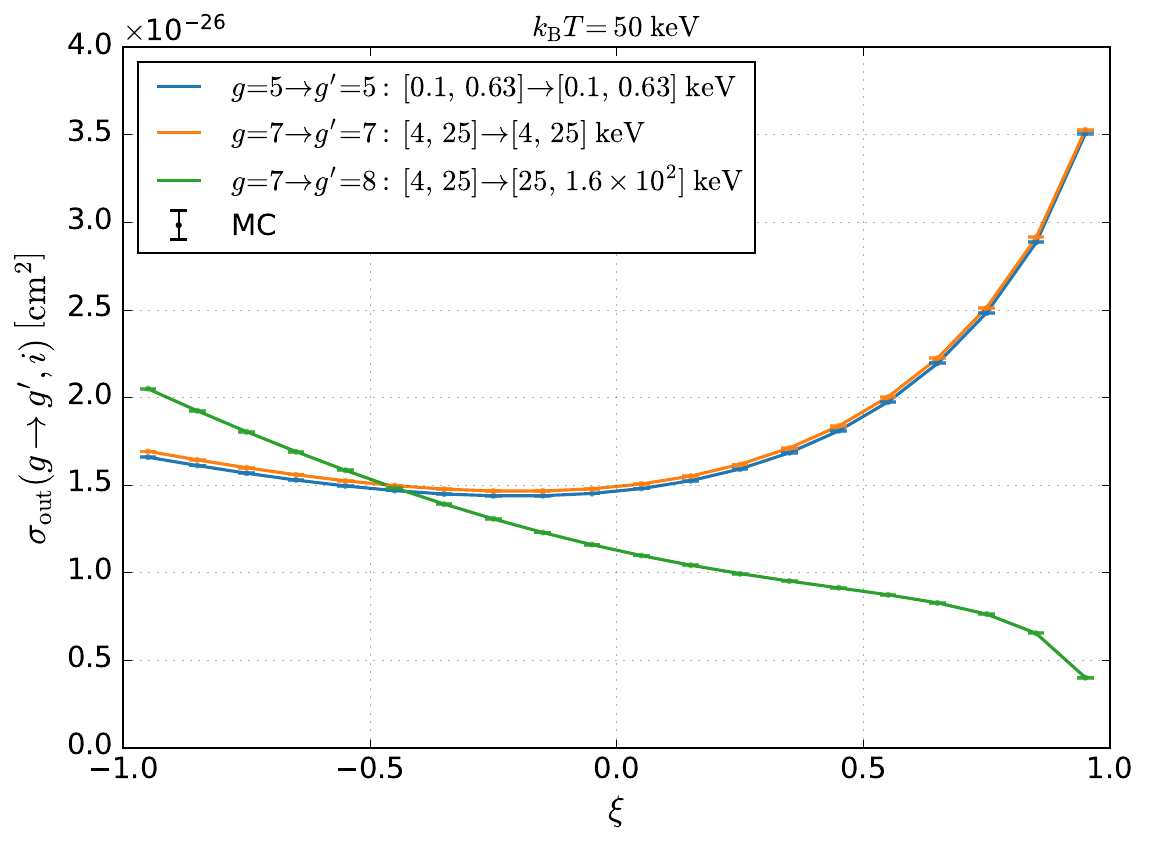}}
  \hfill
  \subfloat[$k_{\rm B}T=100~\mathrm{keV}$]{%
    \includegraphics[width=0.49\textwidth]
      {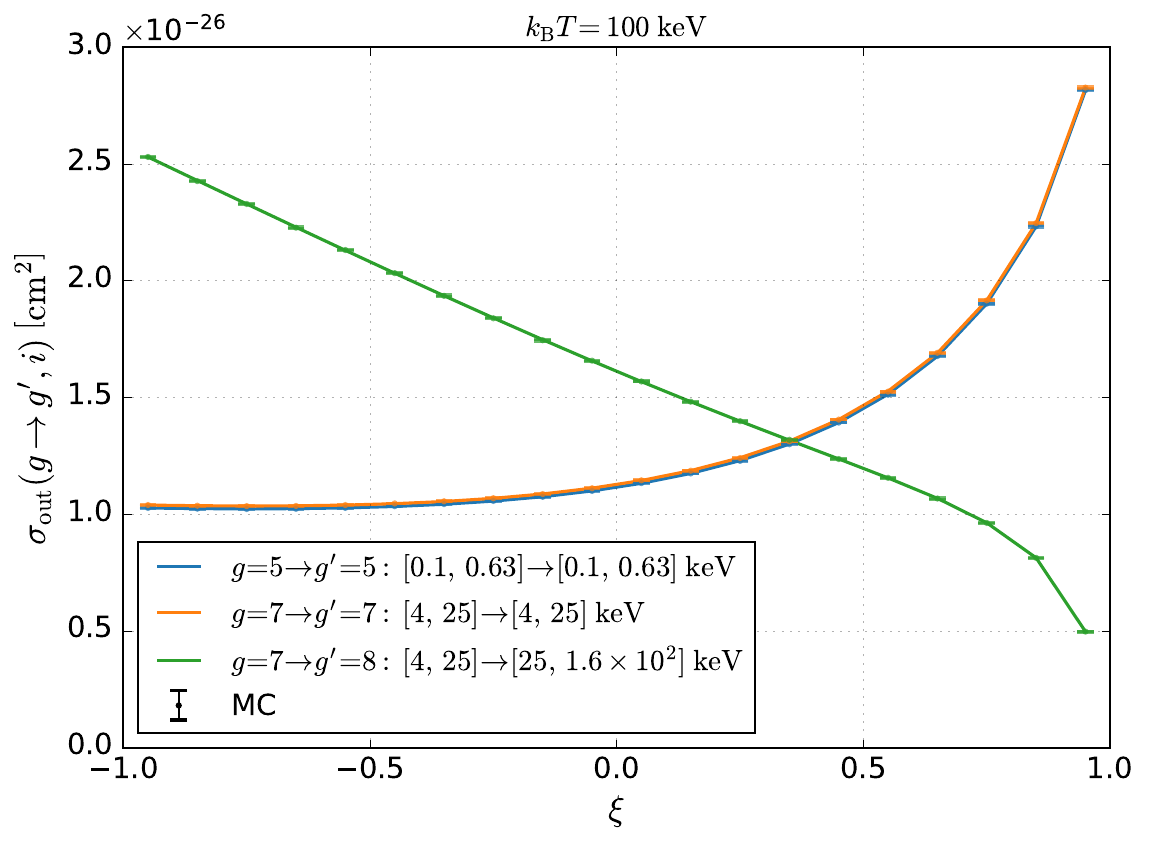}}
  \caption{
  Angular profiles of the multigroup--multiangle outscattering cross
  section $\sigma_{\rm out}$ as a function of $\xi$
  with the capped Planck weighting of
  Eq.~\eqref{eq:capped_planck_weight}.
  Solid lines show the deterministic calculation, while markers with
  error bars corresponding to twice the standard error of the mean show
  the Monte Carlo mean over 16 independent realizations of $10^7$
  samples.
  The grid is a 10-group geometric energy grid spanning
  $10^{-5}$--$10^{3}~\mathrm{keV}$ with 20 uniform angular bins.}
  \label{fig:sigma_xi_profiles}
\end{figure*}

\begin{figure*}[!htbp]
  \centering
  \subfloat[$k_{\rm B}T=1~\mathrm{keV}$]{%
    \includegraphics[width=0.49\textwidth]
      {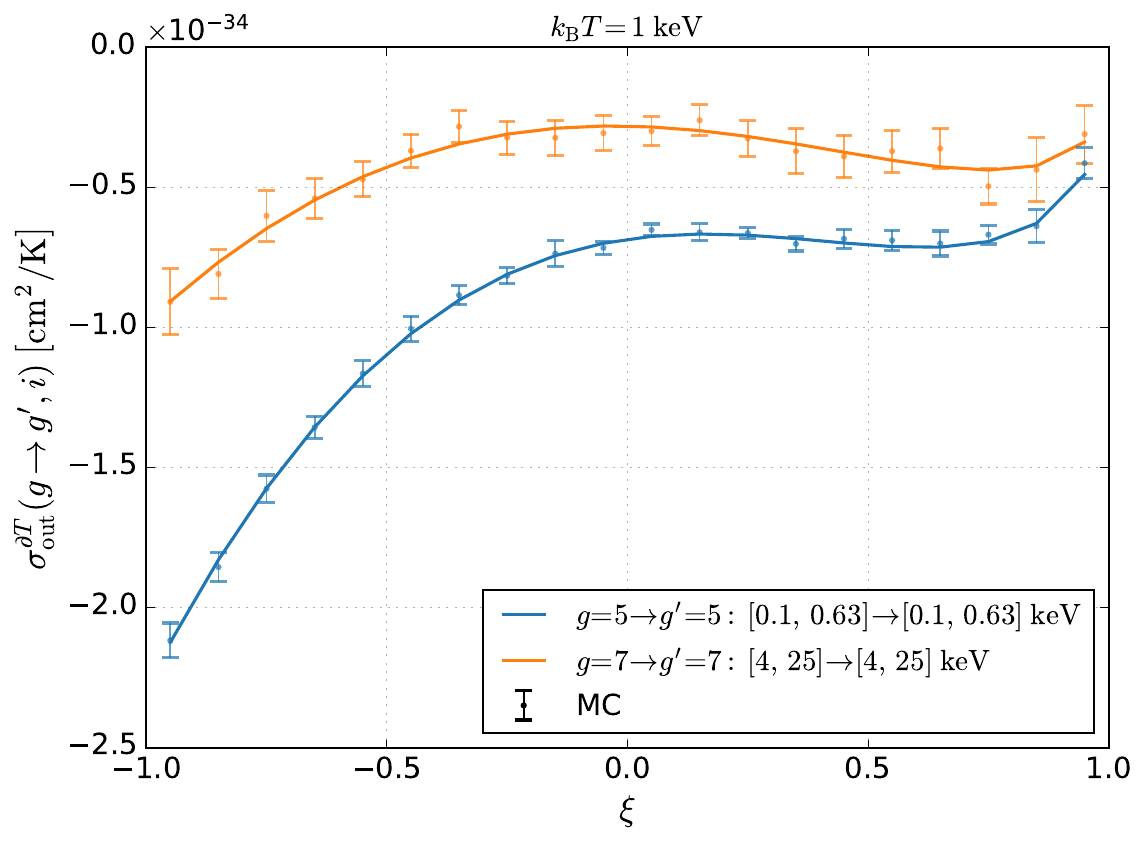}}
  \hfill
  \subfloat[$k_{\rm B}T=10~\mathrm{keV}$]{%
    \includegraphics[width=0.49\textwidth]
      {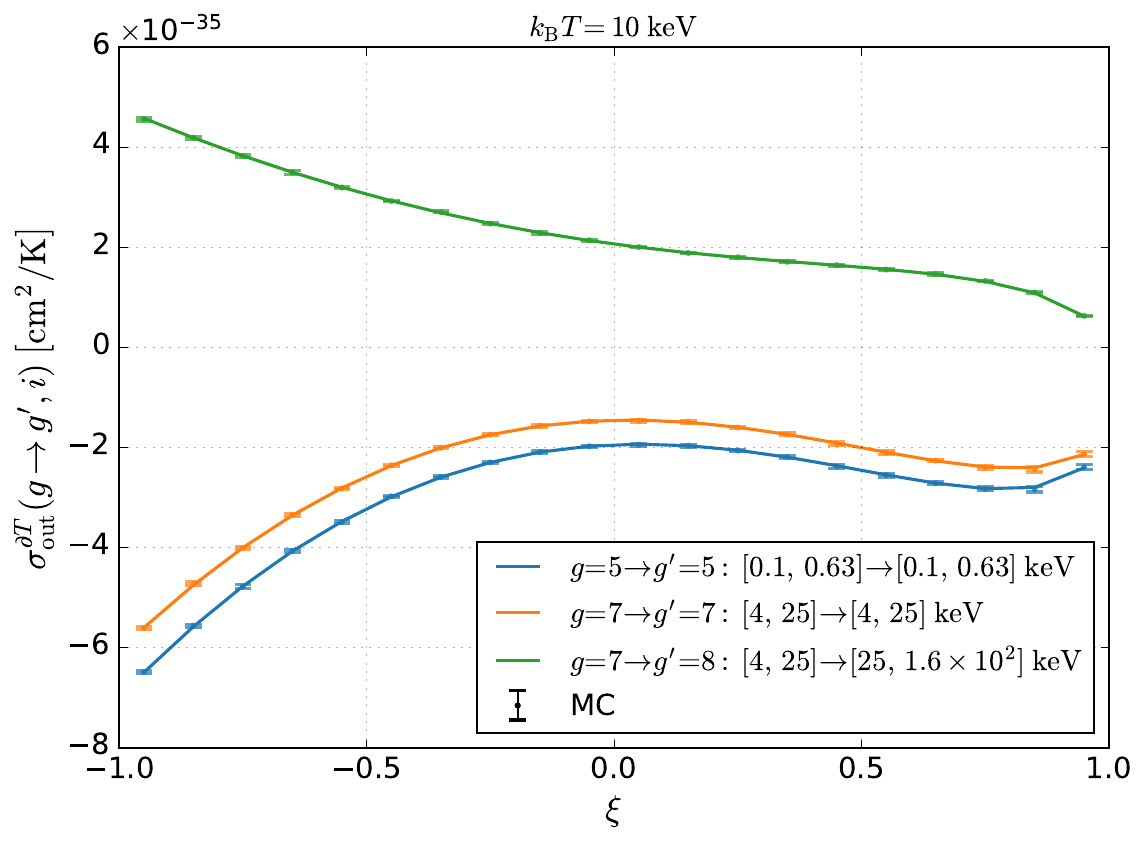}}\\[6pt]
  \subfloat[$k_{\rm B}T=50~\mathrm{keV}$]{%
    \includegraphics[width=0.49\textwidth]
      {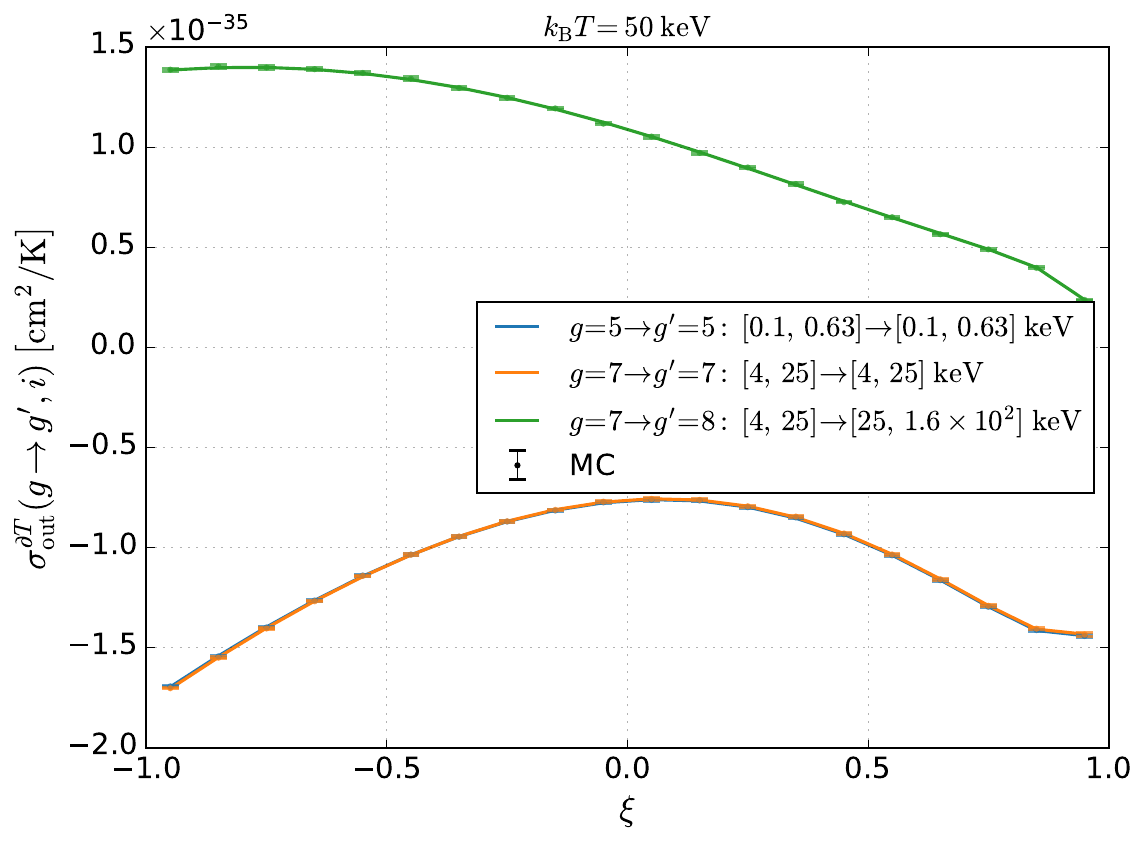}}
  \hfill
  \subfloat[$k_{\rm B}T=100~\mathrm{keV}$]{%
    \includegraphics[width=0.49\textwidth]
      {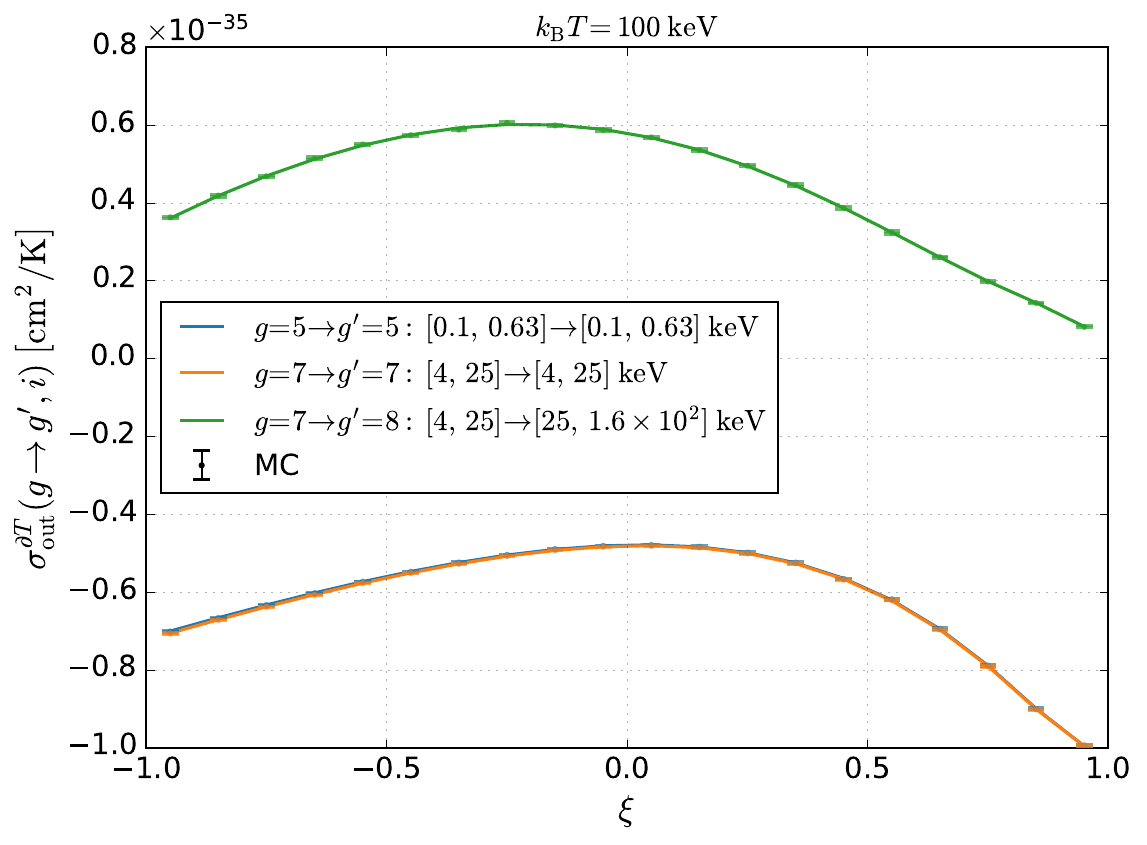}}
  \caption{
  Angular profiles of the kernel-derivative contribution density
  $\sigma^{\partial T}_{\mathrm{out}}$ for the same group transitions,
  temperatures, and Monte Carlo realizations as
  Figure~\ref{fig:sigma_xi_profiles}. Solid lines show the
  deterministic calculation, while markers with error bars
  corresponding to twice the standard error of the mean show the
  Monte Carlo mean.}
  \label{fig:dsigma_dT_xi_profiles}
\end{figure*}

Figures~\ref{fig:sigma_xi_profiles} and
\ref{fig:dsigma_dT_xi_profiles} compare the angular profiles of the
multigroup--multiangle outscattering cross section and its kernel-derivative
contribution for three representative group transitions on a 10-group,
20-angle-bin grid spanning $10^{-5}$--$10^{3}~\mathrm{keV}$. The deterministic and Monte Carlo calculations use the capped Planck weighting of Eq.~\eqref{eq:capped_planck_weight} with $x_{\rm cap}=25$.

\begin{figure*}[!htbp]
  \centering
  \subfloat[{$g=31\;[0.91,\,1.3]~\mathrm{keV}$}]{%
    \includegraphics[width=0.49\textwidth]
      {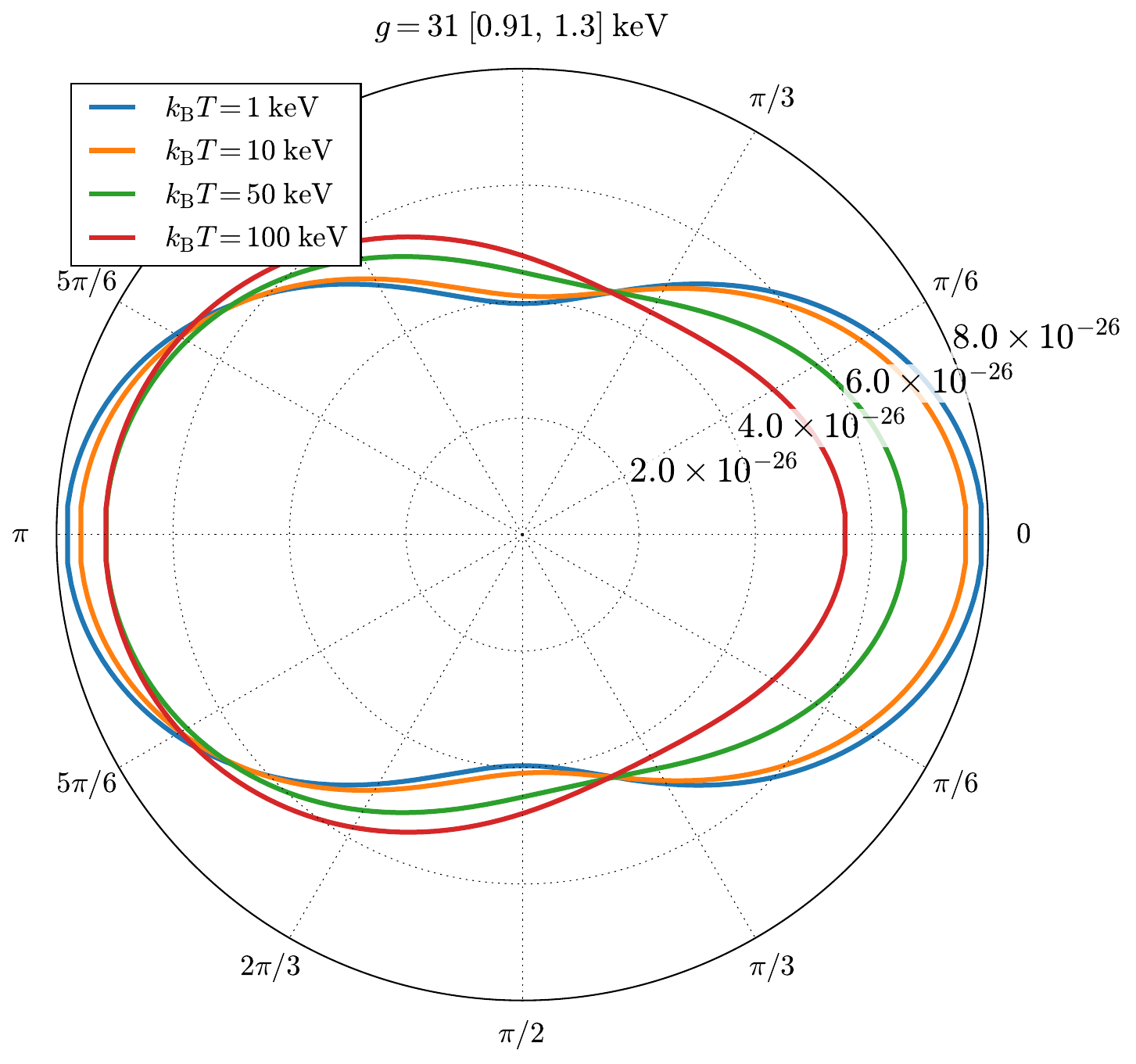}}
  \hfill
  \subfloat[{$g=37\;[8.3,\,12]~\mathrm{keV}$}]{%
    \includegraphics[width=0.49\textwidth]
      {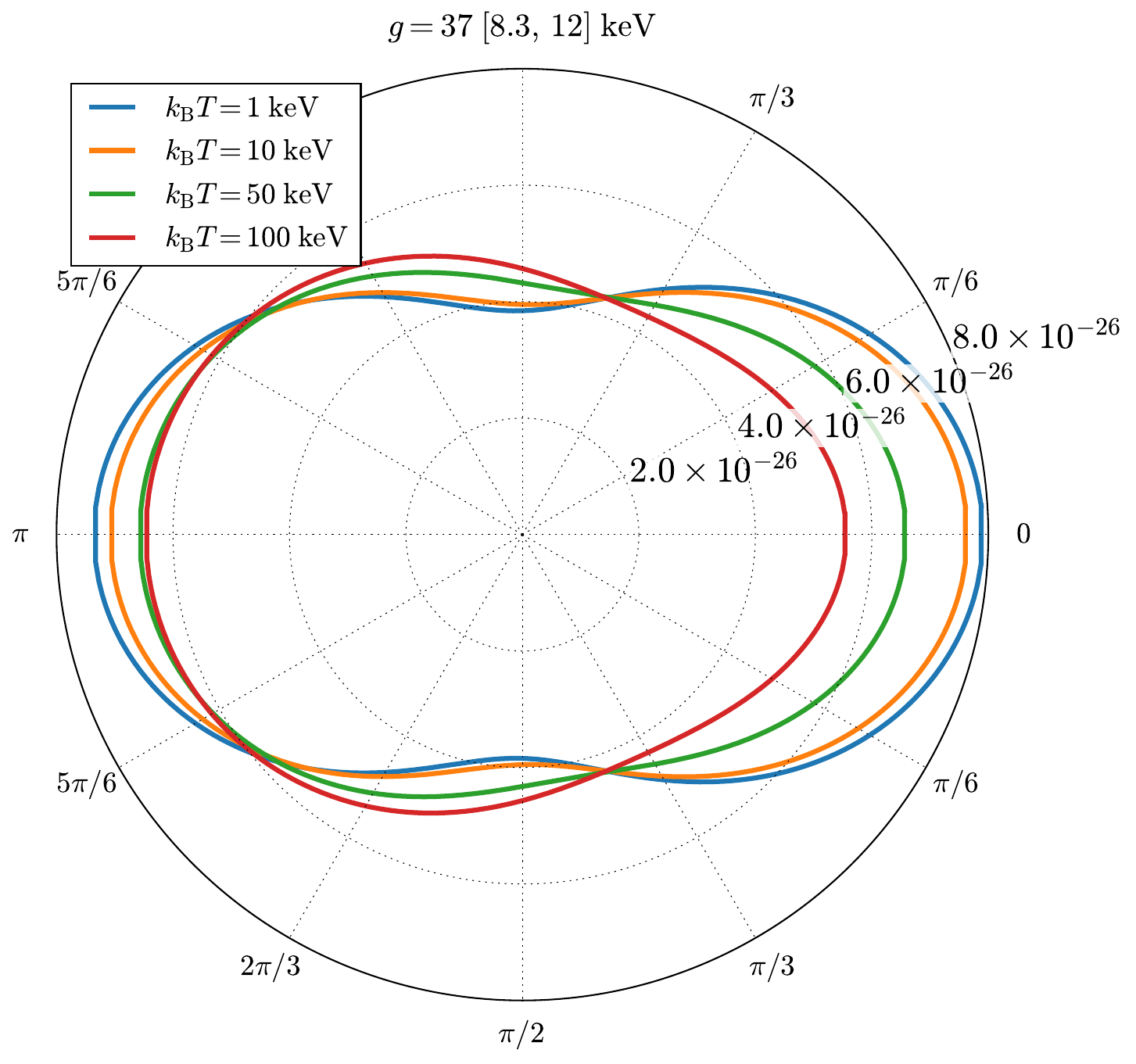}}
  \caption{
    $\dfrac{d\sigma_g}{d\Omega}(\theta)
    = \dfrac{1}{2\pi\,\Delta\xi}
    \displaystyle\sum_{g'}\sigma_{\mathrm{out}}(g\!\to\!g',\xi;T)$
    $[\mathrm{cm}^2\,\mathrm{sr}^{-1}]$,
  polar angular distribution of the total outscattering cross
  section per unit solid angle,
  where $\theta=\arccos\xi$ is the scattering angle, summed over
  all destination groups.}
  \label{fig:polar_angular}
\end{figure*}

Figure~\ref{fig:polar_angular} shows polar
representations of the total angular scattering distribution,
\begin{equation}
  \dfrac{d\sigma_g}{d\Omega}=\dfrac{1}{2\pi\,\Delta\xi}
  \displaystyle\sum_{g'}\sigma_{\mathrm{out}}(g\!\to\!g',\xi;T)
\end{equation}
for source groups centered near $1$ and $10~\mathrm{keV}$.
These are computed on the 50-group grid with 500 uniform
$\xi$-bins and the capped Planck weighting of
Eq.~\eqref{eq:capped_planck_weight}.

\begin{figure*}[!htbp]
  \centering
  \subfloat[{$g=18:\;\mathcal{E}\in[7.6\times10^{-3},\,0.011]~\mathrm{keV}$}]{%
    \includegraphics[width=0.49\textwidth]
      {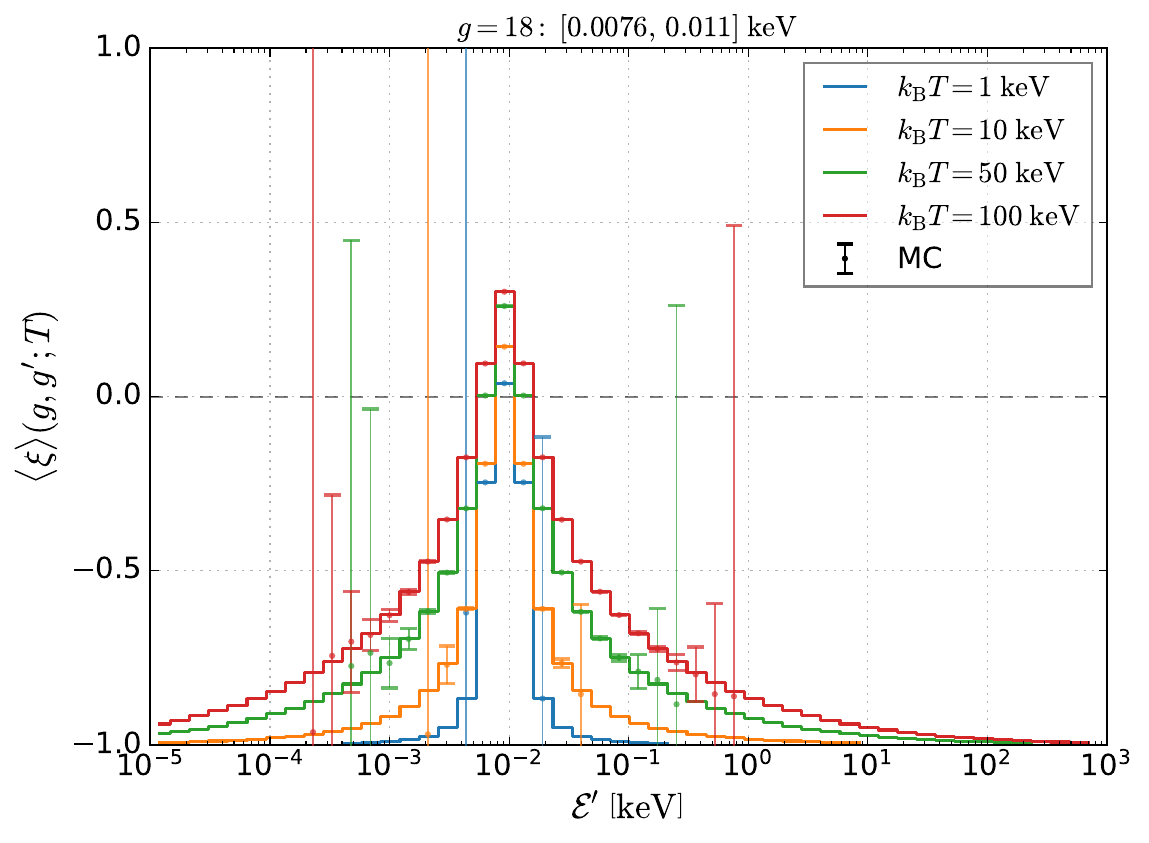}}
  \hfill
  \subfloat[{$g=31:\;\mathcal{E}\in[0.91,\,1.3]~\mathrm{keV}$}]{%
    \includegraphics[width=0.49\textwidth]
      {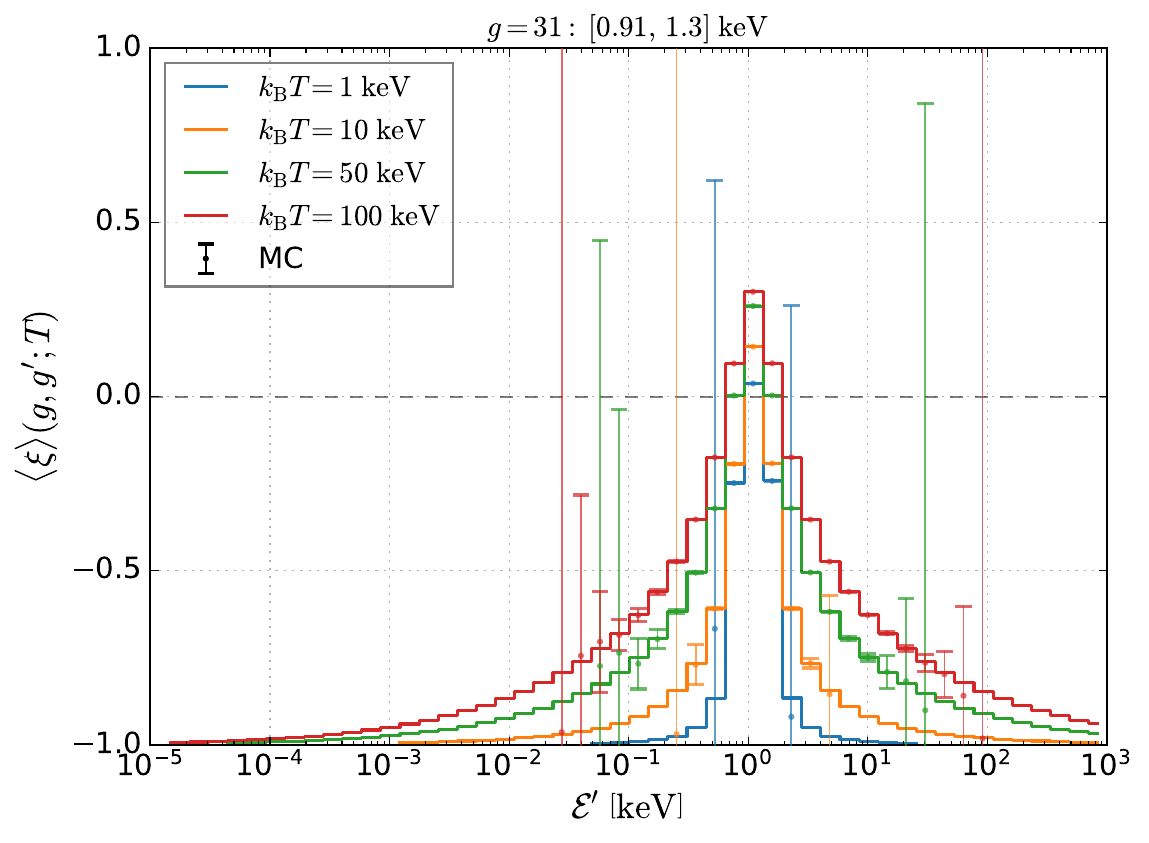}}\\[6pt]
  \subfloat[{$g=37:\;\mathcal{E}\in[8.3,\,12]~\mathrm{keV}$}]{%
    \includegraphics[width=0.49\textwidth]
      {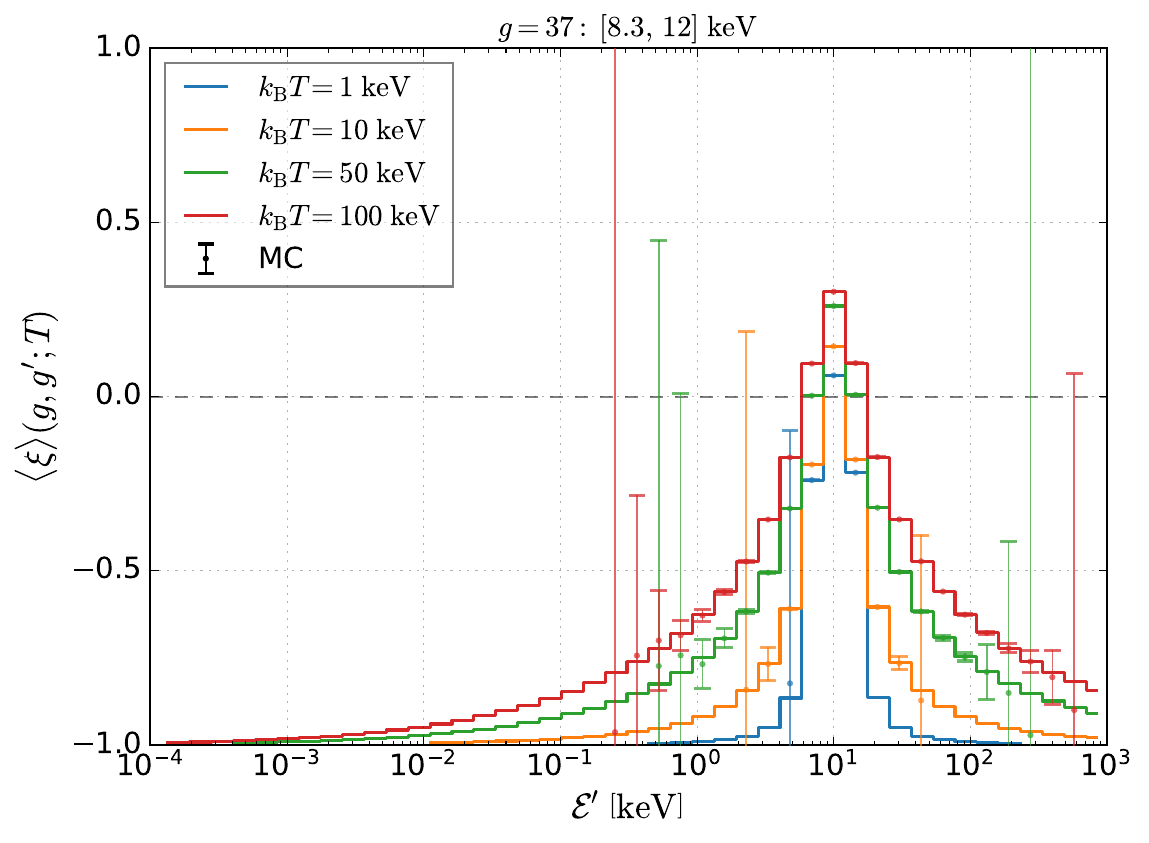}}
  \hfill
  \subfloat[{$g=43:\;\mathcal{E}\in[76,\,110]~\mathrm{keV}$}]{%
    \includegraphics[width=0.49\textwidth]
      {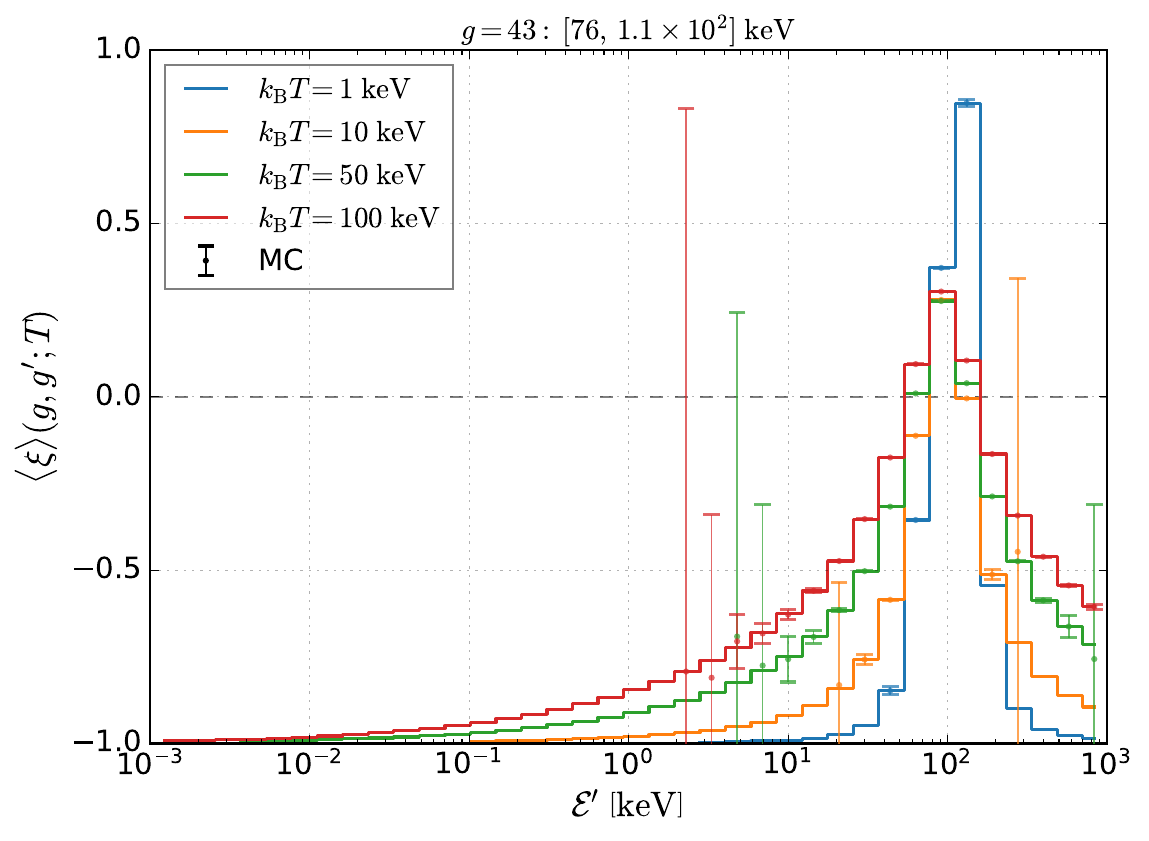}}
  \caption{
  Outgoing-energy profiles of
  $\left\langle \xi\right\rangle(g, g'; T)$ for four
  source groups at $k_{\rm B}T=1$, $10$, $50$, and $100~\mathrm{keV}$.
  Solid step lines show the multigroup result with the capped Planck
  weight of Eq.~\eqref{eq:capped_planck_weight}.
  Evaluated on a 50-group geometric grid spanning
  $10^{-5}$--$10^{3}~\mathrm{keV}$ (angle-integrated). Visible noise in the far wings reflects groups where the kernel is small
  and few Monte Carlo samples contribute to the estimate.}
  \label{fig:xi_weighted_profiles}
\end{figure*}

\begin{figure*}[!htbp]
  \centering
  \subfloat[$k_{\rm B}T=1~\mathrm{keV}$]{%
    \includegraphics[width=0.49\textwidth]
      {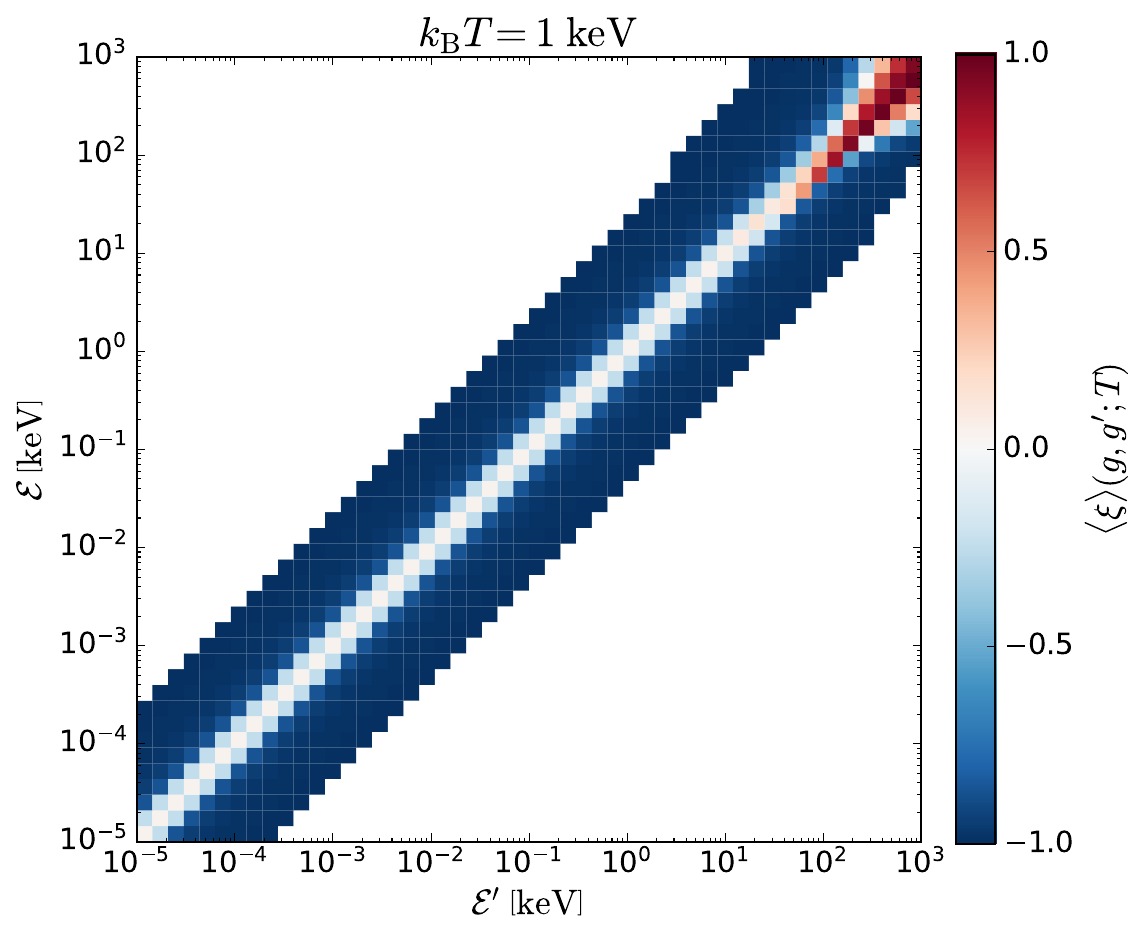}}
  \hfill
  \subfloat[$k_{\rm B}T=10~\mathrm{keV}$]{%
    \includegraphics[width=0.49\textwidth]
      {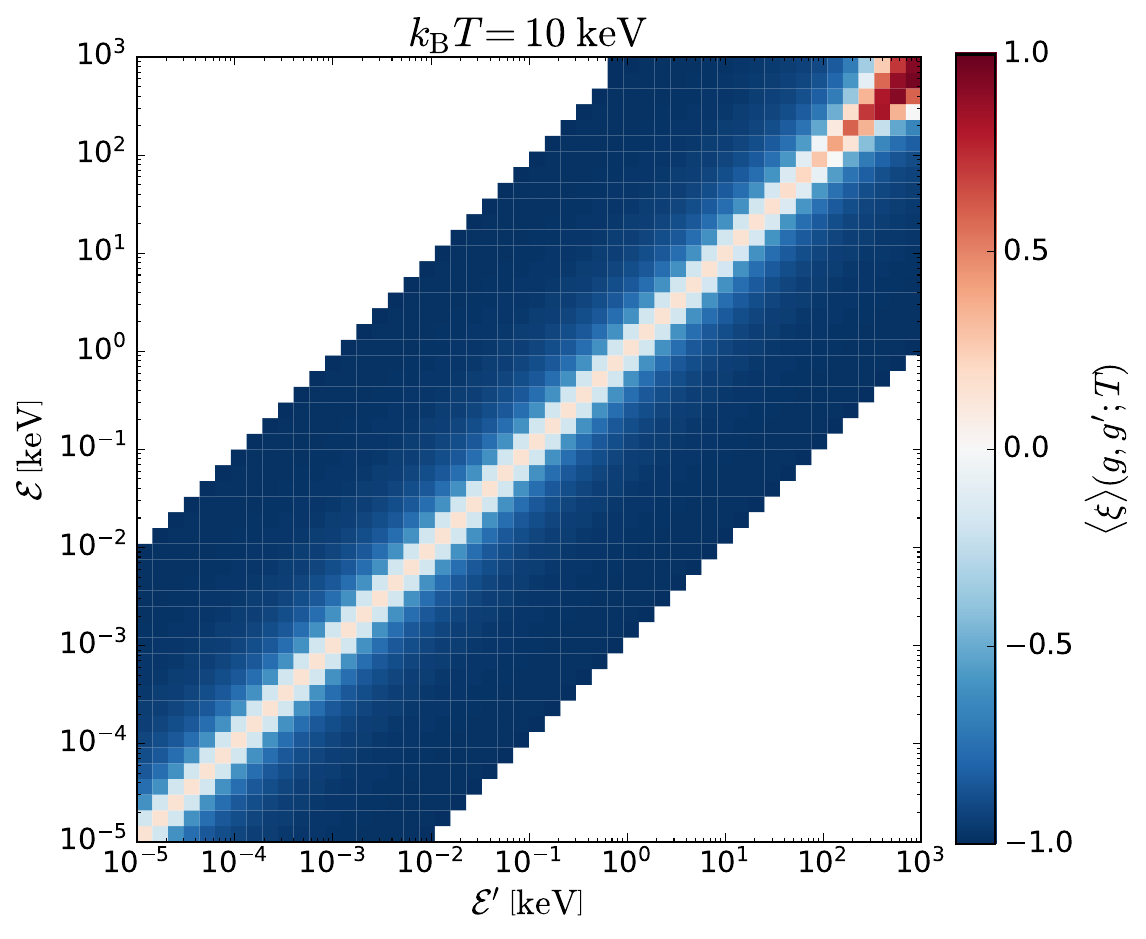}}\\[6pt]
  \subfloat[$k_{\rm B}T=50~\mathrm{keV}$]{%
    \includegraphics[width=0.49\textwidth]
      {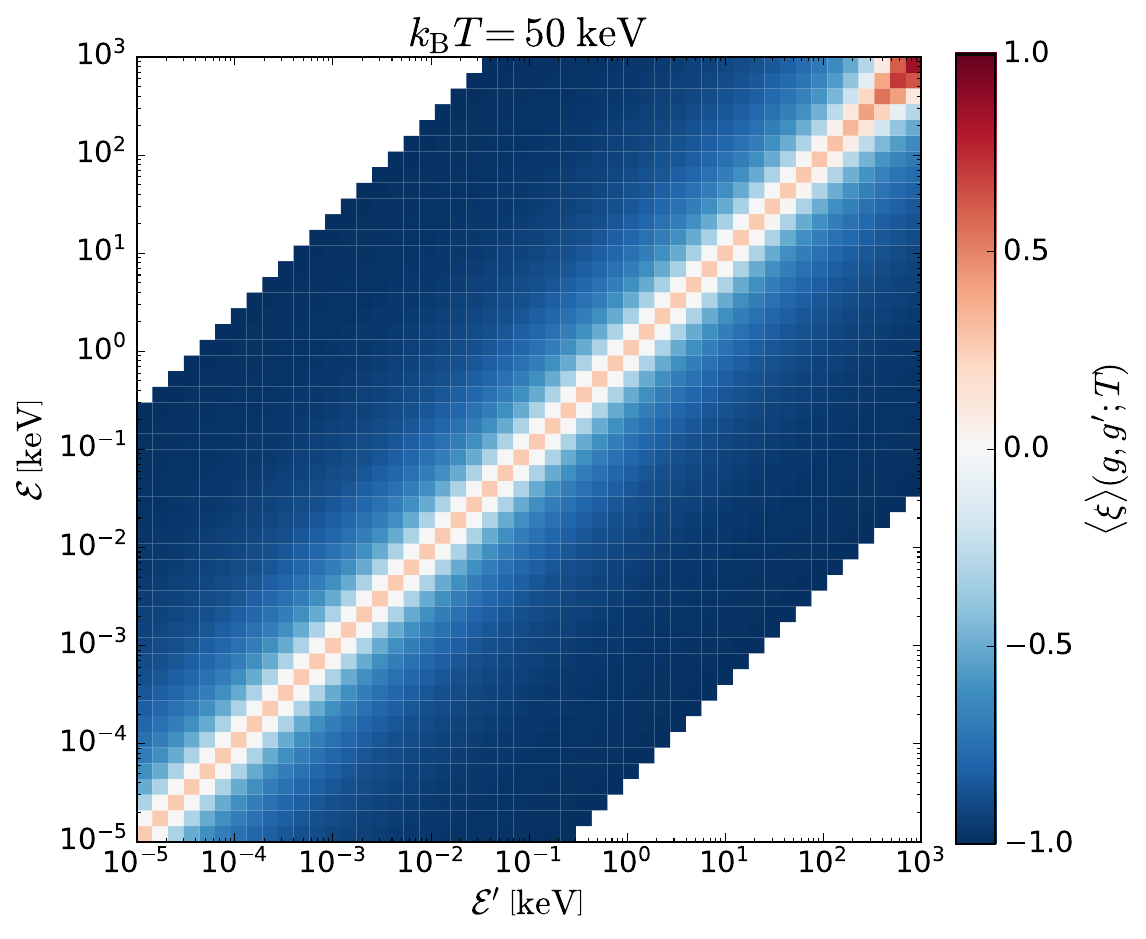}}
  \hfill
  \subfloat[$k_{\rm B}T=100~\mathrm{keV}$]{%
    \includegraphics[width=0.49\textwidth]
      {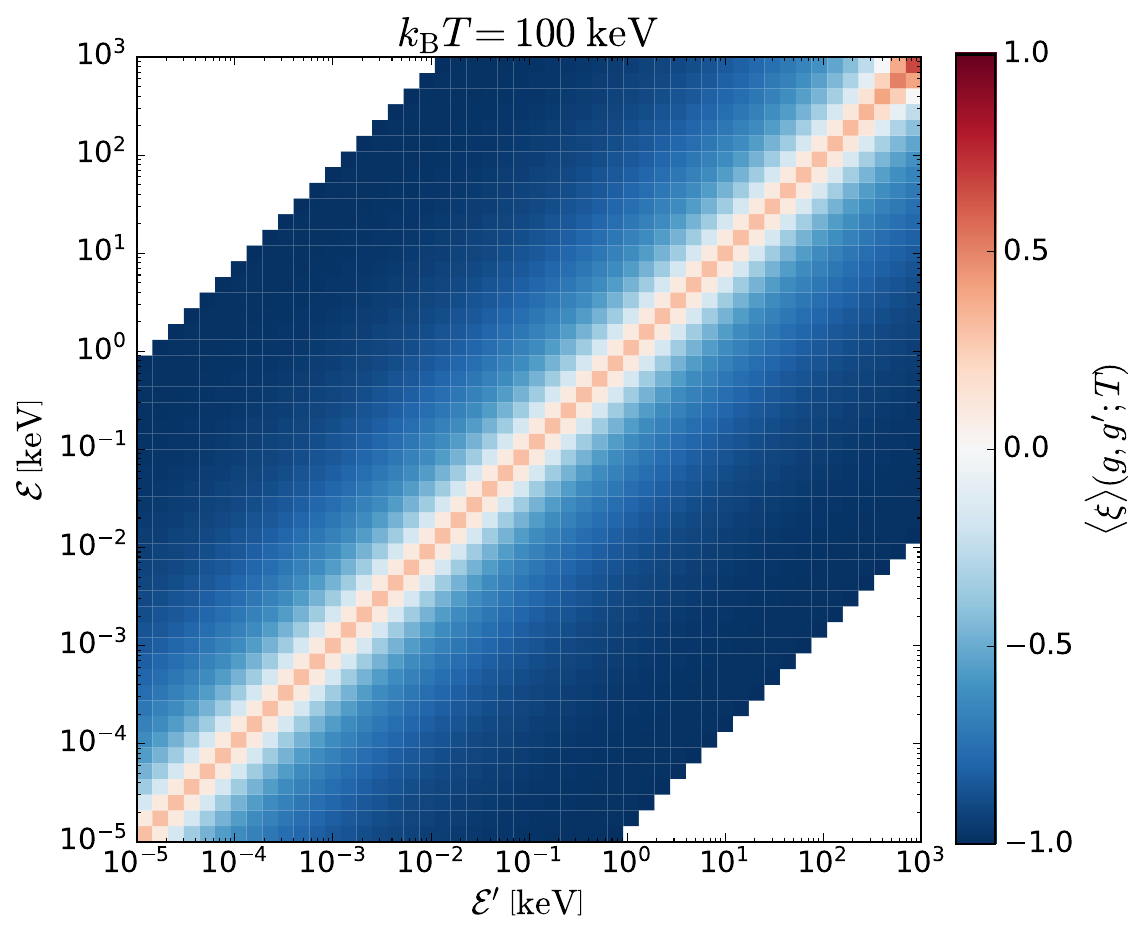}}
  \caption{
  $\langle\xi\rangle(\mathcal{E},\mathcal{E}',T)$ in the
  $\mathcal{E}$--$\mathcal{E}'$ plane.
  Evaluated on a 50-group geometric grid spanning
  $10^{-5}$--$10^{3}~\mathrm{keV}$ (angle-integrated)
  with the capped Planck weighting of Eq.~\eqref{eq:capped_planck_weight}.}
  \label{fig:xi_colormap}
\end{figure*}

\begin{figure*}[!htbp]
  \centering
  \includegraphics[width=0.75\textwidth]{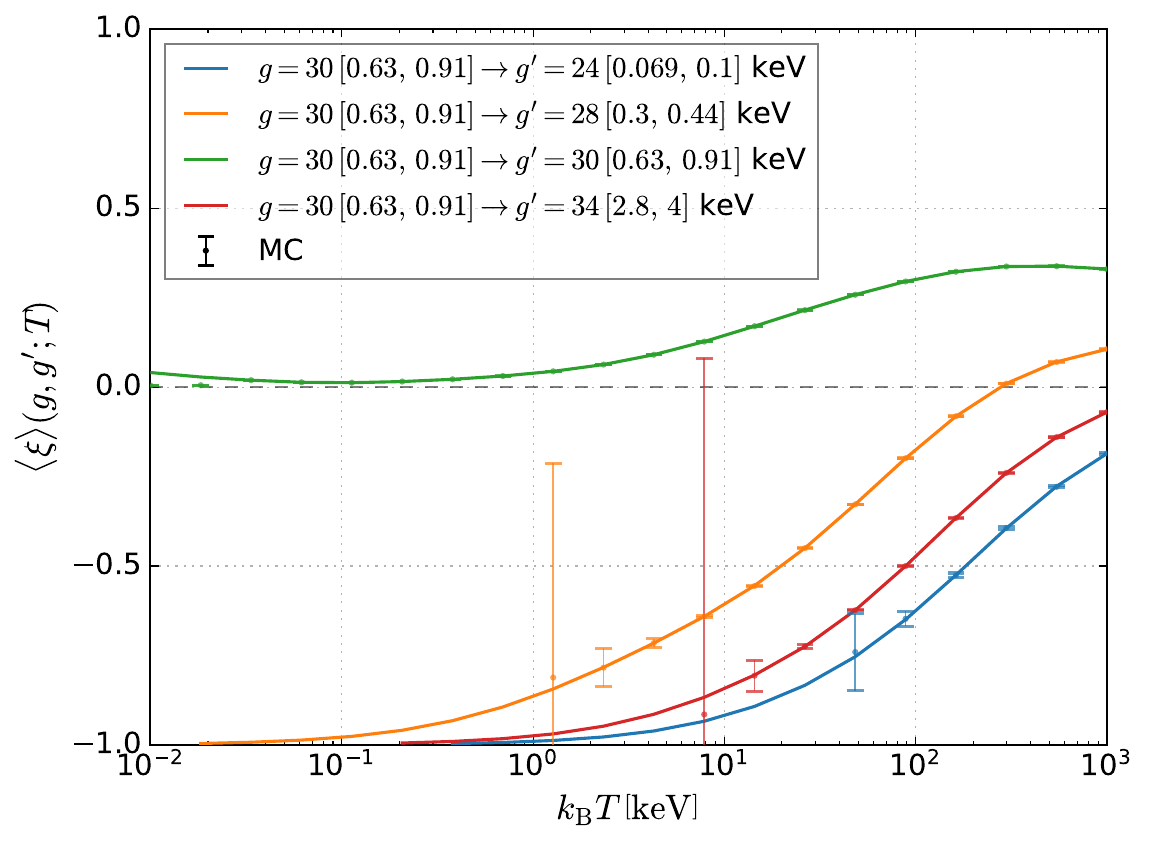}
  \caption{
  $\langle\xi\rangle$ as a function of temperature for several
  $(g,g')$ pairs (solid lines), evaluated at 20
  temperatures geometrically spaced from $10^{-2}$ to
  $10^{3}~\mathrm{keV}$.
  Computed on a 50-group geometric grid spanning
  $10^{-5}$--$10^{3}~\mathrm{keV}$ (angle-integrated)
  with the capped Planck weighting of
  Eq.~\eqref{eq:capped_planck_weight} and $x_{\rm cap}=500$. Monte Carlo error bars grow at lower temperatures, where fewer
  samples contribute to each $(g,g')$ pair, and are absent at the
  lowest temperatures where scattering is too rare for any samples
  to land in the group.}
  \label{fig:xi_vs_T}
\end{figure*}

For the next comparison, we consider the conditional mean scattering
cosine, obtained as a special case of
Eq.~\eqref{eq:multigroup_weighted_average} with $X=\xi$ and
$\Delta\xi_i=[-1,1]$,
\begin{multline}
\left\langle \xi\right\rangle(g, g'; T)
=\frac{2\pi}{\sigma_{\mathrm{out}}(g\rightarrow g';T)\,W_g}\times\\
\int_{\Delta\mathcal{E}_g}
w(\mathcal{E},T)
\int_{\Delta\mathcal{E}_{g'}}
\int_{-1}^{1}
\xi\,
\mathcal{K}(\mathcal{E}\rightarrow\mathcal{E}',\xi;\tau)
\,d\xi\,d\mathcal{E}'\,d\mathcal{E}.
\label{eq:xi_mean_definition}
\end{multline}
Figure~\ref{fig:xi_weighted_profiles} shows outgoing-energy profiles of
$\left\langle \xi\right\rangle(g, g'; T)$.
Figure~\ref{fig:xi_colormap} show
$\langle\xi\rangle(g, g'; T)$ as a function of the
incident and scattered photon energies.
Figure~\ref{fig:xi_vs_T} shows how the mean scattering
cosine varies with temperature for several representative
$(g,g')$ pairs.

\begin{figure*}[!htbp]
  \centering
  \subfloat[$k_{\rm B}T=1~\mathrm{keV},\;\mathcal{E}=1~\mathrm{keV}$]{%
    \includegraphics[width=0.49\textwidth]
      {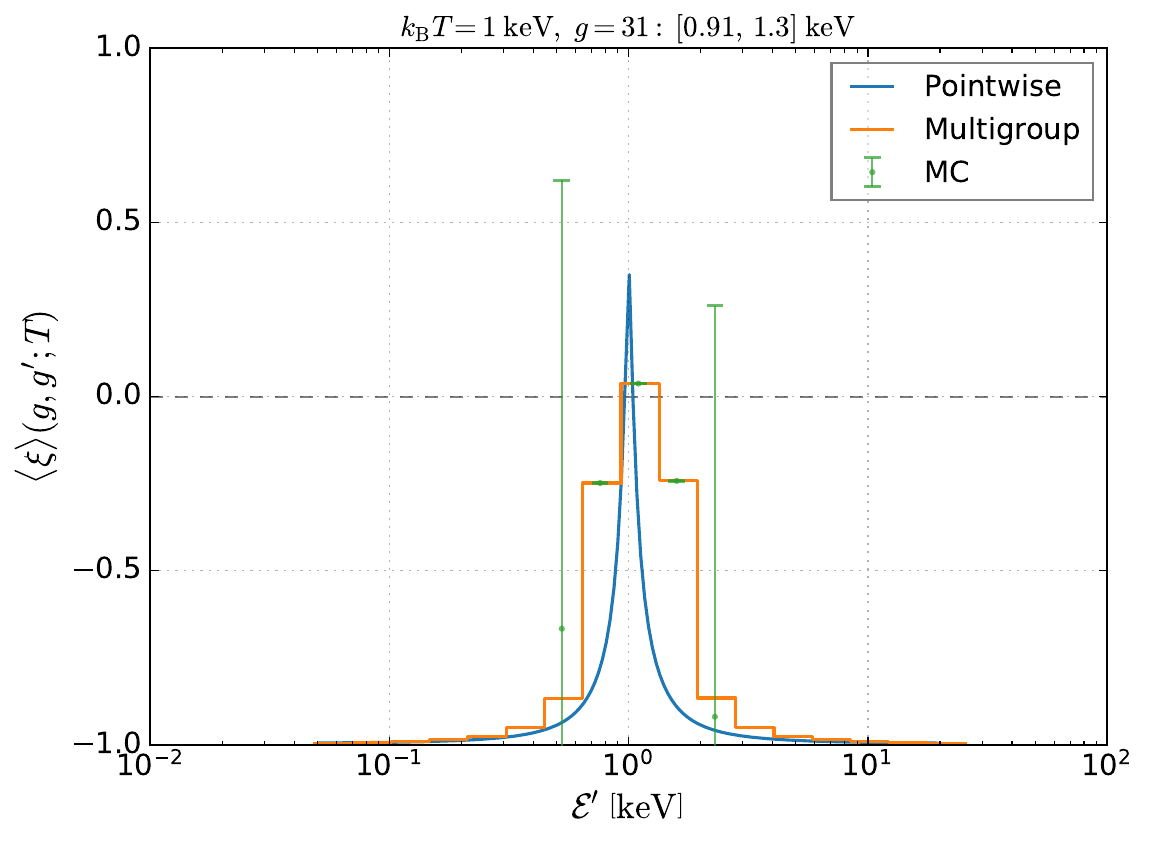}}
  \hfill
  \subfloat[$k_{\rm B}T=10~\mathrm{keV},\;\mathcal{E}=1~\mathrm{keV}$]{%
    \includegraphics[width=0.49\textwidth]
      {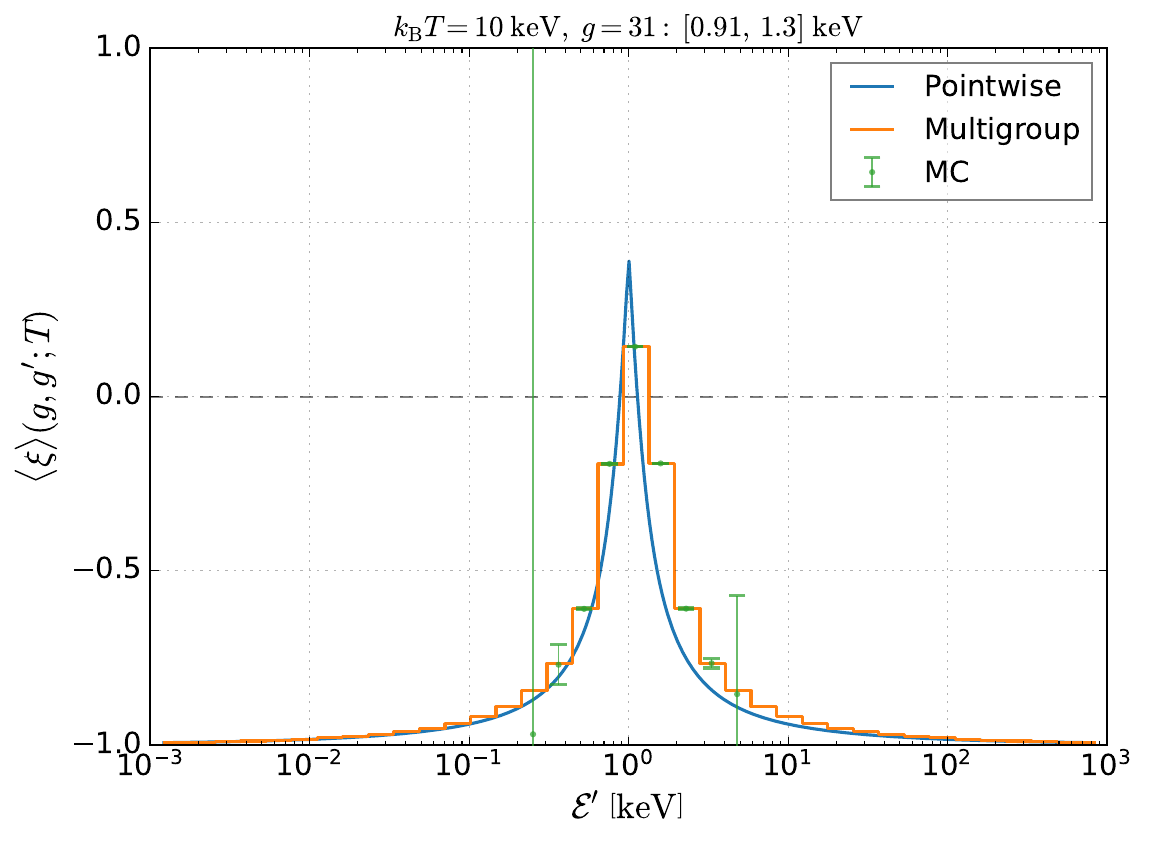}}\\[6pt]
  \subfloat[$k_{\rm B}T=10~\mathrm{keV},\;\mathcal{E}=10~\mathrm{keV}$]{%
    \includegraphics[width=0.49\textwidth]
      {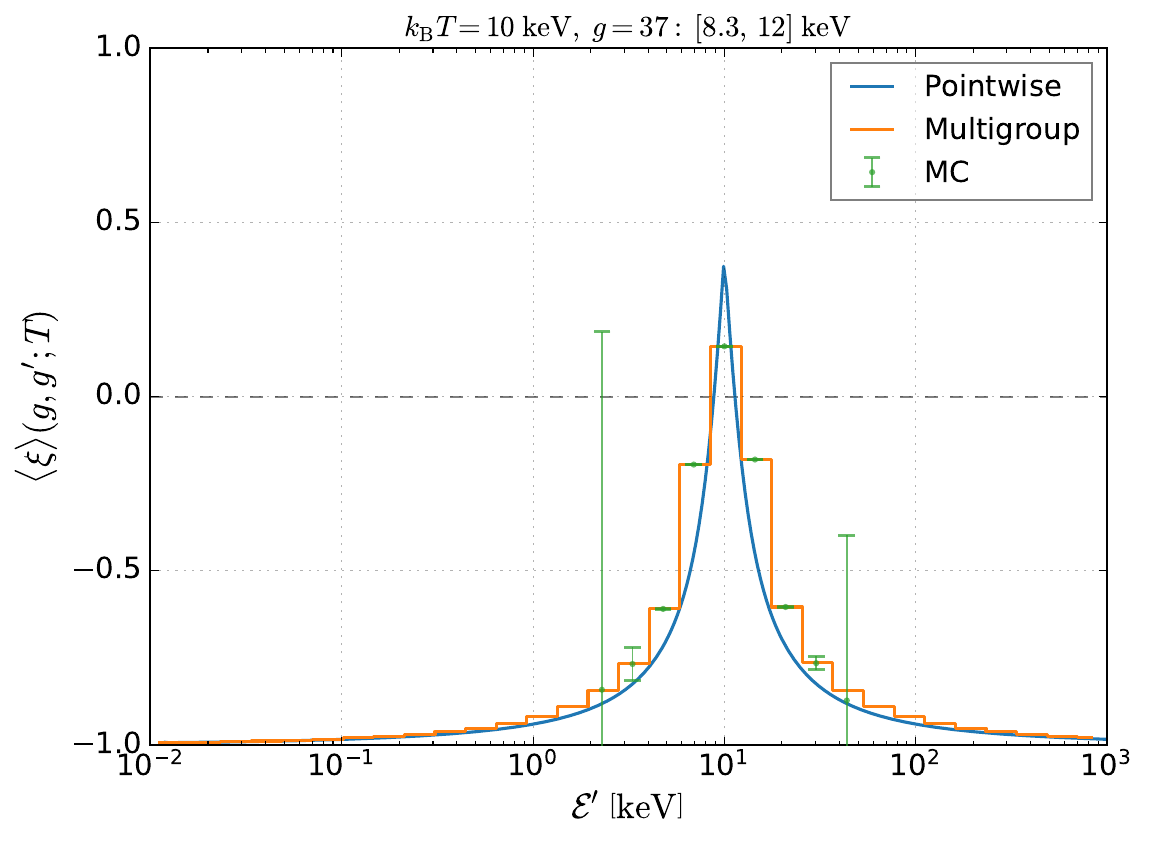}}
  \hfill
  \subfloat[$k_{\rm B}T=100~\mathrm{keV},\;\mathcal{E}=10~\mathrm{keV}$]{%
    \includegraphics[width=0.49\textwidth]
      {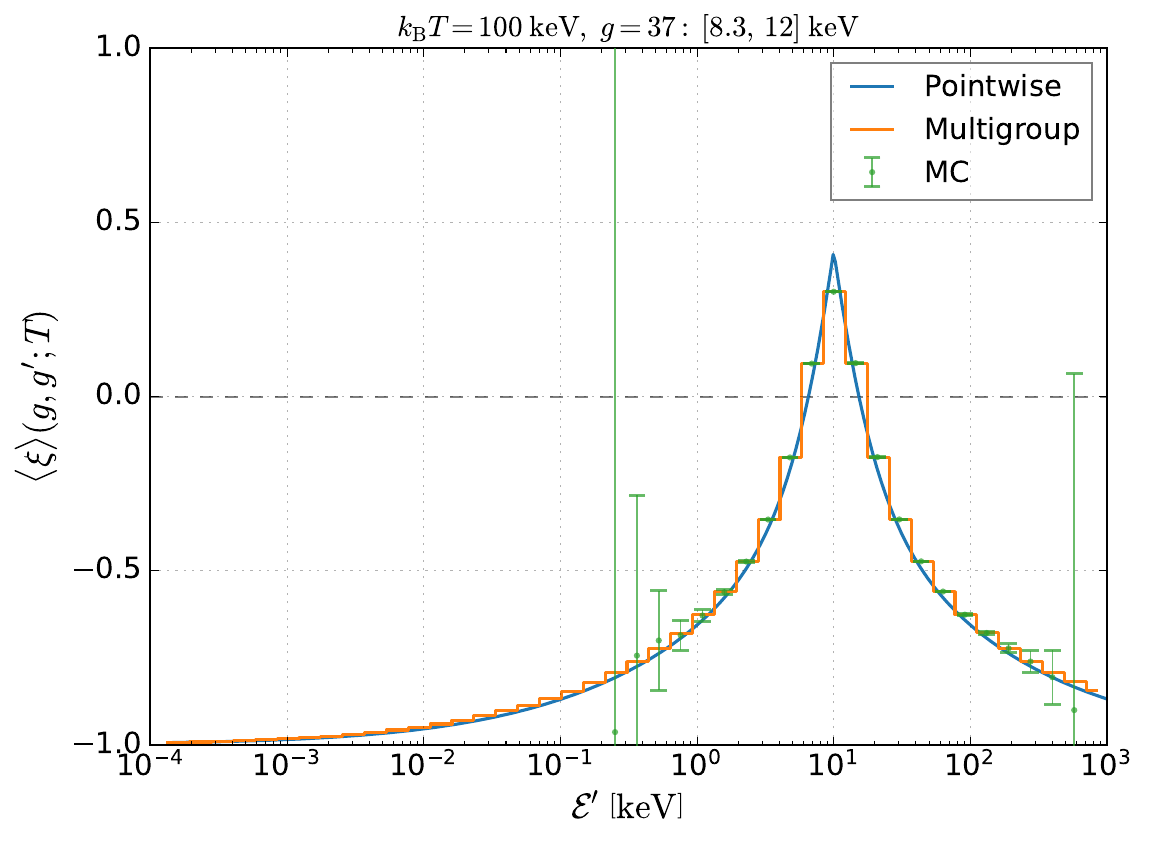}}
  \caption{
  $\left\langle \xi\right\rangle(g, g'; T)$ (step profiles)
  from the 50-group matrices of Figure~\ref{fig:xi_weighted_profiles},
  as a function of destination group $g'$ for fixed source group $g$
  and temperature $T$. Solid lines show the corresponding pointwise
  result (Eq.~\eqref{eq:xi_pointwise}) for comparison. Visible noise in the far wings reflects groups where the kernel is small
  and few Monte Carlo samples contribute to the estimate.}
  \label{fig:xi_vs_Ep}
\end{figure*}

Figure~\ref{fig:xi_vs_Ep} show the pointwise and multigroup mean
scattering cosine as a function of the scattered photon energy
$\mathcal{E}'$ for several combinations of temperature and incident
energy. The pointwise value is defined as the ratio
\begin{equation}
\langle\xi\rangle_\mathrm{pointwise}
=
\frac{\mathcal{I}_{[-1,1]}[\xi](\mathcal{E},\mathcal{E}';\tau)}
     {\mathcal{I}_{[-1,1]}[1](\mathcal{E},\mathcal{E}';\tau)},
\label{eq:xi_pointwise}
\end{equation}
evaluated on a fine angular grid spanning $[-1,1]$.

\begin{figure*}[!htbp]
  \centering
  \subfloat[Cross section $\sigma_{\mathrm{out}}$]{%
    \includegraphics[width=0.49\textwidth]{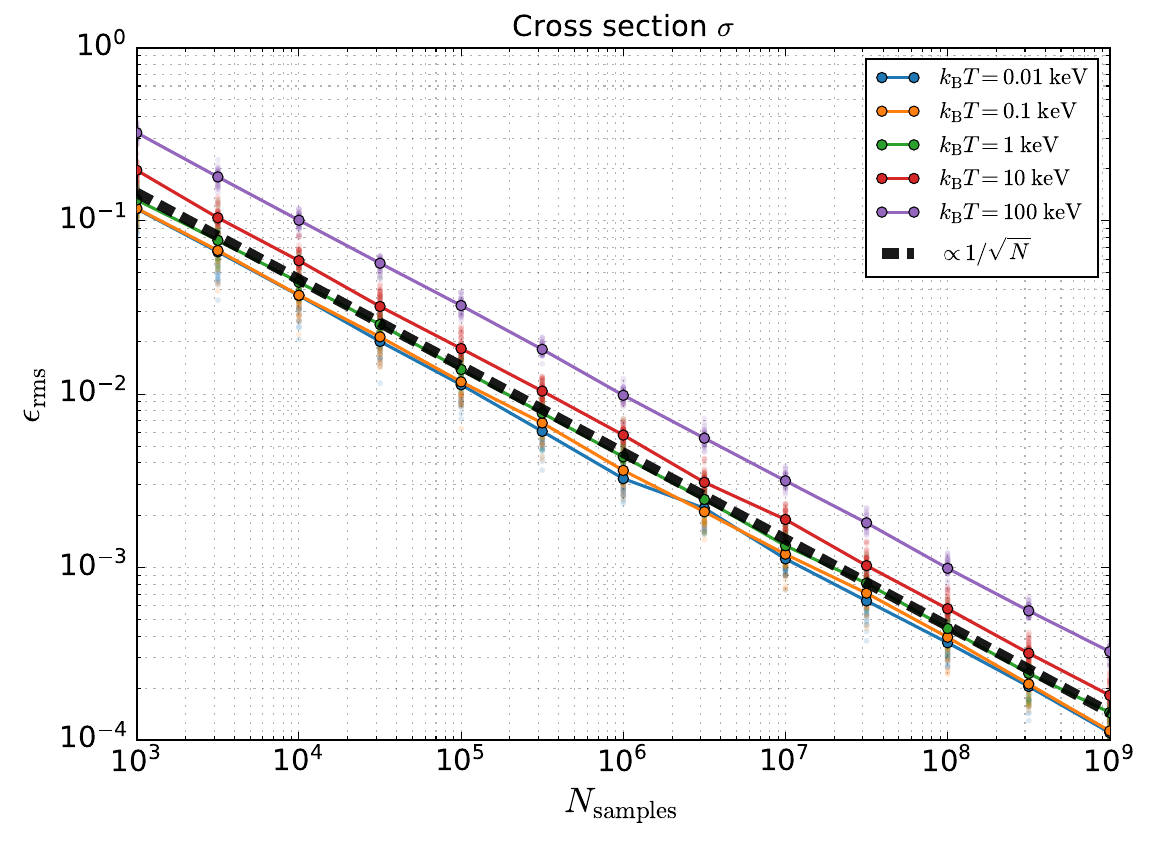}}
  \hfill
  \subfloat[Kernel-derivative contribution $\sigma_{\mathrm{out}}^{\partial T}$]{%
    \includegraphics[width=0.49\textwidth]{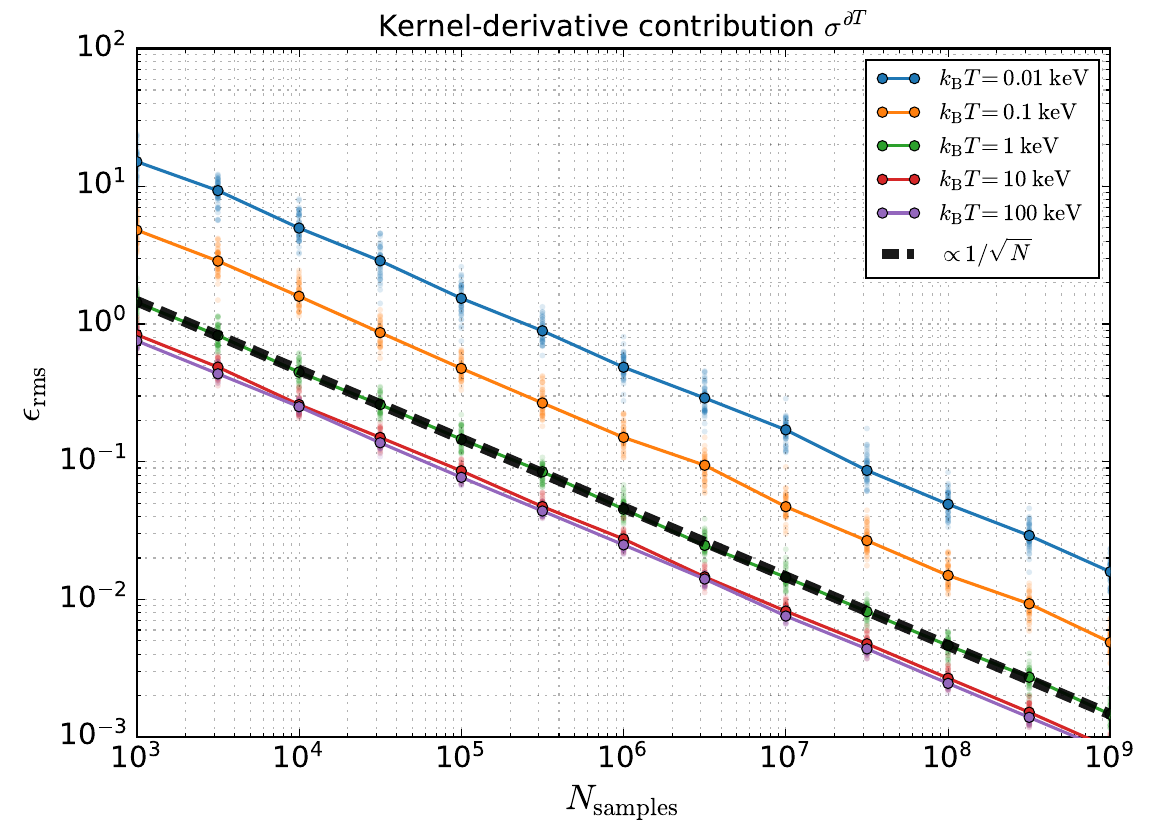}}
  \caption{
  Monte Carlo convergence of
  $\epsilon_{\mathrm{rms}}$ (Eq.~\eqref{eq:relative_rmse}) for the
  complete multigroup--multiangle matrices with the capped Wien
  weighting of Eq.~\eqref{eq:capped_wien_weight}. 
  Each point represents one of 50 independent Monte Carlo
  realizations at the indicated sample count. Solid curves show the
  median error at each temperature, and the dashed line shows the
  expected $N^{-1/2}$ scaling. The paper's quadrature is used as
  the reference.
  The matrices are evaluated on a 30-group geometric grid spanning
  $0.1$--$100~\mathrm{keV}$ with 10 uniform angular bins.}
  \label{fig:mc_convergence}
\end{figure*}

Finally, Figure~\ref{fig:mc_convergence} examines the statistical
convergence of the Monte Carlo estimator for the complete
multigroup--multiangle matrices. The relative root-mean-square error we used is 
\begin{equation}
\epsilon_{\mathrm{rms}}
=
\frac{
\left[\displaystyle\sum_{g,g',i}
\left(Q^{\mathrm{MC}}_{g\rightarrow g',i}
      - Q_{g\rightarrow g',i}\right)^{2}\right]^{1/2}
}{
\left[\displaystyle\sum_{g,g',i}
Q_{g\rightarrow g',i}^{2}\right]^{1/2}
},
\label{eq:relative_rmse}
\end{equation}
where $Q$ stands for either the outscattering cross section or its
kernel-derivative contribution, and the sums run over all group pairs
$(g,g')$ and angular bins $i$. The error is evaluated on a 30-group,
10-angle-bin grid spanning $0.1$--$100~\mathrm{keV}$, with the capped
Wien weighting
\begin{equation}
w(\mathcal{E},T)
=
\begin{cases}
\dfrac{\mathcal{E}^3}{(k_{\rm B}T)^3}\,
\exp\!\left[-\dfrac{\mathcal{E}}{k_{\rm B}T}\right],
& \mathcal{E}/(k_{\rm B}T) < x_{\rm cap},\\[6pt]
x_{\rm cap}^3\,e^{-x_{\rm cap}},
& \mathcal{E}/(k_{\rm B}T) \ge x_{\rm cap},
\end{cases}
\label{eq:capped_wien_weight}
\end{equation}
(Appendix~\ref{sec:group_denominators}), with $x_{\rm cap}=25$. At each sample count between $10^3$ and $10^9$, the error
is measured for 50 independent realizations at
$k_{\rm B}T=0.01$, $0.1$, $1$, $10$, and $100~\mathrm{keV}$, using the
paper's quadrature with elevated orders
($N_{\rm peak}=96$, $N_{\rm tail}=32$,
$N_{\rm edge}=N_{\rm int}=48$, $N_{\mathcal{E}}=24$) as the reference.
Both the outscattering cross
section and its kernel-derivative contribution exhibit the expected
$N^{-1/2}$ Monte Carlo convergence.

\begin{figure*}[!htbp]
  \centering
  \subfloat[$k_{\rm B}T=1~\mathrm{keV}$]{%
    \includegraphics[width=0.49\textwidth]
      {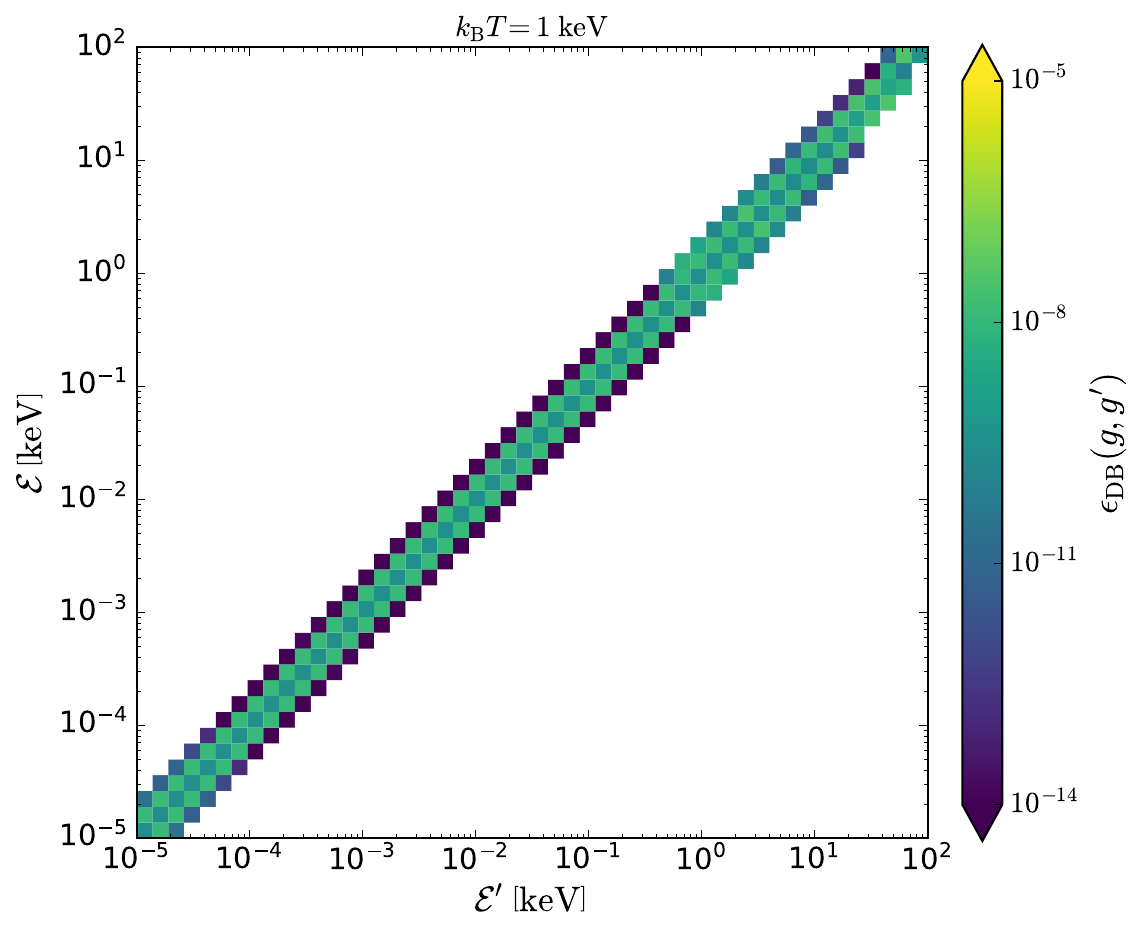}}
  \hfill
  \subfloat[$k_{\rm B}T=100~\mathrm{keV}$]{%
    \includegraphics[width=0.49\textwidth]
      {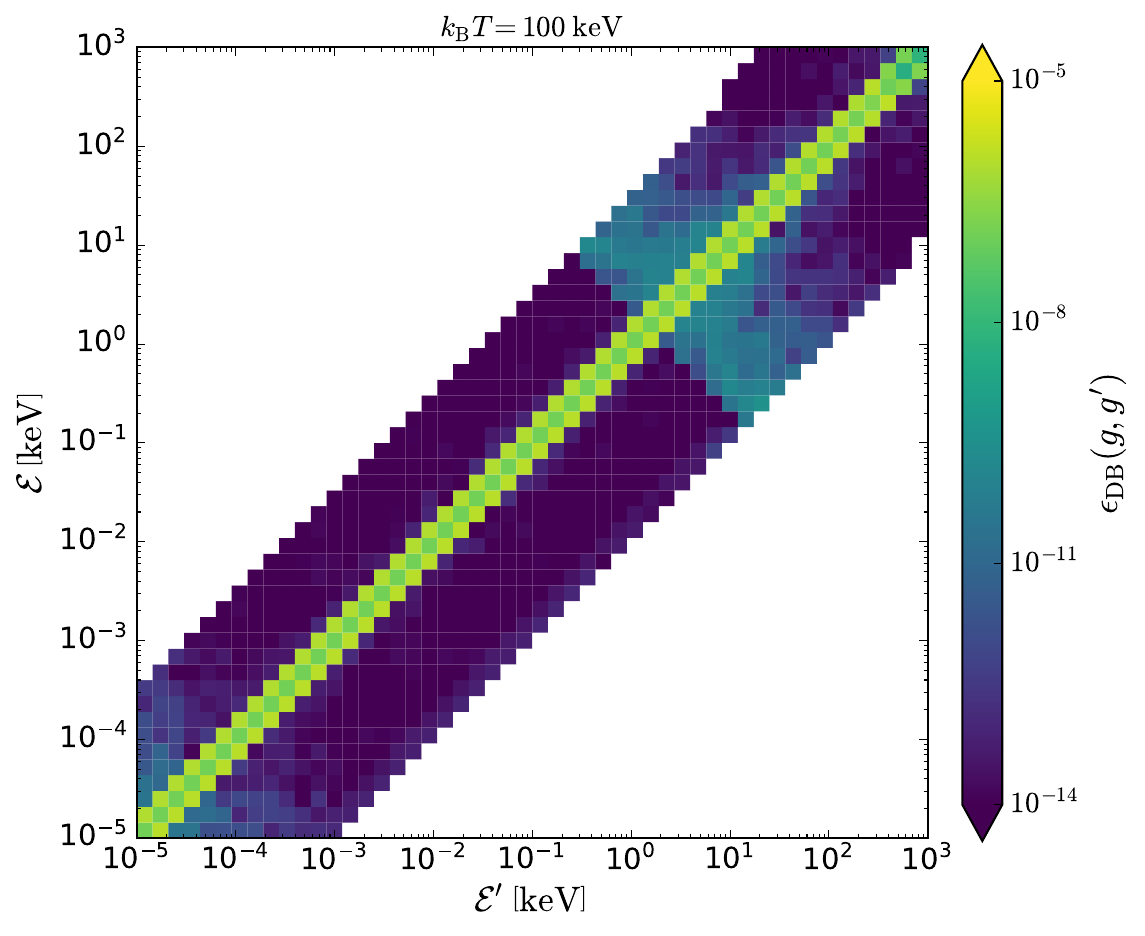}}
  \caption{Relative residual $\epsilon_{\rm DB}(g,g')$ (Eq.~\eqref{eq:detailed_balance_residual}) of the multigroup detailed-balance identity, Eq.~\eqref{eq:multigroup_detailed_balance}, for the angle-integrated matrices with the Wien weight of Eq.~\eqref{eq:capped_wien_weight} without a cap, at the default quadrature orders on 50 logarithmically spaced groups. The vertical axis gives the incident energy and the horizontal axis the scattered energy. White cells denote entries of $\sigma_{\rm out}$ below $10^{-8}$ of the matrix maximum.}
  \label{fig:detailed_balance}
\end{figure*}

Because the electrons are in thermal equilibrium, the kernel satisfies the detailed-balance relation
\begin{multline}
\mathcal{E}^2 e^{-\mathcal{E}/k_{\rm B}T}\,\mathcal{K}(\mathcal{E}\rightarrow\mathcal{E}',\xi;\tau)
=\\
\mathcal{E}'^2 e^{-\mathcal{E}'/k_{\rm B}T}\,\mathcal{K}(\mathcal{E}'\rightarrow\mathcal{E},\xi;\tau)
\label{eq:pointwise_detailed_balance}
\end{multline}
at every scattering angle \citep[Eq.~13 of][see also \citealt{Sarkar2019}]{KERSHAW1986273}. For the Wien weight of Eq.~\eqref{eq:capped_wien_weight} without a cap, multiplying Eq.~\eqref{eq:pointwise_detailed_balance} by $\mathcal{E}$ and integrating over a pair of groups and an angular bin gives an exact identity between the outscattering and inscattering matrices,
\begin{equation}
W_g\,\sigma_{\rm out}(g\rightarrow g',i) = W_{g'}\,\sigma_{\rm in}(g'\rightarrow g,i),
\label{eq:multigroup_detailed_balance}
\end{equation}
where $\sigma_{\rm in}(g\rightarrow g',i)$ is the angle-resolved form of Eq.~\eqref{eq:sigma_in}. The two sides are separate integrals, with panels built around different incident groups, so the residual of Eq.~\eqref{eq:multigroup_detailed_balance} directly measures the numerical error of the scheme. We evaluate both sides at the default orders on 50 logarithmically spaced groups from $10^{-5}$ to $10^{3}$\,keV ($10^{2}$\,keV at $k_{\rm B}T=1$\,keV), at $k_{\rm B}T=1$, 10, 50, and 100\,keV, angle-integrated and with 10 uniform angular bins. Figure~\ref{fig:detailed_balance} shows the relative residual
\begin{equation}
\epsilon_{\rm DB}(g,g')
=
\frac{\left|W_g\,\sigma_{\rm out}(g\rightarrow g')-W_{g'}\,\sigma_{\rm in}(g'\rightarrow g)\right|}
{W_g\,\sigma_{\rm out}(g\rightarrow g')}
\label{eq:detailed_balance_residual}
\end{equation}
of the angle-integrated matrices at 1 and 100\,keV. For every entry above $10^{-8}$ of the matrix maximum, the residual is at most $5\times10^{-7}$ at 1 to 50\,keV and $1.3\times10^{-6}$ at 100\,keV, both angle-integrated and in each angular bin.

\subsection{Computational Cost} \label{sec:timing}

To place the cost of the method in context, we compare it with a general-purpose adaptive integrator applied to the same integrals. The test matrix has 5 logarithmically spaced groups between 0.1 and 100\,keV and 5 uniform angular bins, with the Wien weight of Eq.~\eqref{eq:capped_wien_weight} without a cap, at $k_{\rm B}T=0.1$, 1, 10, and 100\,keV. The adaptive integrator nests three calls of the QUADPACK routine \texttt{qags} \citep{Piessens1983} from the GNU Scientific Library \citep{Galassi2009}, with $\xi$ innermost and $\mathcal{E}$ outermost, as \texttt{scipy.integrate.nquad} does \citep{Virtanen2020}. It evaluates the same pointwise kernel in the same compiled program, and its only setting is the requested relative tolerance $\epsilon_{\rm rel}$, which we vary from $10^{-2}$ to $10^{-8}$. The present method runs at its default orders (Table~\ref{tab:quadrature_orders}). Both methods are scored with $\epsilon_{\rm rms}$ of Eq.~\eqref{eq:relative_rmse}, with the tested matrix in place of $Q^{\mathrm{MC}}$ and the present method at four times its default orders as the reference $Q$. This reference agrees with six times the default orders to $4\times10^{-10}$ or better. An adaptive run counts as converged when every entry finishes within the evaluation and time limits. All times are single-core CPU times on the same hardware.

\begin{figure*}[!htbp]
  \centering
  \includegraphics[width=0.75\textwidth]{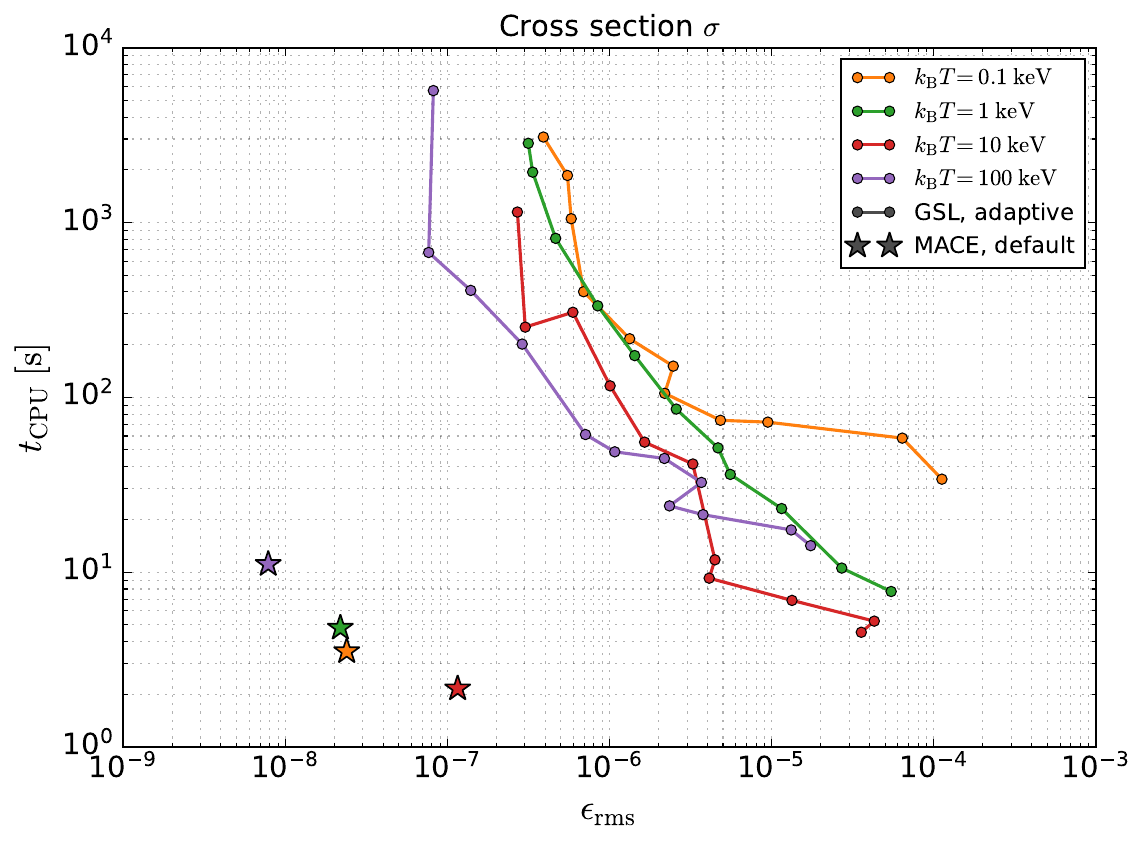}
  \caption{Single-core CPU time against $\epsilon_{\rm rms}$ for the 5-group, 5-angle-bin test matrix of Section~\ref{sec:timing}. Lines connect the runs of the nested adaptive integrator at requested tolerances from $10^{-2}$ (right) to $10^{-7}$, and to $3\times10^{-8}$ at 100\,keV. Stars show the present method at its default orders.}
  \label{fig:gsl_work_precision}
\end{figure*}

Figure~\ref{fig:gsl_work_precision} shows the result. The present method computes the 125 entries in about 2 to 11\,s, with $\epsilon_{\rm rms}$ of $10^{-8}$ to $10^{-7}$. The adaptive error levels off up to an order of magnitude higher, so no converged adaptive run reaches the same accuracy, and the most accurate one takes about 60 to 900 times as long. Even at $\epsilon_{\rm rms}=10^{-6}$ the adaptive integrator is about 5 to 100 times slower.

\section{Conclusions} \label{sec:conclusions}

We have developed a deterministic method for integrating the thermal Compton scattering kernel over arbitrary incident-energy groups, scattered-energy groups, and scattering-angle bins. The same framework evaluates kernel-weighted moments with general smooth multipliers. Its purpose is to convert the exact relativistic redistribution kernel into the finite-dimensional scattering data required by multigroup and angularly resolved radiation-transport calculations.

The pointwise calculation combines complementary representations of the thermal kernel with robust numerical safeguards. The multidimensional integral is then evaluated through nested one-dimensional quadratures designed around the local structure of the kernel. The angular and energy panels follow the Compton recoil ridge, its thermal broadening, and the nearly elastic forward-scattering feature. This concentrates numerical resolution where the integrand varies most rapidly while retaining coverage of the full integration domain.

The method was validated against published angle-integrated profiles, Legendre moments, energy-transfer moments, mean free paths, and multigroup redistribution data. The complete multigroup-multiangle calculation was also compared with an independent Monte Carlo estimator that samples the underlying scattering process. The agreement across these tests supports both the pointwise kernel evaluation and the nested quadrature procedure. In the test of Section~\ref{sec:timing}, the method at its default settings is also more accurate than every converged run of a general-purpose nested adaptive integrator, and about 60 to 900 times faster than the most accurate of them.

The resulting matrices describe both upscattering and downscattering while preserving the relation between energy redistribution and scattering angle. The method can therefore generate Compton scattering data for a range of multigroup transport methods.

By providing a validated and general procedure for constructing full-kernel multigroup Compton scattering coefficients, this work offers a practical foundation for incorporating accurate Compton scattering treatment into astrophysical radiation-transport codes. A multigroup diffusion implementation of the method has already been incorporated into \textsc{rich} \citep{Giron_2026_compton}, where it forms a fundamental component of the code's ongoing development for radiation-hydrodynamic simulations of tidal disruption events with Compton scattering.

The present treatment assumes unpolarized photons scattering from free electrons with an isotropic relativistic Maxwell-Jüttner distribution. It does not include polarization, anisotropic or nonthermal electrons, electron degeneracy, or bound-electron effects. Within these assumptions, the method provides a validated foundation for full-kernel Compton scattering in multigroup-multiangle astrophysical radiation transport.

\section{Software and Data Availability} \label{sec:software}

The MACE (Multigroup and Angle-resolved Compton Evaluator) implementation described in this paper is publicly available as an open-source GitHub repository at \url{https://github.com/ItamarShmelo/MACE}. 

Pre-computed multigroup scattering matrices on configurable energy, angle, and temperature grids can be downloaded through the MACE web interface at \url{https://itamarshmelo.github.io/MACE-Website/}.

\section{acknowledgments}
The simulations were performed on the ICPL cluster at the Racah
Institute of Physics.
\vspace{0.3cm}

During the preparation of this work the authors used ChatGPT \citep{openai_chatgpt} and Claude \citep{anthropic_claude} to assist with writing and refactoring the solver code, writing code to generate the plots for the paper, and improving phrasing and wording. After using these tools, the authors reviewed and edited the content as needed and take full responsibility for the content of the published article.

\appendix

\section{Power Series Representation} \label{sec:power_series_appendix}

This appendix presents the power series representation of the Compton kernel derived by \citet{KERSHAW1986273} (Eq.~15) and its temperature derivative. The quantities appearing in the kernel are
\begin{equation}
  a = 1-\xi,\quad q = \sqrt{(\gamma'-\gamma)^2 + 2\gamma\gamma'a},\quad s = \frac{1}{\gamma} + \frac{1}{\gamma'},\quad \omega^2 = \frac{1 + \xi}{a}
  \label{eq:def_a_q}
\end{equation}
\begin{equation}
  \lambda_+ = \frac{(\gamma'-\gamma) + q\sqrt{1 + 2/(\gamma\gamma'a)}}{2}
  \label{eq:def_lambda_+}
\end{equation}
\begin{equation}
  \rho_+ = \lambda_+ + \gamma,\quad \rho_- = \lambda_+ - \gamma',\quad \alpha_\pm = \frac{1}{\sqrt{\rho_\pm^2+\omega^2}}
  \label{eq:def_rho_pm_omega2}
\end{equation}
\begin{equation}
  G = -\gamma\gamma' + \frac{2}{a} + \frac{2}{\gamma\gamma'a^2},\quad \Lambda_\pm = G \mp \frac{s}{\tau a^2},\quad b = \frac{\omega}{2\tau},\quad \theta_\pm = \operatorname{arcsinh}(\rho_\pm/\omega)
  \label{eq:def_G}
\end{equation}
\begin{equation}
  \Sigma_\mathcal{E} = \frac{N_e r_e^2 m_e c^2}{4\mathcal{E}^2\tau} \times \frac{\exp(-\lambda_+/\tau)}{K_2(1/\tau)}
  \label{eq:def_Sigma_E}
\end{equation}
\begin{equation}
  \Psi = \frac{2\tau\gamma\gamma'}{q} + \frac{s}{a^2}\left(\alpha_+ + \alpha_-\right) + \frac{\rho_+\alpha_+ - \rho_-\alpha_-}{a}
  \label{eq:Psi}
\end{equation}
\begin{equation}
  P_\pm = e^{\rho_\pm / \tau}\sum_{n=0}^\infty \frac{b^n}{n!}e^{-n\theta_\pm}\Bigg[\Lambda_\pm + \frac{2n}{a}\Bigg]E_{n+1}(be^{\theta_\pm})
\end{equation}
where $E_{n+1}(x)$ denotes the generalized exponential integral. The power series is
\begin{equation}
  \mathcal{K}(\mathcal{E}\rightarrow\mathcal{E}', \xi\ ; \tau) = \Sigma_\mathcal{E}\times (\Psi + P_+ - P_-)
  \label{eq:power_series}
\end{equation}

\subsection{Temperature Derivative} \label{sec:power_series_derivative}

The temperature derivative is obtained by differentiating Eq.~\eqref{eq:power_series} with respect to $T$,
\begin{equation}
  \frac{\partial \mathcal{K}}{\partial T} = \frac{k_{\rm B}}{m_ec^2}\Sigma_\mathcal{E} \Bigg\{\frac{2\gamma\gamma'}{q} + \frac{\partial P_+}{\partial \tau}  - \frac{\partial P_-}{\partial\tau} + \Big[\frac{\lambda_+ - \kappa}{\tau^2} - \frac{3}{\tau}\Big](\Psi + P_+ - P_-)\Bigg\}
  \label{eq:dpower_series_dT}
\end{equation}
where $\kappa(\tau) = K_1(\tau^{-1})/K_2(\tau^{-1})$ and
\begin{equation}
  \frac{\partial P_\pm}{\partial \tau} = e^{\rho_\pm/\tau}\sum_{n=0}^\infty \frac{b^n}{n!}e^{-n\theta_\pm}\times \Bigg\{\Big[\pm\frac{s}{\tau^2 a^2} - \Big(\frac{\rho_\pm}{\tau^2} + \frac{n}{\tau}\Big)\Big(\Lambda_\pm + \frac{2n}{a}\Big)\Big]E_{n+1}(be^{\theta_\pm}) + \frac{be^{\theta_\pm}}{\tau}\Big(\Lambda_\pm + \frac{2n}{a}\Big)E_n(be^{\theta_\pm})\Bigg\}
  \label{eq:dP_pm_dtau}
\end{equation}

\subsection{Roundoff Error Estimates} \label{sec:power_series_roundoff}

For the power series we estimate the relative roundoff contribution using the cancellation error estimate
\begin{equation}
  \epsilon_{\rm ro} = N_{terms}\epsilon_{machine}\epsilon_P\epsilon_\Psi
  \label{eq:epsilon_cancel}
\end{equation}
Here $\epsilon_{machine}$ is the machine tolerance\footnote{$\sim10^{-16}$ for double precision, $\sim 10^{-32}$ for double-double precision}
\begin{equation}
  \epsilon_P = \frac{|P_+| + |P_-|}{|P_+ - P_-|},\quad \epsilon_\Psi = \frac{|\Psi| + |P_+ - P_-|}{|\Psi + P_+ - P_-|}
  \label{eq:epsilon_P_epsilon_PSI}
\end{equation}
The derivative cancellation error is determined in a similar way, accounting for the two independent cancellation paths that arise from the product rule. The normalized derivative carries its own chain of cancellation factors,
\begin{equation}
  \epsilon_{\partial P} = \frac{|\partial P_+/\partial\tau| + |\partial P_-/\partial\tau|}{|\partial P_+/\partial\tau - \partial P_-/\partial\tau|},\quad \epsilon_{\partial\Psi} = \frac{|2\gamma\gamma'/q| + |\partial P_+/\partial\tau - \partial P_-/\partial\tau|}{|2\gamma\gamma'/q + \partial P_+/\partial\tau - \partial P_-/\partial\tau|}
  \label{eq:epsilon_dP_epsilon_dPsi}
\end{equation}
and the combined derivative cancellation error is
\begin{equation}
  \epsilon_{\rm ro}^{\partial T} = N_{terms}\epsilon_{machine}\frac{k_{\rm B}\Sigma_\mathcal{E}}{m_ec^2}\times \Bigg( \Big|\frac{2\gamma\gamma'}{q} + \frac{\partial P_+}{\partial\tau} - \frac{\partial P_-}{\partial\tau}\Big|\,\epsilon_{\partial P}\epsilon_{\partial\Psi} + \Big|\Big[\frac{\lambda_+ - \kappa}{\tau^2} - \frac{3}{\tau}\Big](\Psi + P_+ - P_-)\Big|\,\epsilon_P\epsilon_\Psi\Bigg) \Bigg/ \left|\frac{\partial\mathcal{K}}{\partial T}\right|
  \label{eq:epsilon_cancel_deriv}
\end{equation}
where $\epsilon_P$ and $\epsilon_\Psi$ are as defined in Eq.~\eqref{eq:epsilon_P_epsilon_PSI}.

\section{Asymptotic Series Representation} \label{sec:asymptotic_series_appendix}

This appendix presents the asymptotic series representation of the Compton kernel (\citealt{KERSHAW1986273}, Eq.~16), its temperature derivative, the rearranged accumulation that avoids catastrophic cancellation at low photon energy, and the associated roundoff error estimates. The additional quantities (beyond those defined in Appendix~\ref{sec:power_series_appendix}) appearing in the series are
\begin{equation}
  \eta_\pm = \alpha_\pm\left(\pm\frac{s}{a^2}+\frac{\rho_\pm}{a}\right),\quad \chi_\pm = \rho_\pm \alpha_\pm
  \label{eq:def_eta_pm}
\end{equation}
The per-sign terms are
\begin{equation}
  T_n^+ = (-\tau\alpha_+)^{n+1} \bigg[ \left(-Gn!+\frac{(n+1)!}{a}\right) P_n(\chi_+) -\eta_+(n+1)!\,P_{n+1}(\chi_+) \bigg]
  \label{eq:raw_Tplus}
\end{equation}
\begin{equation}
  T_n^- = (-\tau\alpha_-)^{n+1} \bigg[ \left(Gn!-\frac{(n+1)!}{a}\right) P_n(\chi_-) +\eta_-(n+1)!\,P_{n+1}(\chi_-) \bigg]
  \label{eq:raw_Tminus}
\end{equation}
where $P_n$ are the Legendre polynomials. The series is not convergent, it must be truncated when the terms begin to increase. The asymptotic series is
\begin{equation}
  \mathcal{K}(\mathcal{E}\rightarrow\mathcal{E}', \xi\ ; \tau) = \Sigma_\mathcal{E} \times \Bigg(\frac{2\tau\gamma\gamma'}{q} + \sum_{n=0}^{\infty}\big(T_n^+ + T_n^-\big)\Bigg)
  \label{eq:asymptotic_series}
\end{equation}

\subsection{Temperature Derivative} \label{sec:asymptotic_derivative}

The asymptotic series derivative is 
\begin{equation}
  \frac{\partial \mathcal{K}}{\partial T} = \frac{k_{\rm B}}{m_ec^2}\Sigma_\mathcal{E}\times\Bigg(\frac{2\gamma\gamma'}{q}\left[\frac{\lambda_+-\kappa}{\tau}-2\right] + \sum_{n=0}^{\infty}\left[\frac{\lambda_+-\kappa}{\tau^2}+\frac{n-2}{\tau}\right](-\tau\alpha_+)^{n+1} C_n^+ + \sum_{n=0}^{\infty}\left[\frac{\lambda_+-\kappa}{\tau^2}+\frac{n-2}{\tau}\right](-\tau\alpha_-)^{n+1} C_n^-\Bigg)
  \label{eq:dasymptotic_dT}
\end{equation}
where 
\begin{equation}
  C^\pm_n =\left[\mp Gn! \pm \frac{(n+1)!}{a}\right] P_n(\chi_\pm) \mp \eta_\pm\,(n+1)!\, P_{n+1}(\chi_\pm)
\end{equation}

\subsection{Grouped Rearrangement} \label{sec:grouped_rearrangement}

The $G$-dependent contributions to $T_n^+$ and $T_n^-$ nearly cancel in the
soft-photon limit, so accumulating the two sums separately loses significant
digits. The cancellation is avoided by combining $T_n^+$ and $T_n^-$ at each
order before accumulation. Write
$T_n^+$ and $T_n^-$ in terms of a common power
\begin{equation}
S_n^\pm=(-\tau\alpha_\pm)^{n+1}P_n(\chi_\pm).
\label{eq:sign_terms}
\end{equation}
Collecting the coefficients of $G\,n!$, $(n+1)!/a$, and $(n+1)!\,\eta_\pm$
from Eqs.~\eqref{eq:raw_Tplus}--\eqref{eq:raw_Tminus} gives
\begin{equation}
T_n^++T_n^-
=
G\,n!\left(S_n^--S_n^+\right)
+\frac{(n+1)!}{a}\left(S_n^+-S_n^-\right)
-(n+1)!\Big(\eta_+(-\tau\alpha_+)^{n+1}P_{n+1}(\chi_+)-\eta_-(-\tau\alpha_-)^{n+1}P_{n+1}(\chi_-)\Big).
\label{eq:raw_combined_step}
\end{equation}
Introduce the difference quantities
\begin{equation}
D_n=S_n^--S_n^+
=
(-\tau\alpha_-)^{n+1}P_n(\chi_-)
-(-\tau\alpha_+)^{n+1}P_n(\chi_+),
\label{eq:def_Dn}
\end{equation}
and
\begin{equation}
F_{n+1}
=
\eta_-(-\tau\alpha_-)^{n+1}P_{n+1}(\chi_-)
-\eta_+(-\tau\alpha_+)^{n+1}P_{n+1}(\chi_+).
\label{eq:def_Fn}
\end{equation}
Then Eq.~\eqref{eq:raw_combined_step} becomes
\begin{equation}
T_n^++T_n^-
=
\left(G-\frac{n+1}{a}\right)n!\,D_n
+(n+1)!\,F_{n+1}.
\label{eq:rearranged_combined}
\end{equation}
The full rearranged asymptotic kernel is
\begin{equation}
  \mathcal{K} = \Sigma_\mathcal{E} \times \Bigg\{\frac{2\tau\gamma\gamma'}{q} + \sum_{n=0}^\infty\Bigg[\Big(G-\frac{n+1}{a}\Big)n! D_n + (n+1)!F_{n+1}\Bigg]\Bigg\}
  \label{eq:asymptotic_rearranged}
\end{equation}

\subsection{Roundoff Error Estimates} \label{sec:asymptotic_roundoff}

For the asymptotic series \eqref{eq:asymptotic_rearranged} the cancellation occurs within $D_n$ and $F_{n+1}$. The per-term roundoff from each subtraction is
\begin{equation}
  \delta D_n = |(-\tau\alpha_+)^{n+1}P_n(\chi_+)| + |(-\tau\alpha_-)^{n+1}P_n(\chi_-)|
  \label{eq:delta_Dn}
\end{equation}
\begin{equation}
  \delta F_{n+1} = |\eta_+(-\tau\alpha_+)^{n+1}P_{n+1}(\chi_+)| + |\eta_-(-\tau\alpha_-)^{n+1}P_{n+1}(\chi_-)|
  \label{eq:delta_Fn}
\end{equation}
These propagate through the series coefficients, giving an accumulated roundoff after $N$ terms
\begin{equation}
  \epsilon_{\rm ro} = \epsilon_{machine}\Sigma_\mathcal{E}\times \sum_{n=0}^{N_{\mathrm{terms}}}\left(\left|\left(G-\frac{n+1}{a}\right)n!\right|\delta D_n + \left|(n+1)!\right|\delta F_{n+1}\right)\Bigg/|\mathcal{K}|
  \label{eq:asymptotic_roundoff}
\end{equation}
For the derivative \eqref{eq:dasymptotic_dT}, the weight factor $w_n = (\lambda_+ - \kappa)/\tau^2 + (n-2)/\tau$ multiplies each term, so
\begin{equation}
  \epsilon_{\rm ro}^{\partial T} = \frac{k_{\rm B}}{m_ec^2}\,\epsilon_{machine}\Sigma_\mathcal{E}\times \sum_{n=0}^{N_{\mathrm{terms}}}|w_n|\Bigg(\left|\left(G-\frac{n+1}{a}\right)n!\right|\delta D_n + \left|(n+1)!\right|\delta F_{n+1}\Bigg) \Bigg/ \left|\frac{\partial \mathcal{K}}{\partial T}\right|
  \label{eq:asymptotic_roundoff_deriv}
\end{equation}

\section{Monte-Carlo Integration} \label{sec:monte_carlo_derivation_appendix}

The Monte Carlo estimator used in Section~\ref{sec:monte_carlo_comparison}
samples the kernel-weighted moment
$\langle X\rangle(g, g', i; T)\,\sigma_{\mathrm{out}}(g\rightarrow g',i; T)$
(Eqs.~\eqref{eq:sigma_out_angle_resolved}
and~\eqref{eq:multigroup_weighted_average}) directly from its defining
integral rather than evaluating the closed-form kernel.
The procedure draws an electron Lorentz factor $\lambda$ from the Maxwell-Juttner distribution
(Eq.~\eqref{eq:def_MJ}), boosts the incident photon to the electron rest
frame, performs Klein-Nishina scattering, and boosts the scattered photon back
to the lab frame. The resulting $\xi$ and $\mathcal{E}'$
determine the target group $g'$ and angular bin $i$ for the tally.

\subsection{Coordinate Setup} \label{sec:mc_coordinate_setup}

The incoming photon direction in the lab frame is fixed along the $z$ axis,
$\vec{\Omega}=(0,0,1)$. The electron velocity direction is placed in the
$x$-$z$ plane,
$\vec{\Omega}_e=(\sin\theta_e,0,\cos\theta_e)$, where $\theta_e$ is
the polar angle of the electron direction relative to the photon. With this
choice, the incoming photon direction in the electron rest frame also lies in
the $x$-$z$ plane.

\subsection{Lorentz Transforms} \label{sec:mc_lorentz_transforms}

The Doppler factors $D$ and $D'$ are defined in Eq.~\eqref{eq:def_D}
\citep{POMRANING19721047,pomraning1973equations}. 
The rest-frame incident photon energy is
\begin{equation}
\mathcal{E}_0=\lambda D\,\mathcal{E}.
\label{eq:mc_rest_frame_energy}
\end{equation}
The rest-frame incident direction satisfies
\begin{equation}
\cos\theta_0
=
\frac{1}{\lambda D}
\left(
1-\frac{\lambda}{\lambda+1}(\lambda D+1)\beta\mu_e
\right).
\label{eq:mc_rest_frame_direction}
\end{equation}
where $\mu_e=\cos\theta_e=\vec{\Omega}\cdot\vec{\Omega}_e$.
After scattering, the scattered photon direction in the electron rest frame is
transformed back to the lab frame by
\begin{equation}
\vec{\Omega}'
=
\frac{1}{\lambda\widetilde{D}'}
\left(
\vec{\Omega}'_0
+\frac{\lambda}{\lambda+1}(\lambda\widetilde{D}'+1)\vec{\beta}
\right),
\label{eq:mc_inverse_direction_transform}
\end{equation}
where $\widetilde{D}'=1+\vec{\Omega}'_0\cdot\vec{\beta}$ and the lab-frame
scattered energy is
\begin{equation}
\mathcal{E}'=\lambda\widetilde{D}'\,\mathcal{E}'_0.
\label{eq:mc_lab_frame_outgoing_energy}
\end{equation}

\subsection{Electron Sampling} \label{sec:mc_electron_sampling}

The Maxwell-Juttner distribution (Eq.~\eqref{eq:def_MJ}) expressed with
respect to $d\lambda$ is
\begin{equation}
\mathrm{MJ}_\lambda(\lambda,\tau)
=
\frac{\beta\lambda^2\exp(-\lambda/\tau)}
{\tau K_2(1/\tau)},
\label{eq:mc_mj_lambda}
\end{equation}
Direct sampling is complicated by the $\beta$ factor. Instead, $\lambda$ is
drawn from the proposal distribution
\begin{equation}
p(\lambda)\propto \lambda^2\exp(-\lambda/\tau),
\qquad
\lambda\in[1,\infty),
\label{eq:mc_proposal}
\end{equation}
and each sample carries a correction weight $w_\beta=\beta$.
The electron direction cosine is sampled uniformly, $\mu_e=1-2r$, with
$r$ drawn from $(0,1)$.

\subsection{Rest-Frame Scattering} \label{sec:mc_rest_frame_scattering}

In the electron rest frame, the scattering-angle cosine $\xi_0$ and azimuthal
angle $\varphi_0$ are sampled uniformly on the unit sphere, $\xi_0\in[-1,1]$
and $\varphi_0\in[0,2\pi)$. The rest-frame scattered energy follows from the
recoil relation (Eq.~\eqref{eq:compton_recoil_rest}),
\begin{equation}
\mathcal{E}'_0
=
\frac{\mathcal{E}_0}
{1+(1-\xi_0)\gamma_0},
\label{eq:mc_compton_recoil}
\end{equation}
where $\gamma_0=\mathcal{E}_0/(m_ec^2)$.
The scattered direction in the rest frame is constructed from $\xi_0$ and
$\varphi_0$ in the local frame of $\vec{\Omega}_0$, then
rotated into the lab-frame electron coordinates.

Because the scattering angles are sampled uniformly rather than from the
Klein-Nishina distribution, each sample carries a Klein-Nishina tally weight.
Define
\begin{equation}
A=\frac{\mathcal{E}'_0}{\mathcal{E}_0}
=
\frac{1}{1+(1-\xi_0)\gamma_0}.
\label{eq:mc_energy_ratio_rest}
\end{equation}
The Klein-Nishina contribution for the sample is
\begin{equation}
w_{\rm KN}
=
D\,
A^2
\left(
A+\frac{1}{A}-(1-\xi_0^2)
\right).
\label{eq:mc_klein_nishina_weight}
\end{equation}
The factor $D$ arises from the Lorentz transform of the
single-electron cross section to the lab frame
\citep{POMRANING19721047,KERSHAW1986273}.

\subsection{Group Tally and Normalization} \label{sec:mc_normalization}

For each source group $g$, the incident photon energy
$\mathcal{E}$ is sampled uniformly within
$[\mathcal{E}_{g-1/2},\mathcal{E}_{g+1/2}]$ with a weight
\begin{equation}
w_{\mathcal{E}}
=
w(\mathcal{E},T),
\label{eq:mc_energy_weight}
\end{equation}
where $w(\mathcal{E},T)$ is the incident-energy weight function appearing in
Eq.~\eqref{eq:sigma_out_angle_resolved}.
After the lab-frame scattered energy $\mathcal{E}'$ and $\xi$
are determined, the sample is tallied into target group $g'$ and angle
bin $i$. The accumulated contribution from $N$ samples is
\begin{equation}
\left(\langle X\rangle\,\sigma_{\mathrm{out}}\right)_{\rm MC}(g\rightarrow g',i)
=
\frac{3}{4}\sigma_{\rm Th}
\frac{1}{N}
\frac{
\displaystyle\sum_{j\in(g',i)}
X_j\,w_{\beta,j}\,w_{\mathcal{E},j}\,w_{{\rm KN},j}
}{
\overline{w}_\beta\;\overline{w}_{\mathcal{E},g}
},
\label{eq:mc_group_cross_section}
\end{equation}
where $X_j=X(\mathcal{E}_j,\mathcal{E}'_j,\xi_j)$ is the multiplier evaluated
at the $j$-th sample,
$\overline{w}_\beta=N^{-1}\sum_j w_{\beta,j}$ is the mean velocity weight, and
$\overline{w}_{\mathcal{E},g}=N^{-1}\sum_{j\in g}w_{\mathcal{E},j}$ is the
mean energy weight for samples in source group $g$. The prefactor
$\frac{3}{4}\sigma_{\rm Th}$ is the Thomson cross-section normalization
corresponding to the Klein-Nishina cross-section in
Eq.~\eqref{eq:mc_klein_nishina_weight}. The sum in the numerator runs over
samples that land in the target bin $(g',i)$.

\subsection{Kernel Temperature-Derivative Integral} \label{sec:mc_derivative}

The temperature derivative of the kernel enters only through the
Maxwell-Juttner distribution. Differentiating the $\mathrm{MJ}_\lambda$ factor
in the integrand produces an additional per-sample weight
\begin{equation}
w_{\partial\tau,j}
=
\left[
\frac{\lambda_j-\kappa(\tau)}{\tau^2}
-\frac{3}{\tau}
\right]
w_{\beta,j}\,w_{\mathcal{E},j}\,w_{{\rm KN},j},
\label{eq:mc_derivative_weight}
\end{equation}
where $\kappa(\tau)=K_1(1/\tau)/K_2(1/\tau)$. Replacing
$w_{\beta}\,w_{\mathcal{E}}\,w_{\rm KN}$ with $w_{\partial\tau}$ in
Eq.~\eqref{eq:mc_group_cross_section} gives the kernel-derivative
contribution.
\section{Quadrature Rule Definitions} \label{sec:quadrature_rules_appendix}

Throughout the algorithms in this paper, several quadrature rules are
used. Each rule is written as $R(f,N,a,b)$, where $f$ is the integrand,
$N$ is the number of quadrature nodes, and $[a,b]$ is the integration
interval. All rules share the convention that a call with $b\le a$
contributes zero.

\textbf{GL}: Standard $N$-point Gauss--Legendre quadrature on $[a,b]$,
where $\{x_i\}$ and $\{w_i\}$ are the reference nodes and weights on
$[-1,1]$.
\begin{equation}
  \mathrm{GL}(f,N,a,b) = \frac{b-a}{2}\sum_{i=1}^{N} w_i\,f\!\left(\frac{a+b}{2} + \frac{b-a}{2}\,x_i\right).
  \label{eq:gl_formula}
\end{equation}

\textbf{LogGL}: Logarithmically mapped Gauss--Legendre rule. The
substitution $u=\ln x$ transforms the integral as
\begin{equation}
  \int_a^b f(x)\,dx
  = \int_{\ln a}^{\ln b} f(e^u)\,e^u\,du,
  \label{eq:loggl_substitution}
\end{equation}
and standard GL is applied in $u$-space on $[\ln a,\ln b]$. This
concentrates quadrature nodes near the lower bound $a$.

\textbf{RLogGL}: Reflected-log Gauss--Legendre rule. The variable is
first reflected, $y=a+b-x$, and then the logarithmic map $u=\ln y$ is
applied, giving
\begin{equation}
  \int_a^b f(x)\,dx
  = \int_{\ln a}^{\ln b} f(a+b-e^u)\,e^u\,du,
  \label{eq:rloggl_substitution}
\end{equation}
with GL applied in $u$-space on $[\ln a,\ln b]$. This concentrates
quadrature nodes near the upper bound $b$. Requires $a>0$.
When the original interval $[a_0,b_0]$ has $a_0\le 0$ (as in the
angular integration of Algorithm~\ref{alg:xi_integration}), the
substitution $s=x-a_0$ shifts the domain to $[0,\,b_0-a_0]$ and
RLogGL is applied on that positive interval, preserving the
node clustering near the upper endpoint.

\textbf{TailGL}: A dispatch rule used for the high-energy tail in the
scattered-energy quadrature (Algorithm~\ref{alg:Eprime_integration}).
For a nonempty panel, it selects the logarithmic rule when the interval
spans more than a factor of two:
\begin{equation}
\mathrm{TailGL}(f,N,a,b)
=
\begin{cases}
\mathrm{LogGL}(f,N,a,b), & b/a>2,\\
\mathrm{GL}(f,N,a,b), & b/a\leq 2.
\end{cases}
\label{eq:tailgl_definition}
\end{equation}

\textbf{MapGL}: A dispatch rule used in the incident-energy
quadrature (Algorithm~\ref{alg:E_integration}). Given the
incident-spectrum weight function $w(\mathcal{E},T)$ and a panel
$[a,b]$, MapGL selects among the three rules above:
\begin{itemize}
  \item If $b/a \le 2$ (narrow panel), use $\mathrm{GL}$.
  \item If $b/a > 2$ and $w(a,T) \ge w(b,T)$, use $\mathrm{LogGL}$.
  \item Otherwise, use $\mathrm{RLogGL}$.
\end{itemize}

\textbf{Middle}$(f,N,a,b)$: A helper used in the incident-energy
quadrature (Algorithm~\ref{alg:E_integration}). If the spectral peak
energy $\mathcal{E}_{\mathrm{pk}}(T)$ of the weight function lies inside
$[a,b]$, the interval is split at $\mathcal{E}_{\mathrm{pk}}$ and each
half is integrated with $\mathrm{MapGL}(f,N,a,\mathcal{E}_{\mathrm{pk}})$
and $\mathrm{MapGL}(f,N,\mathcal{E}_{\mathrm{pk}},b)$. Otherwise, the
full interval is integrated with a single MapGL call.

\section{Planck and Wien Group Integrals} \label{sec:group_denominators}

The group weight normalization appearing in Eq.~\eqref{eq:sigma_out_angle_resolved} is
\begin{equation}
W_g(T)=\int_{\mathcal{E}_{g-1/2}}^{\mathcal{E}_{g+1/2}} w(\mathcal{E},T)d\mathcal{E}.
\label{eq:group_denominator}
\end{equation}
For Planck weighting,
$w(\mathcal{E},T)=\mathcal{E}^3/(\exp(\mathcal{E}/k_{\rm B}T)-1)$.
Writing $x=\mathcal{E}/(k_{\rm B}T)$ so that $d\mathcal{E}=k_{\rm B}T\,dx$,
\begin{equation}
W_g(T)=(k_{\rm B}T)^4\int_{x_{g-1/2}}^{x_{g+1/2}}\frac{x^3}{e^x-1}dx.
\label{eq:planck_Wg_integral}
\end{equation}
The integral is evaluated using truncated polylogarithmic and Taylor series
\citep{CLARK1987311,Giron_2026,LEWIS2025111374}. Define the piecewise antiderivative
\begin{equation}
\Phi(x)\!=\!
\begin{cases}
\displaystyle
-\!\sum_{l=1}^{L} e^{-lx}\!\left(\frac{x^3}{l}+\frac{3x^2}{l^2}
+\frac{6x}{l^3}+\frac{6}{l^4}\right), & x > x_s,\\[8pt]
\dfrac{x^3}{3}-\dfrac{x^4}{8}+\dfrac{x^5}{60}
-\dfrac{x^7}{5040}+\dfrac{x^9}{272160}
-\dfrac{x^{11}}{13305600}+\dfrac{x^{13}}{622702080}, & x \le x_s,
\end{cases}
\label{eq:planck_antiderivative}
\end{equation}
where $x_s=2$ is the switching point. The large-$x$ branch truncates
the polylogarithm series of the exact antiderivative at $L$ terms,
while the small-$x$ branch is the term-by-term integral of the Taylor
expansion of $x^3/(e^x-1)$. Because the two branches carry different
integration constants that differ by $\pi^4/15$, the group weight is
\begin{equation}
W_g(T)=(k_{\rm B}T)^4\!
\begin{cases}
\Phi(x_{g+1/2})-\Phi(x_{g-1/2})
  +\dfrac{\pi^4}{15},
  & x_{g-1/2}\leq 2 < x_{g+1/2},\\[4pt]
\Phi(x_{g+1/2})-\Phi(x_{g-1/2}),
  & \text{else}.
\end{cases}
\label{eq:planck_Wg_eval}
\end{equation}
With $L=5$ and $x_s=2$ this achieves accuracy better than $10^{-5}$
\citep{CLARK1987311}.

For Wien weighting,
$w(\mathcal{E},T)=\mathcal{E}^3\,e^{-\mathcal{E}/k_{\rm B}T}$, the same
change of variables gives
\begin{equation}
W_g(T)=(k_{\rm B}T)^4
\left[
\mathcal{W}(x_{g+1/2})-\mathcal{W}(x_{g-1/2})
\right],
\label{eq:wien_denominator}
\end{equation}
where
\begin{equation}
\mathcal{W}(x)=6-e^{-x}(x^3+3x^2+6x+6).
\label{eq:wien_antiderivative}
\end{equation}
At small $x$, the closed form in Eq.~\eqref{eq:wien_antiderivative} suffers
from cancellation. The Taylor form
\begin{equation}
\mathcal{W}(x)=x^4\sum_{n=0}^{\infty}\frac{(-x)^n}{(n+4)n!}
\label{eq:wien_taylor}
\end{equation}
is used instead. Truncating at seven terms ($n=0,\ldots,6$) and switching at $x=0.1$ gives relative accuracy better than $10^{-9}$.

\section{Default Parameter Values} \label{sec:production_parameters}

Tables~\ref{tab:dispatch_tolerances} and~\ref{tab:quadrature_orders}
collect the numerical parameters used to produce all results in this
paper. The dispatch tolerances in
Table~\ref{tab:dispatch_tolerances} control the precision cascade in
Algorithm~\ref{alg:solver_dispatch}. The quadrature orders in
Table~\ref{tab:quadrature_orders} are the default values used in all
multigroup integrations (Algorithms~\ref{alg:xi_integration}--\ref{alg:E_integration}).
Where individual comparisons require elevated orders, the values are
stated in the corresponding text.

\begin{table}[!htbp]
\centering
\caption{Kernel solver dispatch tolerances
(Algorithm~\ref{alg:solver_dispatch}).
\label{tab:dispatch_tolerances}}
\begin{tabular}{lcc}
\hline\hline
Symbol & Value & Description \\
\hline
$\epsilon_{\rm double}$ & $10^{-7}$ & Double-precision acceptance threshold \\
$\epsilon_{\rm dd}$   & $0.5$     & Double-double acceptance threshold \\
\hline
\end{tabular}
\end{table}

\begin{table}[!htbp]
\centering
\caption{Default quadrature orders for the multigroup integration.
\label{tab:quadrature_orders}}
\begin{tabular}{lcc}
\hline\hline
Symbol & Order & Integration role \\
\hline
$N_{\rm peak}$ & 48 & $\xi$ peak panel (Algorithm~\ref{alg:xi_integration}) \\
$N_{\rm tail}$ & 48 & $\xi$ tail panels (Algorithm~\ref{alg:xi_integration}) \\
$N_{\rm edge}$ & 24 & $\mathcal{E}'$ edge/tail panels (Algorithm~\ref{alg:Eprime_integration}) \\
$N_{\rm int}$  & 24 & $\mathcal{E}'$ interior/ridge panels (Algorithm~\ref{alg:Eprime_integration}) \\
$N_{\rm el}$   & 8  & $\mathcal{E}'$ elastic-core panel (Algorithm~\ref{alg:Eprime_integration}) \\
$N_{\mathcal{E}}$ & 16 & $\mathcal{E}$ sub-panels (Algorithm~\ref{alg:E_integration}) \\
\hline
\end{tabular}
\end{table}

\section{Explicit Fifth-Order Thermal Compton Kernel}
\label{sec:g5_kernel_appendix}

Building on the endpoint-expansion framework of \citet{Sazonov2000}, this appendix derives an explicit fifth-order thermal kernel by isolating the exact Boltzmann suppression factor and expanding the remaining smooth amplitude in powers of $\tau$. With the kinematic quantities $a$, $q$, $\lambda_+$, $\rho_\pm$, and $\omega^2$ from Eqs.~\eqref{eq:def_a_q}--\eqref{eq:def_rho_pm_omega2}, define the shorthand $A=\gamma\gamma'a$.
\begin{equation}
  \mathcal{K}_{G5}
  =\frac{r_e^2\gamma'}{4\mathcal{E}}
   \sqrt{\frac{2}{\pi\tau}}\,
   \mathcal{N}(\tau)\mathcal{A}_{G5}(\tau)
   \exp\!\left[-\frac{\lambda_+-1}{\tau}\right],
  \qquad
  \label{eq:g5_kernel}
\end{equation}
The remaining prefactor is a smooth function of temperature, composed of a rational Maxwell--J\"uttner normalization $\mathcal{N}(\tau)$ and a pole-suppressed amplitude $\mathcal{A}_{G5}(\tau)$,
\begin{equation}
  \mathcal{N}(\tau)
  =\frac{1+\frac{141}{208}\tau-\frac{441}{3328}\tau^2}
         {1+\frac{531}{208}\tau+\frac{6519}{3328}\tau^2},
  \qquad
  \mathcal{A}_{G5}
  =\frac{D_{32}^{\,3}P_{32}+D_{23}^{\,3}P_{23}}
         {D_{32}^{\,4}+D_{23}^{\,4}} .
  \label{eq:g5_normalization_amplitude}
\end{equation}
To extend the local fifth-order expansion beyond the strictly cold limit, the thermal series is continued by two complementary Pad\'e approximants, combined in the symmetric, pole-suppressing blend of Eq.~\eqref{eq:g5_normalization_amplitude}. The $[3/2]$ continuation has
\begin{equation}
  D_{32}=1+b_1\tau+b_2\tau^2,
\end{equation}
\begin{equation}
  P_{32}=C_0+(C_1+b_1C_0)\tau
    +(C_2+b_1C_1+b_2C_0)\tau^2
    +(C_3+b_1C_2+b_2C_1)\tau^3,
\end{equation}
\begin{equation}
  \begin{pmatrix}C_3&C_2\\C_4&C_3\end{pmatrix}
  \begin{pmatrix}b_1\\b_2\end{pmatrix}
    =-\begin{pmatrix}C_4\\C_5\end{pmatrix},
  \label{eq:g5_pade_32}
\end{equation}
and the $[2/3]$ continuation has
\begin{equation}
  D_{23}=1+e_1\tau+e_2\tau^2+e_3\tau^3,
\end{equation}
\begin{equation}
  P_{23}=C_0+(C_1+e_1C_0)\tau
    +(C_2+e_1C_1+e_2C_0)\tau^2,
\end{equation}
\begin{equation}
  \begin{pmatrix}
  C_2&C_1&C_0\\C_3&C_2&C_1\\C_4&C_3&C_2
  \end{pmatrix}
  \begin{pmatrix}e_1\\e_2\\e_3\end{pmatrix}
    =-\begin{pmatrix}C_3\\C_4\\C_5\end{pmatrix}.
  \label{eq:g5_pade_23}
\end{equation}
Both ratios reproduce the same fifth-order series,
\begin{equation}
  \frac{P_{32}}{D_{32}}
  =\frac{P_{23}}{D_{23}}
  =\sum_{n=0}^{5}C_n\tau^n+\mathcal{O}(\tau^6).
  \label{eq:g5_series_matching}
\end{equation}
The coefficients $C_n$ are obtained by expanding the endpoint integrand (Eq.~11 of \citealt{KERSHAW1986273}) about the minimum-energy electron configuration and evaluating the resulting Laplace moments term by term. Each is a finite combination of kinematic quantities,
\begin{equation}
  \begin{aligned}
  C_n={}&\frac{2\delta_{n0}}{q}
  +B\!\left[F_n^{(1)}(\rho_-)-F_n^{(1)}(\rho_+)\right]\\
  &+\frac{d_-F_n^{(3)}(\rho_-)+d_+F_n^{(3)}(\rho_+)}{A^2}
  +\frac{n(\gamma+\gamma')}{A^2}
   \left[F_{n-1}^{(3)}(\rho_-)+F_{n-1}^{(3)}(\rho_+)\right],
  \quad n=0,\ldots,5,
  \end{aligned}
  \label{eq:g5_coefficients}
\end{equation}
where
\begin{equation}
  d_-=\frac{\rho_+^2-\rho_-^2-q^2}{2},\qquad
  d_+=\frac{\rho_+^2-\rho_-^2+q^2}{2},\qquad
  B=\frac{A^2-2A-2}{A^2},\qquad
  F_{-1}^{(3)}=0,
  \label{eq:g5_auxiliary}
\end{equation}
and
\begin{equation}
  F_n^{(p)}(\rho)
  =n!\sum_{j=\lceil n/2\rceil}^{n}
  \binom{-p/2}{j}\binom{j}{n-j}
  (2\rho)^{2j-n}
  (\rho^2+\omega^2)^{-p/2-j},
  \qquad p\in\{1,3\},\quad 0\leq n\leq5 .
  \label{eq:g5_finite_sum}
\end{equation}

Because the kernel factorizes into temperature-dependent and temperature-independent parts, the analytic temperature derivative follows from the logarithmic derivatives of the Pad\'e blend $\mathcal{A}_{G5}$, the normalization $\mathcal{N}$, the factor $\tau^{-1/2}$, and the exact Boltzmann exponential,
\begin{equation}
  \frac{\partial\mathcal{K}_{G5}}{\partial\tau}
  =\mathcal{K}_{G5}\left[
    \frac{\mathcal{A}_{G5}'}{\mathcal{A}_{G5}}
    +\frac{\mathcal{N}'}{\mathcal{N}}
    -\frac{1}{2\tau}
    +\frac{\lambda_+-1}{\tau^2}\right],
  \label{eq:g5_kernel_derivative}
\end{equation}

\begin{figure*}[!htbp]
  \centering
  \subfloat[Multigroup matrix error vs.\ temperature\label{fig:g5_max_error_vs_T}]{%
    \includegraphics[width=0.49\textwidth]{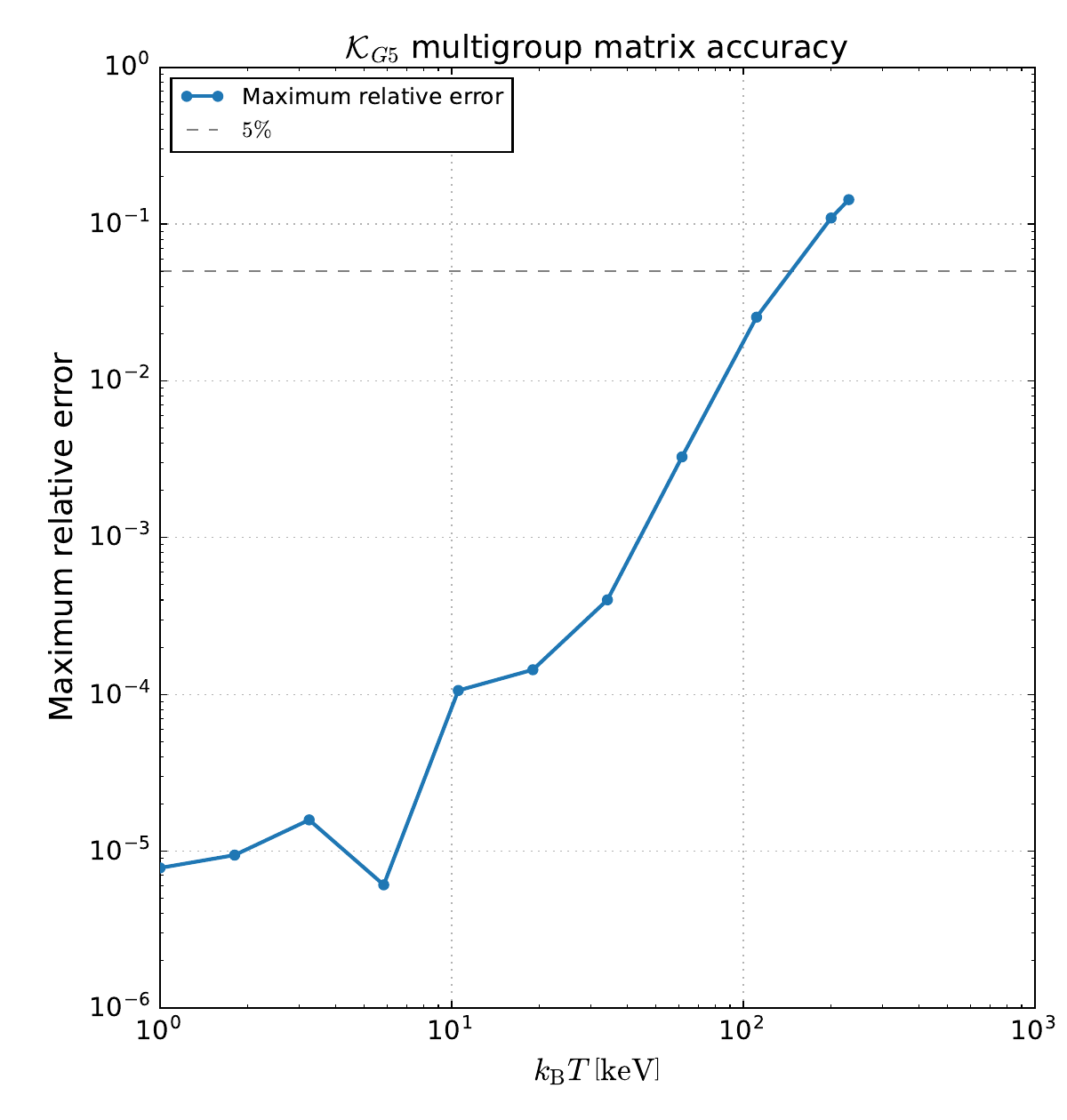}}
  \hfill
  \subfloat[Pointwise kernel comparison at $k_{\rm B}T=100\,{\rm keV}$\label{fig:g5_kernel_profile_comparison}]{%
    \includegraphics[width=0.49\textwidth]{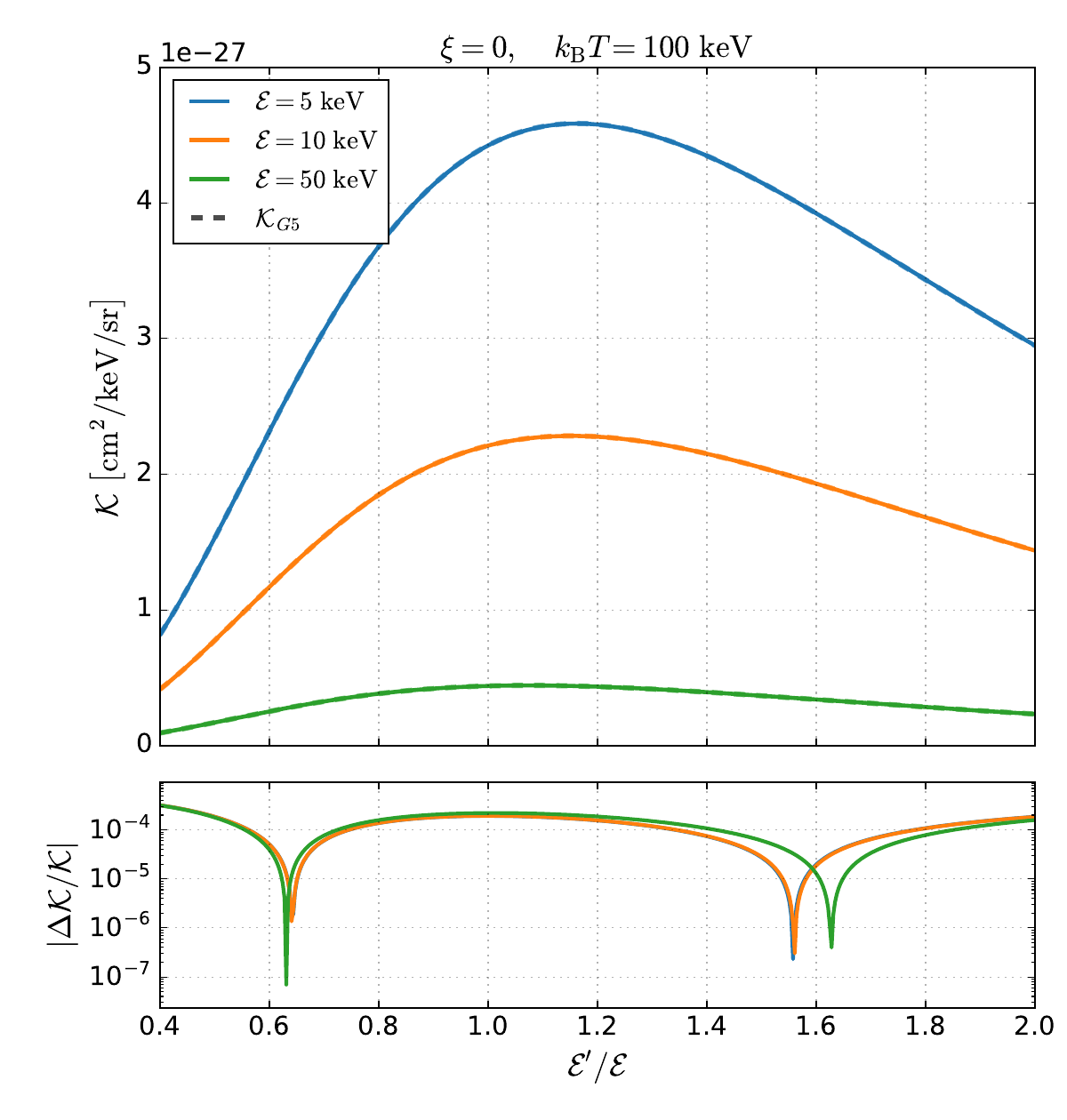}}
  \caption{Accuracy of the fifth-order kernel $\mathcal{K}_{G5}$. (a)~Maximum relative error per matrix entry of the multigroup-multiangle scattering matrix ($32$ logarithmically spaced energy groups from $0.1$ to $1000\,\mathrm{keV}$, eight angle bins, uniform weighting). The horizontal dashed line marks $5\%$. (b)~Pointwise comparison of $\mathcal{K}_{G5}$ (dashed) against the exact evaluation (solid) at $\xi=0$ for three incident energies. The lower panel shows $|\Delta\mathcal{K}/\mathcal{K}|$.}
  \label{fig:g5_accuracy}
\end{figure*}

This closed-form alternative accelerates the multigroup-multiangle integration by $3$--$10\times$ with a maximum relative error per matrix entry below $5\%$ for $k_{\rm B}T \lesssim 100\,{\rm keV}$, tested on a $32$-group geometric grid from $0.1$ to $1000\,{\rm keV}$ with eight angle bins. Figure~\ref{fig:g5_accuracy} shows this error as a function of temperature (left panel) and compares the pointwise kernel against the exact evaluation at $k_{\rm B}T = 100\,{\rm keV}$ for three incident photon energies (right panel).

Table~\ref{tab:g5_stress_calibration} lists the maximum relative error per matrix entry and the row-sum error across ten temperatures up to $200\,{\rm keV}$.

\begin{table}[!htbp]
\centering
\caption{Error metrics of the multigroup-multiangle scattering matrix computed with $\mathcal{K}_{G5}$ on a $32$-group geometric grid ($0.1$--$1000\,\mathrm{keV}$, eight angle bins, uniform weighting).
\label{tab:g5_stress_calibration}}
\begin{tabular}{ccc}
\hline\hline
$k_{\rm B}T$ (keV) & Max relative error & Row-sum error \\
\hline
$1.0$   & $7.82 \times 10^{-6}$  & $1.18 \times 10^{-6}$ \\
$1.8$   & $9.44 \times 10^{-6}$  & $1.73 \times 10^{-6}$ \\
$3.2$   & $1.58 \times 10^{-5}$  & $1.32 \times 10^{-6}$ \\
$5.8$   & $6.10 \times 10^{-6}$  & $1.36 \times 10^{-6}$ \\
$10.5$  & $1.06 \times 10^{-4}$  & $1.39 \times 10^{-5}$ \\
$19.0$  & $1.44 \times 10^{-4}$  & $1.68 \times 10^{-5}$ \\
$34.2$  & $4.01 \times 10^{-4}$  & $1.98 \times 10^{-5}$ \\
$61.6$  & $3.27 \times 10^{-3}$  & $3.78 \times 10^{-5}$ \\
$111.0$ & $2.55 \times 10^{-2}$  & $3.02 \times 10^{-4}$ \\
$200$   & $1.09 \times 10^{-1}$  & $1.99 \times 10^{-3}$ \\
\hline
\end{tabular}
\end{table}

\bibliography{bib}{}
\bibliographystyle{aasjournal}

\end{document}